\documentclass[apj]{emulateapj}
\usepackage{apjfonts}
\usepackage{lscape}
\usepackage{epsfig}
\usepackage{subfigure}
\usepackage{amsmath}
\usepackage{amssymb}
\usepackage{amsfonts}
\usepackage{natbib}          
\lefthead{Paladini et al.}
\righthead{Characterization of dust in evolved \ion{H}{2} regions}

\begin{document}

\title{Spitzer and Herschel multiwavelength characterization of the dust content of evolved \ion{H}{2} regions}

\author{
R. Paladini\altaffilmark{1,*}, 
G. Umana\altaffilmark{2},
M. Veneziani\altaffilmark{3}, 
A. Noriega-Crespo\altaffilmark{3}, 
L. D. Anderson\altaffilmark{4}, 
F. Piacentini\altaffilmark{5}, 
D. Pinheiro Gon\c{c}alves\altaffilmark{6},
D. Paradis\altaffilmark{7},  
C. Tibbs\altaffilmark{8}
J.-P. Bernard\altaffilmark{7}, 
P. Natoli\altaffilmark{9}
}

\altaffiltext{*}{Corresponding author: paladini@ipac.caltech.edu}
\altaffiltext{1}{
Nasa Herschel Science Center, California Institute of Technology,
              1200, E. California Blvd., Pasadena, CA 91125,  USA
}

\altaffiltext{2}{
INAF - Osservatorio Astrofisico di Catania,
         Via S. Sofia 78, 95123 Catania, Italy
}

\altaffiltext{3}{
Infrared Processing and Analysis Center, California Institute of Technology, 
             1200, E. California Blvd., Pasadena, CA 91125,  USA
}

\altaffiltext{4}{
Department of Physics, West Virginia University, 
              Morgantown, WV 26506, USA
}

\altaffiltext{5}{
Dipartimento di Fisica, Universita' di Roma La Sapienza,
          Roma, Italy
}

\altaffiltext{6}{
Department of Astronomy and Astrophysics, University of Toronto
          50 George Street, Toronto, ON, M5S 3H4, Canada
}

\altaffiltext{7}{
Centre d'Etude Spatiale des Rayonnements, 9 Avenue du Colonel Roche, 
31028 Toulouse Cedex 4, France
}

\altaffiltext{8}{
Spitzer Science Center, California Institute of Technology, 
          1200, E. California Blvd., Pasadena, CA 91125,  USA
}

\altaffiltext{9}{
Istituto Nazionale di Fisica Nucleare, Sezione Ferrara, 
      Ferrara, Italy
}

\begin{abstract}

We have analyzed a uniform sample of 16 evolved \ion{H}{2} regions located in a 2$^{\circ}\times$2$^{\circ}$ Galactic field centered at ({{\em{l,b}}}) = (30$^{\circ}$, 0$^{\circ}$) and observed as part of the Herschel Hi-GAL survey.
The evolutionary stage of these  \ion{H}{2} regions was established using ancillary radio continuum data. By combining Hi-GAL PACS (70 $\mu$m, 160 $\mu$m) and SPIRE (250 $\mu$m, 350 $\mu$m and 500 $\mu$m)
measurements with MIPSGAL 24 $\mu$m data, we built Spectral Energy Distributions (SEDs) of the sources and showed that a 2-component grey-body model is a good representation of the data.
In particular, wavelengths $>$ 70 $\mu$m appear to trace a cold dust component, for which we estimated an equilibrium temperature of the
Big Grains (BGs) in the range 20 - 30 K, while for $\lambda <$ 70 $\mu$m, the data indicated the presence of a warm dust component 
at temperatures of the order of 50 - 90 K. This analysis also revealed that dust is present in the interior of  \ion{H}{2} regions, although likely not in a large amount. 
In addition, the data appear to corroborate the hypothesis that
the main mechanism responsible for the (partial) depletion of dust in \ion{H}{2} regions is radiation-pressure-driven drift. In this framework, we speculated that the 
24 $\mu$m emission which spatially correlates with ionized gas might be associated with either Very Small Grain (VSG) or BG replenishment, as recently proposed for the case of Wind-Blown Bubbles (WBB). 
Finally, we found that evolved \ion{H}{2} regions are characterized by distinctive far-IR and sub-mm colors, which can be used
as diagnostics for their identification in unresolved Galactic and extragalactic regions. 
\end{abstract}


\section{Introduction}

 \ion{H}{2} regions are sites of ionized gas surrounding OB associations and represent powerful laboratories for understanding 
the mechanisms regulating massive star formation. Their evolutionary sequence 
spans different phases, from the hyper-compact (HC\ion{H}{2}) phase, in which the newly-formed star is still fully embedded in its dust cocoon, to the 
evolved one, when most of the natal material has been ejected from the birth site. Intermediate
phases are represented by ultra-compact (UC\ion{H}{2}) and compact \ion{H}{2} regions. 
Each of these stages is characterized by critical values of electron density, n$_{e}$, and linear diameter (Kurtz 2005): 
HC\ion{H}{2} have extremely high electron density (n$_{e} > 10^{6}$
cm$^{-3}$) and small sizes (diameters $<$ 0.05 pc), UC\ion{H}{2} and compact \ion{H}{2} regions have lower n$_{e}$
(n$_{e} \sim$ 10$^{3}$ - 10$^{4}$ cm$^{-3}$) and larger diameters ($<$ 0.1 - 0.5 pc), while evolved \ion{H}{2} regions present 
even lower electron densities (n$_{e} \sim$ 10$^{2}$ cm$^{-3}$) and diameters of the order of several pc. 

The emission properties
also vary with the evolutionary stage. The radio emission is dominated, in all cases, by free-free emission associated with ionized gas. However, HC\ion{H}{2},
UC\ion{H}{2} and, occasionally (depending on n$_{e}$) compact \ion{H}{2} regions, have an optically thick spectrum for frequencies from a few to several GHz. On the contrary,
the radio spectrum of an evolved \ion{H}{2} region is optically thin already at $\sim$ 1 GHz. Both young and old sources 
present radio recombination lines (RRL), although HC\ion{H}{2} are characterized by substantially broader lines, likely due to pressure
broadening and gas bulk motions. All evolutionary phases, with the exception of HC\ion{H}{2} (Kurtz 2005), are very bright at FIR and sub-mm wavelengths. This 
emission has often been ascribed to the presence of dust grains. However, it is not clear whether dust is directly associated with \ion{H}{2} regions or rather with their surrounding Photodissociation Regions (PDRs, see Hollenbach $\&$ Tielens (1997) for 
a complete review), since 
the historical lack of resolving power in the existing IR data has long prevented us from distinguishing these two components. 
This elusive separation has led many studies to use large photometric  
apertures which often included both the actual \ion{H}{2} region and the PDR, with the consequence
that the derived properties were a global average over intrinsically widely different regions. In fact, while an \ion{H}{2} region is by definition 
the region in space where the gas is fully ionized, the PDR is mainly neutral, characterized by an A$_{V}$ $>$ 1. In addition, the
intensity of the radiation field, as density, is not the same inside or outside the \ion{H}{2} region, as both scale with distance from the ionizing OB association. 

This situation has dramatically changed with the release of high-sensitivity, high-resolution radio (MAGPIS, Helfand et al. 2006) and IR 
(GLIMPSE, Benjamin et al. 2003; MIPSGAL survey, Carey et al. 2009; Hi-GAL, Molinari et al. 2010) data, which allows the clear identification 
of an \ion{H}{2} region from its PDR, thus offering the unprecedented possibility of carrying out systematic investigations of the specific properties of each of  
these environments. An example of what is now  possible is demonstrated by the work of Povich et al. (2007). These authors used GLIMPSE and MSX (Midcourse Space Experiment, Price et al., 2001) data to perform
a multiwavelength study of the M17 complex, and found that the SED from the \ion{H}{2} region peaks at shorter
wavelengths and has a qualitatively different shape than the SED from the PDR. They also showed that different gas and dust components are mostly distributed 
in mutually exclusive regions:  hot plasma occupies the very inner part of the \ion{H}{2} region; warm ionized gas defines the cavity walls;
warm dust permeats the ionized gas; and Polycyclic Aromatic Hydrocarbons (PAHs) surround the whole star forming complex, tracing the PDR. 

In this paper, we make use of the newly available data to investigate the IR properties of the interior of evolved \ion{H}{2} regions. We also address the question - from the observational point of 
view - whether dust grains survive the extreme conditions (i.e. intense UV photon flux and radiation pressure, stellar winds and high temperatures) which characterize this 
type of environments.  
Remarkably, evolved \ion{H}{2} regions, due to their large physical size, are ideal candidates for carrying out such analysis, as 
they allow us to take full advantage of the high spatial 
resolution (e.g. from a few to tens of arcsecs) of these new data sets. 

The paper is organized as follows. In Section~2 and 3 we illustrate the data base and the source selection criteria. 
In Section~4, we generate source SEDs and describe the modelling procedure. In Section~5, we discuss our expectations, based on observations and 
theoretical predictions, of finding dust associated with a warm (10$^{4}$ K) gas. In Section~6 we show   
the relative spatial distribution of different populations of grains inferred by our multiwavelength data set. 
In Section~7 we compute the IR excess for each source and discuss its implications. In Section~8 we create color-magnitude and color-color plots, and show how these can be used for the identification of unresolved 
\ion{H}{2} regions with respect to other classes of sources. We provide final remarks and conclusions in Section~9.

\section{The Data}

The bulk of the data is from the Herschel Hi-GAL survey. 
The Herschel infrared Galactic Plane Survey (Hi-GAL; Molinari et al. 2010) consists in the 
observations of the first and fourth Galactic quadrants using the PACS (Poglitsch et al. 2010) and SPIRE (Griffin et al. 2010) instruments. 
The observations were carried out in parallel mode 
at 70 $\mu$m and 160 $\mu$m (PACS), and at 250 $\mu$m, 350 $\mu$m and 500 $\mu$m (SPIRE), with an angular resolution from 6" (70 $\mu$m) to  
35" (500 $\mu$m). During the Herschel Science Demonstration Phase (SDP), two Hi-GAL fields were observed.
The fields are centered, for {\em{b}} = 0$^{\circ}$, at {\em{l}} = 30$^{\circ}$ and {\em{l}} = 59$^{\circ}$, and extend
2 degrees in both latitude and longitude. The data were processed with the ROMAGAL pipeline (Traficante et al., 2011), which allows for an accurate 
reconstruction of both compact and extended emission. Since the SDP data were taken early in the mission, corrective factors specific to extended emission were applied. 
In particular, PACS 70 $\mu$m and 160 $\mu$m data were divided by 1.05 and 1.29, while SPIRE 250 $\mu$m, 350 $\mu$m and 500 $\mu$m data were multiplied, respectively, by 1.02, 1.05 and 0.94. 
For the calibration uncertainties, we assumed 20$\%$ for both PACS and 15$\%$ for SPIRE (Bernard et al. 2010). 

We complemented the far-IR and sub-mm Herschel observations,
with MIPS 24 $\mu$m (6") and IRAC 8 $\mu$m (2") data obtained by the MIPS Galactic Plane Survey (MIPSGAL) and 
by the Galactic Legacy Infrared Mid-Plane Survey Extraordinaire (GLIMPSE) survey. We adopted a 10$\%$ calibration error 
at both wavelengths (S. Carey, {\em{private communication}}). To overcome the occasional problem (e.g. the core of G39.8-0.3) posed by hard-saturated pixels in the MIPS 24 $\mu$m data, we replaced the MIPS 24 $\mu$m data with   
the combined MIPS 24 $\mu$m and MSX 21 $\mu$m images produced by the MIPSGAL team and available to the members of the consortium. These images were generated taking into account appropriate color corrections and 
wavelength scaling (S. Carey, {\em{private communication}}). 

We also made use of archival single-dish 6 cm data (see Section~3), which provided the basis for our source selection, and
allowed us to constrain the evolutionary phase of the \ion{H}{2} regions
in the sample. In addition, we used the Multi-Array Galactic Plane Imaging Survey (MAGPIS) high-resolution 20 cm data for a more accurate determination of 
the boundary of the \ion{H}{2} regions and to perform our photometric measurements (Section~4). 

\section{Source selection and definition of the sample} 

Since the launch of Spitzer in August 2003 (Werner et al. 2004) and then of Herschel in May 2009 (Pilbratt et al. 2009), a steady flow 
of papers addressed multi-wavelength high-resolution investigations of Galactic \ion{H}{2} regions. Apart for a few exceptions (e.g. the aforementioned Povich et al. (2007) work on M17, or the investigation of 
Flagey et al. (2011) of M16), all these studies 
(e.g. Watson et al., 2008; Deharveng et al. 2010; Zavagno et al. 2010; etc.) targeted bubble \ion{H}{2} regions only. These sources are typically characterized by a common IR ''structure'': 
8 $\mu$m emission, dominated by PAH features, encloses the bubble; the bubble itself is filled with 24 $\mu$m emission, and this spatially correlates with free-free emission associated 
with ionized gas. In this work, one of our goals is to investigate if it is possible to extend and generalize to other classes of \ion{H}{2} regions the results obtained from the analysis of bubbles. For this purpose, 
we focused our attention on \ion{H}{2} regions spanning a variety of morphological types. In order to assure uniformity of the sample, we required, for the selected sources, the availability of ancillary data allowing 
us to assess their evolutionary stage. 

The starting point of our selection was the catalog of Paladini et al. (2003). While for
29$^{\circ} <$ {\em{l}} $<$ 31$^{\circ}$, $|b| <$ 1$^{\circ}$, the catalog lists 30 \ion{H}{2} regions, for
58$^{\circ} <$ {\em{l}} $<$ 60$^{\circ}$, $|b| <$ 1$^{\circ}$, we found only two sources. Therefore
we concentrated our analysis on the Hi-GAL SDP field centered at {\em{l}} = 30$^{\circ}$. Out of the 30 cataloged sources falling in the l = 30$^{\circ}$ field, we made 
a further selection, by favoring those sources for which coordinates are known at least at the arcmin precision, and an estimate of
the distance is available. We note that the distance information is central to our study, since it not only allows us to 
compute the luminosity of the sources, but also makes it possible to generate the linear 
longitude and latitude profiles discussed in Section~6. The final source selection consists of 16 objects, including well studied  \ion{H}{2} regions such as N49 (G288-0.2, Churchwell et al. 2006; Everett $\&$ Churchwell 2010; Draine 2011) and 
W43 (G30.8-0.3, Bally et al. 2010). Table~1 gives details relative to each source, namely: source name (Column 1); Galactic coordinates 
(Column 2 and 3); 6 cm flux (Column 4); Galactocentric distance (R, Column 5); heliocentric distance (D, Column 6); far distance solution in case of distance ambiguity (D$_{f}$, Column 7); 
electron temperature (T$_{e}$, Column 8); emission measure derived from single-dish (EM$_{6cm}$, Column 9) and MAGPIS measurements (EM$_{20cm}$, Column 10); emission measure obtained assuming shell-like geometry 
of the source (EM$_{shell}$, Column 11); electron density computed from single-dish (n$_{e,6cm}$, Column 12) and MAGPIS measurements (n$_{e,20cm}$, Column 13); electron density computed assuming shell-like geometry 
of the source (n$_{e,shell}$, Column 14). Table~2 provides additional information on angular sizes ($\theta$) and linear diameters (d). 

6 cm fluxes and angular diameters are from Altenhoff et al. (1979),
Downes et al. (1980) and Kuchar $\&$ Clark (1997). Galactocentric and solar distances are, for all the sources with the exception of three cases (G29.0-0.6, G30.2-0.1, G30.5+0.4), from
Anderson $\&$ Bania (2009, hereafter AB09), based on Lockman (1989) RRL observations. For these sources, AB09 solved the kinematic distance ambiguity, using
HI emission/absorption and self-absorption data. We adopted their recommended values. In the case of G29.0-0.6, AB09 suggested that the source is situated at the 
far distance (11.5 kpc), although they obtained contradictory results from the HI emission/absorption data (near solution) with respect to the self-absorption data (far solution). In their paper, 
the authors stated that the HI emission/absorption measurements are usually more reliable to discriminate between near and far solution. However, for this specific 
source, the self-absorption data are of better quality, so they assigned it to the far distance. We visually inspected the $^{13}$CO data from the Galactic Ring Survey (GRS, Simon et al. 2001) and noted that   
G29.0-0.6 appears to form a coherent structure with a companion source, G29.1-0.7, for which the likely source of ionization (the B1 II type star S65-4) has a known photometric distance 
of 3.6 kpc (Forbes 1989). For this reason, we  re-located G29.0-0.6 to the near distance (3.4 kpc). 
For two sources (G30.2-0.1 and G30.5+0.4) AB09 did not solve the distance ambiguity. For these, we computed the Galactocentric and solar distances
using the McClure-Griffiths $\&$ Dickey (2007) rotation model, taking R$_{0}$ = 8.5 kpc and
$\theta_{0}$ = 220 km/sec, combined with the Lockman (1989) RRL measurements. 
The near and far solar distances for these two sources are quoted in Table~1 and both are used for the derivations of d, T$_{e}$, EM and n$_{e}$. 
For four sources the value of T$_{e}$ reported in the table is from Downes et al. (1980 - G30.2-0.1, G30.6-0.1, G30.7-0.3) 
and Quireza et al. (2007 -  G30.8-0.0). For the other sources, 
a crude estimate of T$_{e}$ is obtained by applying the relation (Deharveng et al. 2000): 

\begin{equation}
T_{e} = (372 \pm 38) R + (4260 \pm 350)
\end{equation}

For all the sources EM is computed from Mezger $\&$ Henderson (1967):

\begin{equation}
T_{b} = 8.235 \times 10^{-2} a T_{e}^{-0.35} \nu^{-2.1} EM
\end{equation}

In equation~(2), T$_{b}$ (the observed brightness temperature) and T$_{e}$ are in K, $\nu$ is in GHz, and $a$ is a constant set equal to 1. EM, combined with the linear diameter, is
used to derive n$_{e}$ from:

\begin{equation}
n_{e} = \sqrt{\frac{EM}{d}}
\end{equation}

In the expression above, we assumed that n$_{e}$ is uniform across the source and that this is spherically symmetric. However, most of the sources 
in our sample do not have a spherical morphology. In Section~6 we will return on this point and dicuss its implications. 
Because EM has a weak dependence on T$_{e}$ (EM $\propto$ T$_{e}^{-0.35}$), we do not expect the application of the empirical relation in equation~(1) to bias significantly the 
derivation of this quantity and, in turn, of the electron density. We note that all the emission measures quoted in Table~1 are of the order of 10$^{2}$ cm$^{-3}$, suggesting 
that our sample is homogeneous and consists only of evolved \ion{H}{2} regions. 

Finally, we computed, from the single-dish measurements, the Ly({$\alpha$}) luminosity of each source:

\begin{eqnarray}
L(Ly \alpha) = 3.20 \times 10^{2}  \left(S_{\nu}\over{Jy}\right) \left(T_{e}\over{10^{4} K}\right)^{-0.45} \cdot \\ \nonumber
              \cdot  \left(\nu\over{\rm GHz}\right)^{0.1} \left(D\over{kpc}\right)^{2} L_{\odot}  \nonumber
\end{eqnarray}

(Garay et al. 1993) and the number of ionizing photons needed to excite the \ion{H}{2} region:

\begin{eqnarray}
N_{Lyc} = 6.3 \times 10^{52} \hbox{photons}\,\hbox{s}^{-1} \left(T_{e}\over{10^{4} K}\right)^{-0.45} \cdot \\ \nonumber
 \hspace*{-1truecm} \cdot \left({\nu\over{\rm GHz}}\right)^{0.1} {L_{\nu}\over 10^{27}\,\hbox{erg}\,\hbox{s}^{-1}\,\hbox{Hz}^{-1} } \nonumber
\end{eqnarray}

(Condon 1992). In the relations above, $L_{\nu}$ = $S_{\nu} 4\pi D^{2}$, and $S_{\nu}$, $D$ and $T_{e}$ are, respectively, the 6 cm flux, heliocentric distance and electron temperatures 
which appear in Column 4, 6/7 and 8 of Table~1. L(Ly{$\alpha$}) and N{$_{Lyc}$} are reported in Table~6. The sources in our sample have ionizing luminosities in the 
range $\sim$ 10$^{48}$ -- 10$^{50}$ photons s$^{-1}$, with a mean value of 10$^{49.1 \pm 0.54}$ photons s$^{-1}$. According to Smith, Biermann $\&$ Mezger (1978), \ion{H}{2} regions with 
an ionizing luminosity at least 4 times greater than the Orion Nebula (i.e. 4$\times$10$^{49}$ photons s$^{-1}$ at  R$_{0}$ = 10 kpc) are defined as {\em{giant}}. 
When properly re-scaled by a factor 0.85 for R$_{0}$ = 8.5 kpc (McKee $\&$ Williams 1997), the thereshold for giant \ion{H}{2} regions becomes equal to 3.4$\times$10$^{49}$ photons s$^{-1}$, indicating 
that our sample consists of two giant \ion{H}{2} regions (G29.1+0.4 and G30.8-0.0) and 14 sub-giant, many of which (at least 7) are close to the giant regime.

\section{SEDs}

\subsection{Evaluating the SEDs} 

The single-dish archival data used to establish
the evolutionary stage of our sources have too coarse a spatial resolution for allowing the accurate definition of the 
boundaries of the ionized-gas dominated region within each source. To this end, we made use instead the MAGPIS 20 cm data, 
which are characterized by an angular resolution of 6'', i.e. comparable to the PACS 70 $\mu$m resolution. For each \ion{H}{2} region, we drew MAGPIS contours at 3-$\sigma$ above the 
background level and used these to evaluate the extent of the actual \ion{H}{2} region. 

The IR wavelengths we are considering, as well as being contributed to by dust emission, are contaminated, especially in the PACS and SPIRE bands, by 
additional free-free (e.g. thermal bremstrahlung) emission. To evaluate this contribution, we need to take into account the dependence 
on both frequency and electron temperature of the free-free spectral index, $\beta_{ff}$ (S $\sim$ $\lambda^{\beta_{ff}}$). This is because the 
canonical approximation $\beta_{ff}$ = 0.1 holds true only up to 10 GHz (3cm) and electron temperatures of the order of 8000 K. At higher frequencies (shorter wavelengths) and 
lower electron temperatures, the 
spectral index can differ from this value by a few to tens of a percent. Therefore, to each source we applied the relation (Bennett et al. 1992):

\begin{equation}
\beta_{ff,\lambda} =  {{1}\over{11.68 + 1.5 \ln (T_{e}/8000 K) + \ln (\lambda/m)}}
\end{equation}
 
We used the values of T$_{e}$ reported in Table~1, together with the assumption that the abundance of ionized He (He++) is
negligible with respect to the abundance of ionized H (H+). Table~3 provides, for each source, $\beta_{ff}$ in the range 24 $\mu$m -- 500 $\mu$m. 
We did not compute $\beta_{ff}$ 
at 8 $\mu$m, since in the mid-IR free-free transitions become sub-dominant with respect to free-bound transitions and these, in turn, are negligible
when compared to dust emission (Beckert, Duschl $\&$ Mezger, 2000). From Table~3, we see that, in the range 24 $\mu$m to 500 $\mu$m, $\beta_{ff}$ varies between a minimum of 1.7 and a maximum of 0.2, 
thus the subtraction of the free-free contribution becomes relatively important (of the order of a few percent) 
only at the longer SPIRE wavelengths. For each source, the free-free corrected flux at a given wavelength $\overline{\lambda}$, $\widetilde{S_{\overline{\lambda}}}$, 
was obtained by subtracting from the IR flux, S$_{\overline{\lambda}}$, the MAGPIS 20 cm flux rescaled to $\overline{\lambda}$ using:

\begin{equation}
\widetilde{S_{\overline{\lambda}}} = S_{\overline{\lambda}} \hspace*{0.1truecm} - \sum_{\lambda = 20 cm }^{\overline{\lambda}} {S_{20 cm} * \left(\frac{20 \times 10^{5}}{\lambda}\right)^{\beta_{ff,\lambda}}}
\end{equation}

with $\lambda$ in $\mu$m. Noteworthily, the single-dish fluxes quoted in Table~1, properly rescaled to
20 cm using a canonical free-free spectral index 0.1, are within $\sim$ 20$\%$ of the MAGPIS measurements.   

After free-free subtraction, using the apertures obtained from the 20 cm MAGPIS data, we estimated the flux and built 
the SEDs by combining the MIPS 24 $\mu$m (and when, necessary, the MIPS/MSX data) with the PACS and SPIRE data. We also tried to include the 
1.1 mm data, at 33" resolution, from the Bolocam Galactic Plane Survey (BGPS, Rosolowsky et al. 2010). However, the SED modelling revealed a 
signifcant flux loss at this wavelength (Figure~1 and Figure~2). The origin of this effect can be attributed to the BGPS data processing, which largely (at more than 90$\%$ level) removes 
emission on scales of or greater than 5.9'. This operation is necessary in order to separate the astronomical signal from the atmospheric fluctuations in the data streams. 
For structures larger than
3.8', the attenuation is of the order of 50$\%$ (Aguirre et al. 2011). The 16 \ion{H}{2} regions in our sample have an average angular size
of 3.2', hence still in the range in which the extended emission should be preserved for the most part. Therefore, our result appears to indicate that the amount of filtered flux is 
more than expected, at least on the scales we are considering. We could compensate for this effect by artificially increasing the systematic error. Instead, we opted for a 
more conservative approach and decided to exclude the 1.1 mm data from the fit. We note that 
the loss of signal on large angular scale would have been comparable or even greater if, rather than the BGPS data set, we had used the 19" resolution ATLASGAL (Schuller et al. 2009) data at 870 $\mu$m. In this case, a 
significant fraction of flux is removed from the data already on scales as small as 2.5', that is smaller than the average size of the \ion{H}{2} regions in our sample.

For 24 $\mu$m $< \lambda <$ 500 $\mu$m, the flux at each wavelength and for each \ion{H}{2} region was determined within the defined apertures
after point source subtraction -- performed with he Starfinder algorithm (Diolati et al., 2000) -- and convolution to the SPIRE 500 $\mu$m angular resolution (35''). 
To evaluate the background level, we took the median within a control region, selected to be close to the source under consideration, yet paying attention
to avoid potentially contaminating bright objects. Uncertanties on measured fluxes ($\widetilde{\sigma_{S}}$), prior to free-free subtraction, were obtained by adding in quadrature
the calibration error and the background standard deviation, according to:

\begin{equation}
\widetilde{\sigma_{S}} = \left(\sum_{i=1}^{N}{\widetilde{\sigma_{{S}_{i}}}}^2 + N^2 \sigma_{\overline{back}}^2\right)^{1/2}
\end{equation}

where $\widetilde{S_{i}}$ is the flux in pixel $i$, $N$ is the number of pixels within the aperture, and $\overline{back}$ is the median
background in the control region. To this error, due to the subtraction of the free-free contribution, we added in quadrature an extra 30$\%$ uncertainty associated with the
MAGPIS data (e.g. Povich et al. (2007)). We also removed the zodiacal light contribution from the 24 $\mu$m data. This operation was carried out using the entry 'ZODY$\_$EST' in the FITS header 
of each 24 $\mu$m frame, where the amount of zodiacal light per frame was estimated by the Spitzer Science Center from the predictions on Kelsall et al. (1998). 

The photometric measurements performed after background subtraction retrieved, at every wavelength and for each source, a positive flux (see Table~4). The 
average S/N (Signal-to-Noise) across all the bands and for the entire sample is $\overline{S / N_{\lambda_{all}}}$ = 3.6$\pm$2.1. If we break this number per wavelength, we have:  
 $\overline{S / N_{8 \mu m}}$ = 4$\pm$ 2.3, $\overline{S / N_{24 \mu m}}$ = 6.7$\pm$ 1.9, $\overline{S / N_{70 \mu m}}$ = 3.3$\pm$ 0.9, $\overline{S / N_{160 \mu m}}$ = 2.4$\pm$ 1, 
$\overline{S / N_{250 \mu m}}$ = 2.9$\pm$ 1.6, $\overline{S / N_{350 \mu m}}$ = 2.6$\pm$ 1.4 and $\overline{S / N_{500 \mu m}}$ = 2.7$\pm$ 1.5. We note that, as a general trend, the significance 
of the detection tends to decrese with increasing wavelength. This is likely due to the fact that at $\lambda \sim$ 160 $\mu$m, the emission from the \ion{H}{2} region starts competing with 
the emission of the Interstallar Medium (ISM). At longer wavelengths (i.e. in SPIRE bands), this contribution becomes sub-dominant with respect to the local background. 

It is worth mentioning that the detected IR emission that we ascribed entirely to \ion{H}{2} regions could instead originate from warm/cold material located in the foreground/background along the line of 
sight. Indeed this bias at least partly affects our flux extraction procedure. In fact, in the presence of
a structured, highly-varying background such as the one characterizing the Galactic Plane, background-subtraction does not assure the complete removal of emission unrelated to the source, 
due to the intrinsic difficulty of estimating the background level. However, what provides confidence in our results is the fact that, for the majority of the sources, the MAGPIS 20 cm emission, which traces the 
ionized gas component of our \ion{H}{2} regions, does not (or not completely) overlap with the peak of the IR emission in the other bands (likely associated with the PDR),  
suggesting that the \ion{H}{2} region and its dust content is largely (or partly) exposed to the observer.

\subsection{SED Fitting} 

Figure~1 and 2 show the SEDs derived from the photometric measurements described in the previous section. Since the pioneering works of Chini et al. (1986a, 1986b, 1986c, 1987), it is known that,  
in order to explain observations of \ion{H}{2} regions above and below $\sim$ 100 $\mu$m in the framework of traditional dust models, that is with a constant spectral emissivity index $\beta_{dust}$, 
it is necessary to invoke the existence of a 2-temperature component dust distribution: a warm, low density population
of dust grains situated in the proximity of the ionizing source, and a colder dust population far from the central star (or stars).  
One has to keep in mind that the observations that Chini et al. used to constrain their modelling were taken in the 
early Eighties and, as such, were characterized 
by relatively low angular resolution, with the consequence that the actual \ion{H}{2} region was not resolved with respect to the associated PDR. 

Both laboratory experiments and observations conducted in recent years suggest that traditional dust models assuming a constant spectral emissivity 
index might need to be revised (see discussion below). For this reason, we tried to fit our SEDs in three possible ways: (1) with an isothermal 
modified blackbody with constant spectral emissivity index ($\beta_{dust}$ = 2); (2) with an isothermal modified blackbody 
with variable $\beta_{dust}$; (3) with a 2-temperature component model with $\beta_{dust}$ =2. We did not attempt the fit with a 2-temperature component model with variable $\beta_{dust}$ 
since the measurements would not provide enough constrains (6 data points) with respect to the number of free parameters (6 parameters). We also did not include in the fit the IRAC 8 $\mu$m band, which is 
typically contributed to by emission associated with stochastically heated PAHs. To properly take into account PAH emission, we would need, for each \ion{H}{2} region in our sample, an accurate knowledge of the radiation field
generated by the central source, and this information should be coupled to a sophisticated dust model (e.g. DustEM, Compiegne et al. 2011) able to reproduce, in a coherent way, both the aromatic
features and the continuum thermal emission due to larger grains. Since detailed information on the radiation field is 
not available for most of our \ion{H}{2} regions, we considered only the data at wavelengths longer than 24 $\mu$m. The three functional forms corresponding to our three adopted models are:

\begin{equation}
\widetilde{S_{\lambda,1}} = A_{1,1} \left(\frac{\lambda}{\lambda_{0}}\right)^{-2} B_{\lambda}(T_{dust,1})
\end{equation}

for an isothermal model with constant emissivity index;

\begin{equation}
\widetilde{S_{\lambda,2}} = A_{1,2} \left(\frac{\lambda}{\lambda_{0}}\right)^{-\beta_{dust}} B_{\lambda}(T_{dust,2})
\end{equation}

for an isothermal model with variable $\beta_{dust}$, and:

\begin{equation}
\widetilde{S_{\lambda,3}} = A_{1,3} \left(\frac{\lambda}{\lambda_{0}}\right)^{-2} B_{\lambda}(T_{c,3}) + A_{2,3} \left(\frac{\lambda}{\lambda_{0}}\right)^{-2} B_{\lambda}(T_{w,3})
\end{equation}

for a 2-temperature component model, where T$_{c,3}$ and T$_{w,3}$ are, respectively, the temperatures of the cold and warm components, and $\lambda_{0}$ is set 
to 100 $\mu$m.

The choice of $\beta_{dust}$ = 2 is motivated by the fact that observations of the silicate absorption feature near 10 $\mu$m seem to indicate that BGs are amorphous materials (Kemper et al. 2004), 
and these are expected to behave in a similar fashion to crystalline dielectric materials, for which sub-mm absorption has a temperature-independent quadratic dependence on frequency (Boudet et al. 2005). 

To estimate the parameters A$_{1,1}$, T$_{dust,1}$, A$_{1,2}$, T$_{dust,2}$, A$_{2,3}$, T$_{c,3}$ and T$_{w,3}$ for each model (equation~(9), (10) and (11)), we used a Monte Carlo Markov Chain (MCMC) 
method (Lewis $\&$ Bridle, 2002).
This technique consists in looking for the likelihood maximum by sampling the parameters space with a Metropolis-Hastings algorithm. The a priori probability densities of the parameters were set to be 
as wide as possible to avoid introducing artificial biases in the fit. Color corrections were 
computed iteratively during the fitting procedure using the MIPS, PACS and SPIRE transmission filters. In addition, systematic (i.e. calibration) uncertainties were taken into account 
and included in the parameters error bars by means of a dedicated Monte Carlo run. A straightforward $\chi^{2}$ goodness-of-fit method was applied in parallel with the MCMC method to test the quality of the fits. In general, 
the best-fit values recovered by the two methods (MCMC and $\chi^{2}$ goodness-of-fit) were found in agreement within a few $\%$. 

\begin{figure*}[h]
\centering
 \includegraphics[width=17.5cm,height=16cm,angle=0]{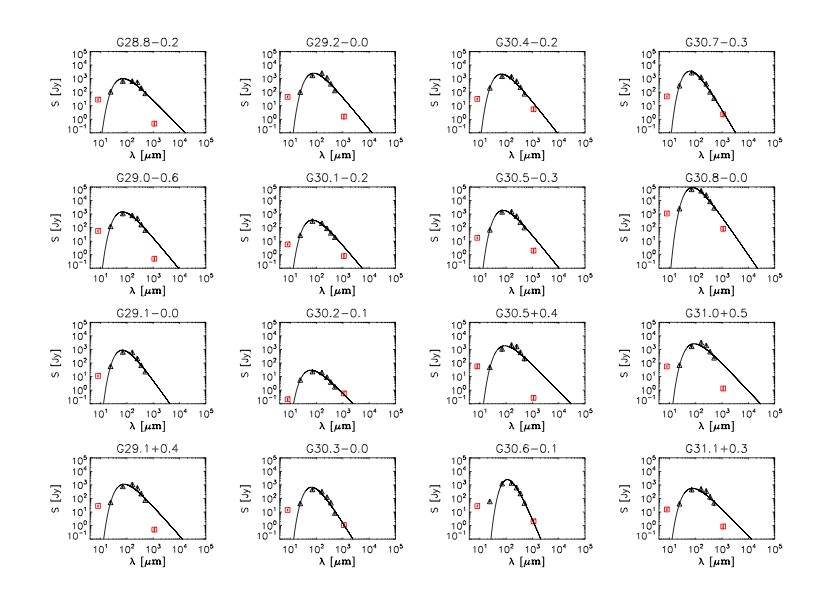}
   \caption{SEDs for all the \ion{H}{2} regions in our sample. Triangles denote photometric measurements at 24 $\mu$m (MIPS), 70 $\mu$m and 160 $\mu$m (PACS), 
250 $\mu$m, 350 $\mu$m and 500 $\mu$m (SPIRE). Red squares indicate the IRAC 8 $\mu$m and Bolocam 1.1 mm data points. Also shown are photometric uncertainties.  
The solid line corresponds to the best-fit isothermal model with varying emissivity spectral index.}
\end{figure*}

\begin{figure*}[h]
\centering
 \includegraphics[width=17.5cm,height=16cm,angle=0]{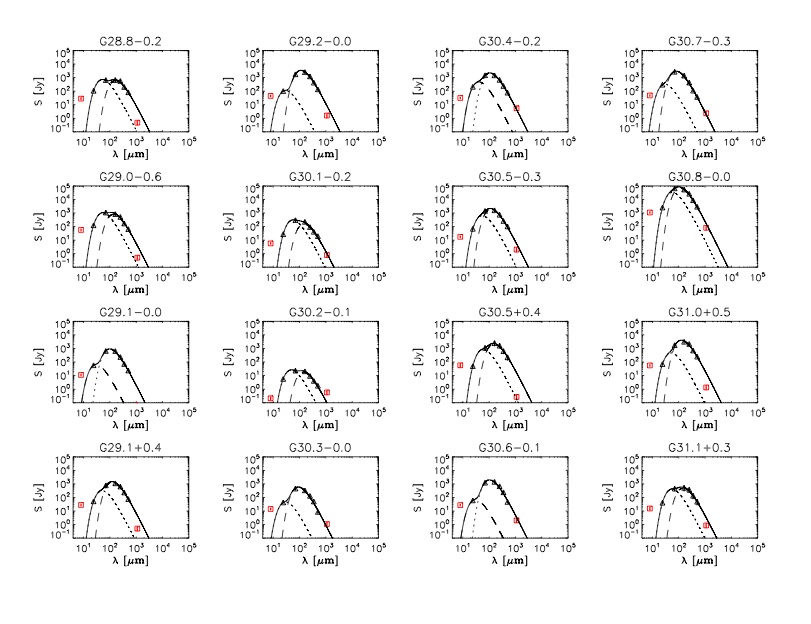}

   \caption{SEDs for all the \ion{H}{2} regions in our sample. Triangles denote photometric measurements at 24 $\mu$m (MIPS), 70 $\mu$m and 160 $\mu$m (PACS), 
250 $\mu$m, 350 $\mu$m and 500 $\mu$m (SPIRE). Also shown are the  IRAC 8 $\mu$m and Bolocam 1.1 mm data points (red squares) and photometric uncertainties. 
The solid line indicates the best-fit 2-temperature dust component model with $\beta_{dust}$ =2. Dashed lines illustrate the warm and cold components.
}
\end{figure*}

The MCMC results of the fitting procedure for the isothermal model with varying spectral emissivity index and the 2-temperature component model are given in Table~5. 
We do not show the results of the isothermal model with fixed $\beta_{dust}$ as this model fails to reproduce simultaneously the 24 $\mu$m - 500 $\mu$m observations. 
For the remaining two models, we made trial fits by including/excluding the 1.1 mm BGPS measurements. In both cases, the best-fit parameters (i.e. the best $\chi^{2}$) 
were obtained  by excluding the 1.1 mm Bolocam extracted flux, as discussed in the previous section. A direct comparison between the isothermal and 2-temperature component models in terms
of $\chi^{2}$ could not be perfomed, since the degrees of freedom are not constant. However, a comparison in terms of the Cumulative Density Functions (CDF){\footnote{The CDF (or {\em{1-p-value}}) 
gives the probability that a random variable $X$ 
with a given probability distribution will be found at a value less or equal than x.} for the $\chi^{2}$ distribution of each model model revealed, for all the 
sources but three, that the 2-temperature component model is significantly more in agreement with the observations (see Column 7 and 13 in Table~5). For G30.1-0.2, the 1-temperature model seems to work 
surprisingly better than the 2-temperature one, although the best-fit $\beta_{dust}$ (0.32) 
lies outside the normally accepted range of values. In the case of G30.3-0.0 and G30.6-0.1 none of the models appears to be a good representation of the measured SEDs. 
For the 13 sources for which the 2-temperature dust component model is in agreement with the data, 
we obtain that the cold component peaks between 100 $\mu$m and 160 $\mu$m and is characterized by a temperature in the range $\sim$ 20 - 30 K, 
while the warm component has a peak around 20 $\mu$m - 70 $\mu$m with a temperature from $\sim$ 50 to $\sim$ 90 K. The average values for the cold and warm components are,
respectively, $\overline{T_{c,3}}$ = 24.53$\pm$5.1 K, and $\overline{T_{w,3}}$ = 68.43$\pm$20.52 K. We note that $\overline{T_{c,3}}$ does not differ significantly from 
temperatures measured in PDRs associated to WBB (e.g. Rodon et al. 2010, Anderson et al. 2010). In principle, both cold and warm components could be attributed to BGs. 
This is indeed possible only if the warm and cold dust emissions are not spatially correlated, 
since a scenario in which two populations of large grains, subject to the same radiative field, thermalize
at very different temperature is rather unrealistic. An alternative solution, for spatially coincident components, would be to assign 
the cold component to a population of BGs, and the warm component to a population of VSGs. This option would be in agreement with the increase
of VSG abundance in ionized environments reported (see Section~5) by Paradis et al. (2011) and Flagey et al. (2011), and which can be explained through partial sputtering
of the BGs in the most inner parts of the \ion{H}{2} regions. Without a more refined dust modeling which can take into
account the radiation field properties of the sources, we cannot unambiguously conclude on this point.

For the 2-temperature component model, we used a constant $\beta_{dust}$. However, an increasing amount of evidence  
obtained both from laboratory experiments (Agladze et al. 1996, Mennella et al. 1998, Boudet
et al. 2005) and observations targeting the Cosmic Microwave Background (Dupac et al. 2003, Desert et al. 2008, Paradis et al. 2010; Veneziani et al. 2010) appears to support a 
temperature dependence (as well as a wavelength dependence, e.g. Paradis et al. 2011b) of the spectral emissivity index. Theoretically, the existence of an inverse relation between
dust temperature and $\beta_{dust}$ could be interpreted in light of the model recently proposed by Meny et al. (2007). This model, which was built over the Two Level System (TLS) 
theory first formulated by Phillips (1972) and Anderson et al. (1972), explains the observed T$_{dust}$ - $\beta_{dust}$ anti-correlation as due to both the disordered charge distribution characterizing amorphous materials 
and TLS (or {\em{tunneling}}) effects which occur on quantum scale in the grains structure.

To analyze the possibility of a variation of $\beta_{dust}$ with temperature, we considered the cold component only, which we fitted with a modified blackbody. 
For the warm component, we do not have enough measurements to constrain the fit with three free parameters ($\beta_{dust}$, A$_{2,3}$, T$_{w,3}$). Since  for the majority
of the sources (10 out of 16) the 70 $\mu$m flux seems to contribute to the cold component (Figure~2), we included all the measurements in the range 70 $\mu$m $< \lambda <$ 500 $\mu$m. 

The major complication in investigating a potential inverse relation between the spectral emissivity index and
dust temperature is that these two quantities are intrinsically degenerate in parameter space. In fact, their 2-D posterior probability has an elongated, slant shape, and is characterzed by the
same functional form of the anti-correlation that we are trying to identify in the data (Shetty et al. 2009). In order to overcome this issue, 
we adopted the following procedure. We first fitted the measured fluxes using the MCMC algorithm, which allowed us to recover the joint posterior
distribution of the parameters, thus keeping under control their intrinsic degeneracy. 
We then applied a technique successfully tested on BOOMERanG data (Veneziani et al. 2010), consisting in estimating the T$_{dust}$ - $\beta_{dust}$ relation through a Monte Carlo of the calibration 
errors. That is, for each iteration {\em{j}} (100 in total): (1) we fitted equation~(10) to each source and derived the best-fit values of the parameters; (2) we then estimated the relation: 

\begin{equation}
\beta_{dust} = A \times \left(T_{dust}\over{20 K}\right)^{\alpha}
\end{equation}

\noindent
by combining the results obtained for all the sources; (3)  at the end of the 100$^{th}$ iteration, we had a set of 100 pairs of $A_{j}$ and $\alpha_{j}$ values, 
one for each realization of the calibration error. By marginalizing 
over the calibration errors\footnote{This operation is performed using the GetDist software of the public CosmoMC package.}, we obtained the final best guess for $A$ and $\alpha$: A = 3.0 $\pm$ 0.2, 
$\alpha$ = -1.5 $\pm$ 0.2. 
At first sight, this result might seem to confirm
the existence of an inverse relation between spectral emissivity index and temperature. However, several authors (Masi et al. 1995, Shetty et al. 2009) 
argued that the observed anti-correlation might be spurious and due to line-of-sight temperature mixing effects. To test
this hypothesis, we performed a simple simulation. We generated a synthetic random distribution of isothermal (30K) \ion{H}{2} regions and ISM, for which we adopted a dust temperature of 17 K 
(Boulanger $\&$ Perault, 1988b). For both the \ion{H}{2} regions and the ISM, we took $\beta_{dust}$ =2. 
With the same strategy described above, consisting in applying the MCMC technique for both fitting the sources SEDs and
estimating the best-fit values A and $\alpha$, we obtained:  A = 3.1 $\pm$ 0.1 , $\alpha$ = -1.1 $\pm$ 0.1. These values are very similar to 
the ones derived from the 16 evolved \ion{H}{2} regions. If, instead of $\beta_{dust}$ = 2, we adopt a slightly lower value, such as $\beta_{dust}$ = 1.7, then
the best-fit anti-correlation for the simulated sample almost perfectly ovelarps with the inverse relation estimated from the real sample (see Figure~3).
We interpret this results as a strong indication that the observed anti-correlation is, at least in this case, likely due to a line-of-sight
temperature mixing effect.

\begin{figure}[h]
\hspace*{-1truecm}
\includegraphics[width=10cm,height=8.5cm,angle=0]{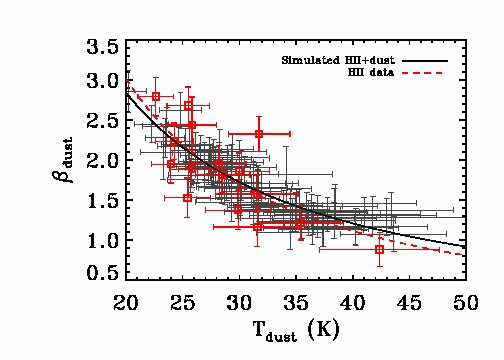}
\caption{Simulated sample of a random distribution of \ion{H}{2} regions and ISM (grey points and error bars) overlaid on the observed \ion{H}{2} region sample (red points and error bars). 
The grey solid line is the best-fit function for the synthetic sample, while the red dashed line illustrates the best-fit model for the 16 \ion{H}{2} regions considered in this work.} 
\end{figure}

\section{Can dust survive in \ion{H}{2} regions ?}

The presence of dust in the warm ionized gas has been extensively debated in the literature, due to its relevance for our understanding of the processes of destruction and formation of dust grains 
in the ISM. Given the extreme conditions characterizing the interior of 
\ion{H}{2} regions, theoretical models predict the depletion or partial destruction of the grains in these environments. Three main mechanisms could be responsible for 
this phenomenon (Inoue 2002): (1) radiation pressure; (2) stellar wind by the central source; (3) dust sublimation.  
The effect of radiation pressure was investigated by Gail $\&$ Sedlmayr (1979b), and more recently by Draine (2011). According to Gail $\&$ Sedlmayr (1979b), radiation 
pressure is capable of generating a central zone of dust low-density (or {\em{cavity}}) of size of the order of 20$\%$ the ionization radius. Gas cavities are instead known to be produced 
by stellar winds (Comeron 1997). However, given the coupling between dust and gas, a corresponding 
dust depletion zone is also expected. Sublimation appears to be the least effective process. In fact, if on one side, 
sublimation of dust grains could indeed occur as a consequence of the intense radiation field characterizing the most inner parts of an \ion{H}{2} region, yet the estimated radius of the cavity 
induced by this mechanism is only of the order of 10$^{-4}$ pc (Mookerjea $\&$ Ghosh 1999). 

From the observational point of view, one of the main findings of Povich et al. (2007) is the significant decrease of PAHs inside M17, which is thought to be due to destruction of the
aromatic molecules due to the extreme ultraviolet (EUV) flux. Indication of PAH depletion associated with ionized gas is also reported by Paradis et al. (2011) for 
the Large Magellanic Cloud. These authors also found that the relative abundance of VSGs with respect to BGs appears to increase when transitioning from
the neutral to the ionized medium. Flagey et al. (2011) arrived at a similar conclusion by analyzing the bright 24 $\mu$m emission
filling up the interior of the M16 \ion{H}{2} region, which they argued was caused by sputtering of BGs into VSGs. Finally, while the existence of BGs in neutral PDRs is supported 
by several studies (e.g. Povich et al. 2007; Compiegne et al. 2007), their survival in the interior of \ion{H}{2} region is rather controversial and so far 
it is corroborated only by statistical analysis, such as cross-correlation of large data set tracing both IR and free-free emission (Lagache et al. 2000; 
Paladini et al. 2007, Planck Collaboration, 2010).

\section{IR Emission Distribution Inside and Around \ion{H}{2} regions}

In this section, we address the question of {\em{if}} and what kind of dust is present in \ion{H}{2} regions, and investigate 
the interplay between the neutral PDR and the inner \ion{H}{2} region. Following the same procedure outlined in 
Section~4, we used the MAGPIS 20 cm contours to trace the ionized gas content of the inner \ion{H}{2} region, and IRAC 8 $\mu$m, MIPS 24 $\mu$m,
PACS 70 $\mu$m, and SPIRE 250 $\mu$m data to trace PAHs, VSGs and BGs potentially associated with both the PDR and the  \ion{H}{2} region.  
With this radio/IR data set, we created two types of 3-color images (Appendix A). The first set of images (left panels in Figure~A1) is obtained by combining MAGPIS 20 cm (red), MIPS 24 $\mu$m (green) 
and IRAC 8 $\mu$m (blue) data, while the second set (right panels in Figure~A1) is generated from the composite MAPGPIS 20 cm (red), SPIRE 250 $\mu$m (green), PACS 70 $\mu$m (blue) data. 
A visual inspection of these figures reveals a wide range of morphologies of the sample \ion{H}{2} regions, namely: bubbles (e.g. G28.8-0.2, G29.0-0.6, G31.1+0.3); 
elongated structures (e.g. G29.1+0.4, G30.5-0.3); diffuse nebulae (e.g. G30.1-0.2, G30.6-0.1); and more complex shapes (e.g. G30.8-0.0, G31.0+0.5).

\begin{figure}[h]
\centering
\includegraphics[width=7cm,height=6.2cm,angle=0]{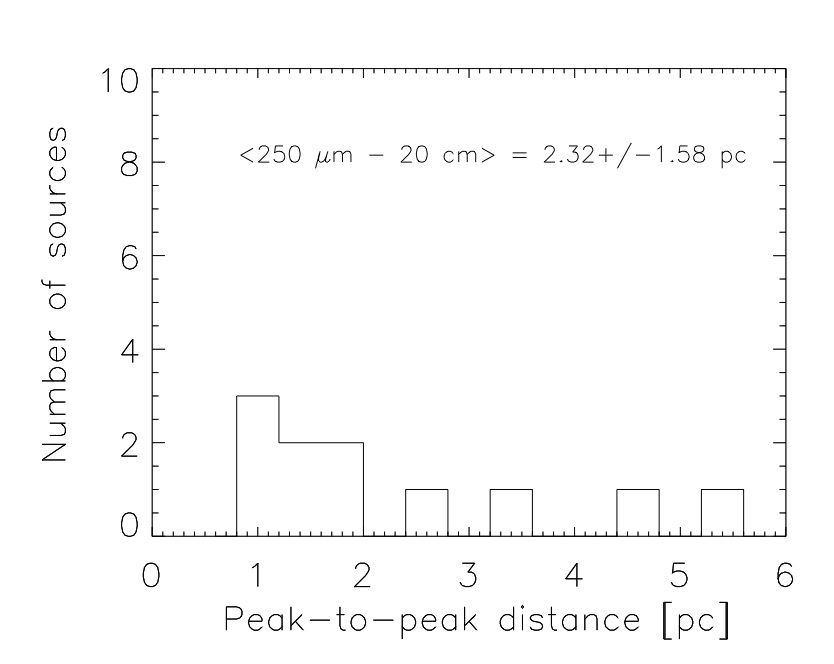}\\
\vspace*{0.3truecm}
\includegraphics[width=7cm,height=6.2cm,angle=0]{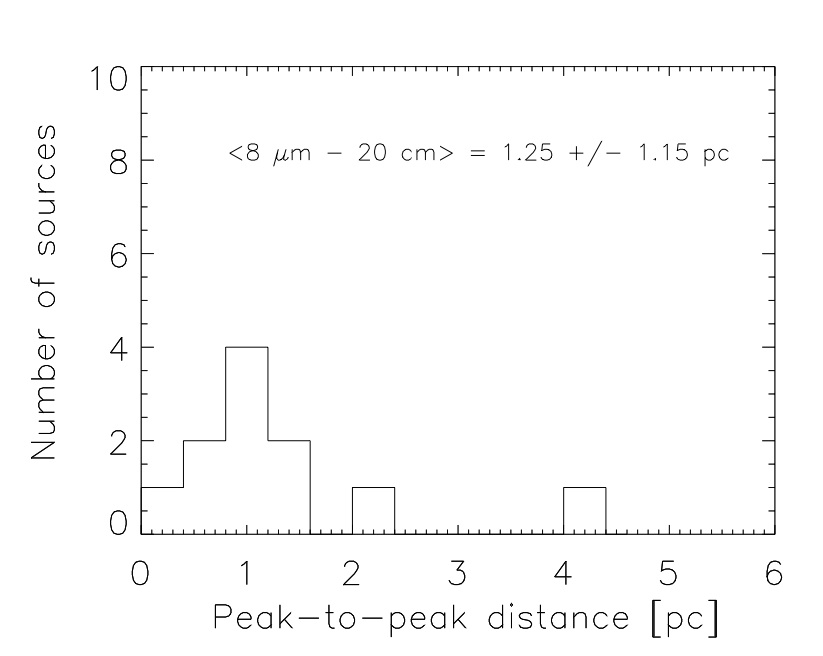}\\
\vspace*{0.3truecm}
\includegraphics[width=7cm,height=6.2cm,angle=0]{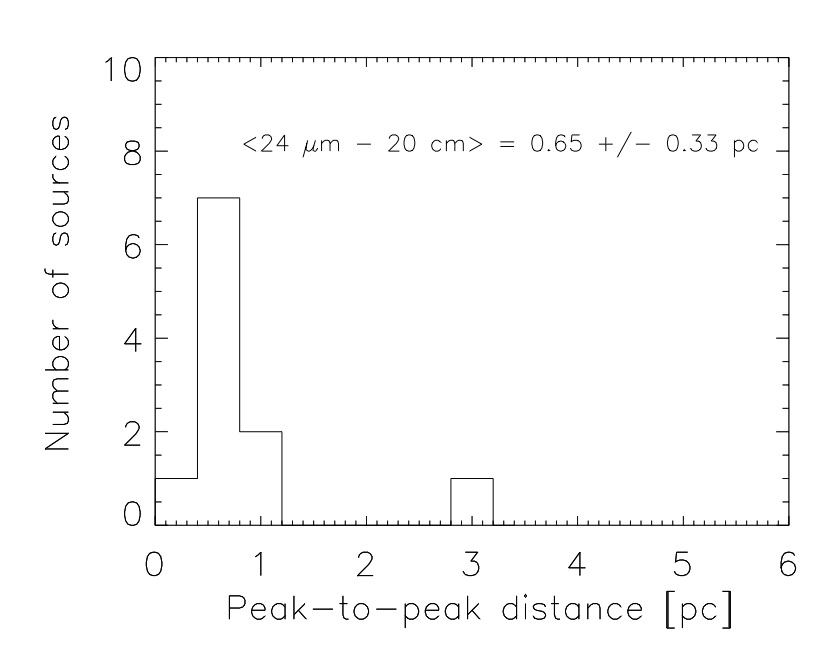}
\vspace*{0.3truecm}
\caption{Top to bottom panel: distributions of the linear separation between the peaks of emission of, respectively, SPIRE 250 $\mu$m, IRAC 8 $\mu$m, MIPS 24 $\mu$m
with respect to MAGPIS 20 cm.
}
\end{figure}

Despite the complexity posed by dealing with different geometries, we identified 3 general trends in the color images: (1) the radio and 24 $\mu$m emission ovelap in several cases; 
(2) the 8 $\mu$m and 70 $\mu$m emission are spatially correlated; (3) the 250 $\mu$m emission distribution is frequently confused with the 
background level. In regions where the ISM emission is more attenuated, the peak of the 250 $\mu$m emission is often displaced with respect to the 20 cm peak, although one 
can find instances of partial overlap. 

To investigate even further the relation between dust and ionized gas, we generated latitude and longitude emission profiles. For this purpose, we used MAGPIS 20 cm, IRAC 8 $\mu$m, MIPS 24 $\mu$m and SPIRE 250 $\mu$m 
data. The profiles were obtained following these guidelines: (1) for each source and at each wavelength, we created a postage-stamp image, either 10'$\times$10', 15'$\times$15' or 
20'$\times$20', depending 
on the angular extent of the source; (2) we convolved the postage-stamp images at the same angular resolution (18''); (3) we regridded the images using a 6" pixel size; (4)  
we generated the actual profiles by computing the median along each column (latitude cut) or row (longitude cut) of the convolved and regridded postage-stamp images; 
(5) at each wavelength we normalized each latitude/longitude profile to its maximum; (6) finally, for every source, we overlaid the profiles at different wavelengths. 
The resulting profiles (see Appendix~B) highlight the relative spatial distributions of the various components of dust emission already evidenced by the 3-color images: the 8 $\mu$m and 250 $\mu$m 
emission peak almost exclusively outside 
the gas-dominated region, while the prominent emission at 24 $\mu$m appears to be closely associated with the \ion{H}{2} region. To quantify this effect, 
we used the latitude and longitude profiles to measure the linear separation between the peaks of emission 
at different wavelengths. To this end, we excluded all ambiguous cases, i.e. \ion{H}{2} regions for which the identification of the source with respect 
to the background, either in the latitude or longitude profile, is not straightforward. This is the case of G29.1-0.0, G29.2-0.0 and 
G31.1+0.3 (latitude profile), G30.1-0.2 (both latitude and longitude profiles), G30.5+0.4 (longitude profile). For the remaining sources, we computed from the profiles 
the angular distance of the IRAC 8 $\mu$m/SPIRE 250 $\mu$m/MIPS 24 $\mu$m peak of emission with respect tp the MAGPIS 20 cm peak. With an estimate of the 
angular separation for each pair of wavelengths, we took the average of the values obtained from all the latitude and longitude profiles, 
and used the distance of the source to convert such an average into a linear distance (in pc). Figure~4 shows that the average linear separation  
between the peak of the 20 cm emission with respect to the peak of the 24 $\mu$m emission is half the 
distance between the peak of the same radio continuum emission and the 8 $\mu$m emission and this, in turn, is also half the separation between the MAGPIS peak and 
the 250 $\mu$m one. If we assume, as customary, that the 8 $\mu$m data trace PAH emission and the 250 $\mu$m data trace BG emission, we can read this result as a strong indication  
that PAHs and BGs are not spatially correlated, with the PAHs statistically distributed along a ridge which appears to be closer to the ionizing source than the 
BGs ridge. This stratification is consistent with the scenario recently described in Draine (2011), in which radiation-pressure-driven drift is very effective in moving grains 
with a size $a >$ 0.01 $\mu$m (BGs) outward, while smaller grains (PAHs) also drift but more slowly. Remarkably, this is also in agreement with Krumholz $\&$ Matzer (2009) who claim 
that radiation pressure is usually unimportant for \ion{H}{2} regions ionized by a small number of stars but  becomes a dominant factor for the expansion dynamics of large \ion{H}{2} regions. 
What remains as very puzzling is the source of the 24 $\mu$m emission. If this wavelength traced a population of VSGs co-eval with respect to the BGs and PAHs at the periphery of the 
\ion{H}{2} regions, we would expect from the predictions of Draine (2011) to find the emission peak between the 8 $\mu$m and the 250 $\mu$m peak. Likewise if the 
24 $\mu$m emission was due to a co-eval distribution of warm BGs. We then tentatively attribute this emission to a new generation of either VSGs or BGs. 
We note that this conclusion is similar to what was advocated by Everett $\&$ Churchwell (2010) from the analysis of the behaviour of the 24 $\mu$m emission in WBB: that is, within bubbles,   
dust grains are re-supplied by destruction of embedded, dense cloudlets of interstellar material that are overrun by the expansion of the bubble itself.

The latitude and longitude profiles also evidence that some of the sample \ion{H}{2} regions are very prominent in emission at all wavelengths, while 
others are barely visible above the background level. This might indicate that different evolutionary stages are present in our sample. Although
all the \ion{H}{2} regions we are considering are evolved, some might be more evolved than others and, in particular, close to the stage where the internal pressure equalizes the external one,
like, for instance, G30.1-0.2. Alternatively, as suggested by the recent MHD simulations by Arthur et al. (2011), highly irregular morphologies
might be associated with highly ionized, density-bounded nebulae, powered by the hottest stars. Certainly, a mixture of both scenarios is also a possibility. 

Both the 3-color images and the profiles clearly
show that the vast majority of the \ion{H}{2} regions we are analyzing are not characterized by spherical symmetry. Therefore, the angular diameters (and corresponding linear diameters)
obtained from gaussian fits of the single-dish data (Column 2 of Table~2) have to be considered only as a rough indication of the actual size of the sources. We then used
the MAGPIS 20 cm contours to measure the angular size of each source along its major ($\theta_{maj}$) and minor ($\theta_{min}$) axis (Colum 4 and 5 of Table~2). The analysis of the latitude and longitude profiles also reveals that the 20 cm emission of 6 \ion{H}{2} regions
(G29.0-0.6, G29.1+0.4, G30.3-0.2, G30.8-0.0, G31.1+0.5, G31.5+0.3) is characterized by a double-peak behavior, suggesting that these objects might present a shell-like geometry. If this was the case, the electron densities  
derived by assuming that a much larger volume is occupied by the ionized gas would be highly underestimated. Therefore, we computed the electron densities from the MAGPIS measurements  
both in the hypothesis of uniformly filled sources (case A) and, by assuming a shell-like structure for the 6 \ion{H}{2} regions showing a double-peak profile (case B). For case A, electron densities were 
obtained using the geometric mean,  $d_{g}${\footnote{$d_{g}$ = $\sqrt{d_{maj}\times d_{min}}$.}}, of the linear
diameters $d_{maj}$ and $d_{min}$ inferred from the MAGPIS angular major and minor axis.  For case B, we estimated the angular size of the shell from one of the peaks of the 20 cm double-peak 
profile (Column 8 of Table~2) and we 
converted this value into a shell thickness (d$_{shell}$, see Column 9 in Table~2) using the heliocentric distance of the source. To evaluate the electron density, we doubled the estimated d$_{shell}$, in order to account 
for the double crossing of the shell with respect to the line of sight. The electron densities obtained for case A and B are provided in Table~1.  

\begin{figure}[h]
\centering
\hspace*{-1truecm}
\includegraphics[width=9cm,height=8cm,angle=0]{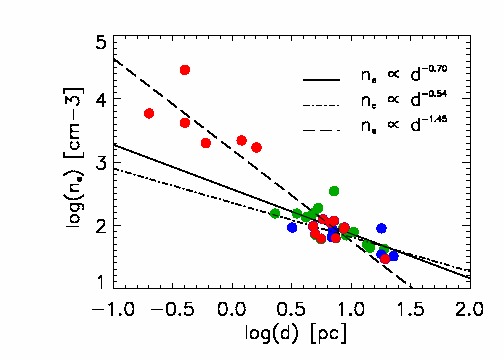}
\caption{
Electron density-size relation derived from: (1) single-dish 6 cm data (green circles, solid line), (2)
MAGPIS 20 cm data (blue circles, dashed-dotted line), (3) MAGPIS 20 cm data (red circles, dashed line) taking into account the potential
shell-type morphology of 6 sample HII regions.
}
\end{figure}

With the derived electron densities and linear sizes, we explored the n$_{e}$-d relation.  For a uniform-density nebula, n$_{e}$ is expected to scale with the linear diameter as d$^{-3/2}$. 
Departures from this behavior can be explained by either invoking a non-uniform, clumpy nature of \ion{H}{2} regions (Garay $\&$ Lizano, 1999; Kim $\&$ Koo 2001), or the presence of dust particles in the ionize gas, which 
would absorb part of the ionizing radiation and would cause the \ion{H}{2} region to be smaller. Here, for completeness, we considered the values of n$_{e}$ and d estimated both from 
the single-dish and MAGPIS data, and from both case A and B. Figure~5 shows the best-fit model for, respectively, (1) the single-dish measurements assuming that the ionized gas uniformly fills the source volume (green dots); (2) the MAGPIS 
measurements with the same approximation as in (1) (blue dots); (3) the MAGPIS 
measurements conidering the potential shell-like geometry of 6 sample \ion{H}{2} regions (red dots). Visibly, when the shell-like geometry is adopted, 
the inferred electron densities (linear diameters) are $\sim$ one hundred times higher ($\sim$ ten times smaller) than the values obtained using correspondingly larger volumes. 
In the absence of these large/small electron densities/linear diameters, the n$_{e}$-d distribution is flatter and can be well represented by a power-law with 
spectral index $>$ -1. However, when the effective size of the source is correctly evaluated, 
the best-fit relation (n$_{e}$ $\sim$ d$^{-1.45}$) steepens significantly, ruling out the possibility that a significant amount of dust is present in the sample \ion{H}{2} regions.

\section{IR Excess}

The presence of dust inside \ion{H}{2} regions can also be investigated in terms of energy budget, through the so-called {\em{Infrared Excess}}  (Garay et al. 1993): 

\begin{equation}
IRE =  \frac{L{_{IR}}}{L(Ly{\alpha})}
\end{equation}

where L$_{IR}$ is the IR luminosity and  L(Ly{$\alpha$}) is the Lyman photon luminosity derived in Section~3. Evaluation of the IRE 
provides an estimate of the fraction of photons emitted by the central source and directly absorbed by dust particles. 
We computed the IRE both for \ion{H}{2} regions only, and for the combined \ion{H}{2} regions and associated PDRs. For \ion{H}{2} regions, 
the IR luminosity was obtained by integration of the best-fit 2-temperature component models ($\widetilde{S_{\lambda,2}}$) derived in Section~4.2:

\begin{equation}
L_{IR, HII} = 4 \pi D^{2} \int_{\lambda_{min}}^{\lambda_{max}} { \widetilde{S_{\lambda,2}}  \hspace*{0.1truecm} d\lambda}
\end{equation}

In the expression above, we set $\lambda_{min}$ = 1 $\mu$m and $\lambda_{max}$ = 10$^{4} \mu$m, and $D^{2}$ is, for each source, the solar distance provided in Table~1. In order to 
estimate the IR luminosities for the (\ion{H}{2} region $+$ PDR) ensembles, we perfomed photometric measurements between 24 $\mu$m and 500 $\mu$m by including in the aperture not only the region identified by the 
MAGPIS 20 cm contours, but also the PDR. For this purpose, instead of using the MAGPIS 20 cm data to generate our apertures, we used the IRAC 8$\mu$m images. Following the same steps described in Section~4, we estimated and 
removed the free-free contribution in each band, and fitted the resulting SEDs with a 2-temperature component model with $\beta_{dust}$ = 2. L$_{IR, HII+PDR}$ for each source was then evaluated by applying a relation 
similar to equation~(14). The derived L$_{IR, HII}$, L$_{IR, HII+PDR}$, IRE$_{HII}$ and IRE$_{HII+PDR}$ are shown in Table~6. 
For all the sources (with one exception, i.e. G30.2-0.1), both IRE$_{HII}$ and IRE$_{HII+PDR}$ are slightly (but consistently) greater than 1. 
However, IRE$_{HII+PDR}$ is systematically higher than IRE$_{HII}$, with  
$\overline{IRE_{HII}}$ = 1.08 $\pm$ 0.14 and $\overline{IRE_{HII+PDR}}$ = 1.17 $\pm$ 0.04. As expected, this result indicates that the amount of radiation absorbed in the PDR is larger than the amount 
absorbed {\em{in situ}}. It also shows that, if dust co-exists with the ionized gas in the interior of  \ion{H}{2} regions, its mass is not significant, in agreement with what we found from the 
analysis of the n$_{e}$-d relation described in the previous section. Finally, we note that the values of IRE obtained for the present sample of \ion{H}{2} regions are substantially lower (a factor 10 on average) 
than the IR excess reported by 
Garay et al. (1993). This is likely a consequence of the different evolutionary stage of the sources in the two samples: the sources we are considering are more evolved and less dusty, while the compact \ion{H}{2} 
regions analyzed by Garay et al. (1993) are still embedded in the natal cocoon. 

The photometric measurements for both the \ion{H}{2} regions and the (\ion{H}{2} region $+$ PDR) systems allowed us, as well as to derive IR luminosities, to carry out a qualitative comparison 
between the average behaviour of their SEDs in the wavelength range 24 $\mu$m $< \lambda <$ 500 $\mu$m. The average SEDs, for the \ion{H}{2} regions and (\ion{H}{2} region $+$ PDR) systems separately,  
were obtained by first normalizing each individual SED to its peak and then by taking the average of these normalized SEDs. The result is illustrated in Figure~6. The peak of the  \ion{H}{2} regions SED is shifted towards shorter 
wavelengths ($\sim$ 70 $\mu$m) with respect to the (\ion{H}{2} regions $+$ PDRs) SED peak ($\sim$ 160 $\mu$m). This trend is in agreement with the finding of Povich et al. (2007) and can be ascribed to the higher temperature of 
the dust grain population in the \ion{H}{2} region compared to the temperature of the grain population located in the PDR.

\section{Color-color plots}

We used the fluxes extracted in the previous section for the (\ion{H}{2} region $+$ PDR) complexes to generate color-magnitude and color-color plots. Goal of this exercise was 
to look for characteristic colors which can allow the identification of this class of objects with respect to other populations of Galactic sources. In particular, we considered sources in either an earlier 
or later evolutionary stage with respect to our sample \ion{H}{2} regions. In the former category (e.g. younger sources) fall the approximately 100 sources listed in the Hi-GAL preliminary catalog
(Elia et al., 2010) and located in our 2$^{\circ}\times$2$^{\circ}$ SDP field. These sources have measured fluxes at all PACS and SPIRE wavelengths and at 24 $\mu$m. Most of them are thought to be Young Stellar
Objects (YSOs) (Elia et al. 2010). For the latter category (e.g. older sources) we made use of the 39 Supernova Remnants (SNRs) cataloged by Pinheiro-Gon\c{c}alves et al. (2011, hereafter PG11), based on GLIMPSE and MIPSGAL data. 

The color-magnitude and color-color diagrams are shown in Figure~7. Panels b), c) and d) highlight the general segregation of the (\ion{H}{2} regions $+$ PDRs) with respect to the YSO population. This effect can be attributed to  
the relatively high dust temperatures ($\sim$ 30 - 50 K) characterizing the (\ion{H}{2} region $+$ PDR) systems, which causes the peak of their SED to occur at wavelengths shorter than $\simeq$ 160 $\mu$m (see Figure~6 and Section~7). 
Conversely, in the very early stages of star formation (e.g. YSOs) typical dust temperatures do not go beyond $\sim$ 15 K, with the consequence that the SED peak is located at $\lambda >$ 160 $\mu$m. In practise, while the SED 
of an (\ion{H}{2} region $+$ PDR) is rising fast between 24 $\mu$m and 70 $\mu$m and dropping rapidly between 160 $\mu$m and 250 $\mu$m (and between 250 $\mu$m and 500$\mu$m), 
that of a YSO is also rising in the 24 $\mu$m $< \lambda <$ 70 $\mu$m but much more slowly, while for 160 $\mu$m $< \lambda <$ 250 $\mu$m (and 250 $\mu$m $< \lambda <$ 500)  it keeps steepening or, if it drops, it drops slowly. 
Since the Hi-GAL preliminary catalog was not band-merged
with the GLIMPSE 8 $\mu$m catalog, we were not able to compare the mid-IR color distribution of the (\ion{H}{2} region $+$ PDR) population with that of YSOs.

Panel a) illustrates the behaviour of the (\ion{H}{2} region $+$ PDR) systems compared to that of SNRs in the 8 $\mu$m/24 $\mu$m vs. 24 $\mu$m/70 $\mu$m color-color space. Both populations appear to span quite a wide range of 
8 $\mu$m/24 $\mu$m ratios. However, while SNRs also strech along the x-axis direction for an order of magnitude, the (\ion{H}{2} regions $+$ PDRs) are confined to a narrow range of 24 $\mu$m $< \lambda <$ 70 $\mu$m values. 
The relatively spread out distribution of SNRs likely reflects the variety of morphologies and emission mechanisms highlighted in PG11. It is worth to mention that a partial overlap of mid-IR colors for the composite 
(\ion{H}{2} region + PDR) systems and SNRs was already reported by Arendt (1989).

A search for Hi-GAL counterparts of the SNRs in the PG11 catalog is currently on-going (Noriega-Crespo et al., 2012 in preparation). Preliminary results suggest that the detection rate in the PACS 70 $\mu$m band 
should match the MIPSGAL one, while only few, very bright SNRs (e.g G11) are detected at 160 $\mu$m, and almost none in the SPIRE bands. This effect is currently interpreted as either
due to an intrinsic lack of emission from these sources at far-IR wavelengths or to a competing background level (Noriega-Crespo, private communication).

\begin{figure}[h]
\centering
\includegraphics[width=7.5cm,height=8cm,angle=90]{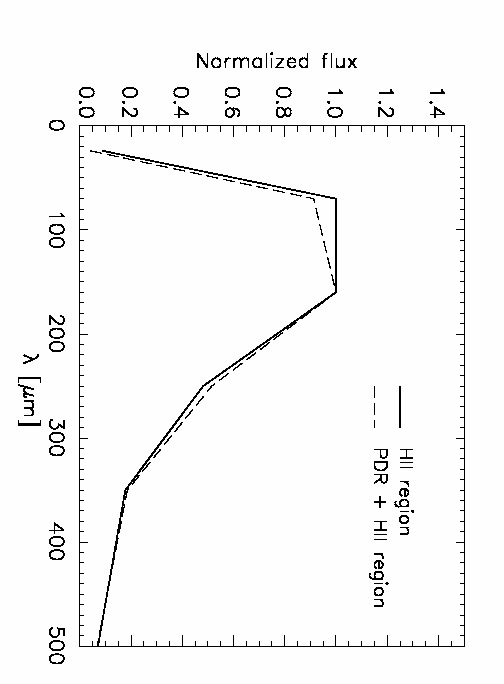}
\caption{Average \ion{H}{2} region (solid line) and PDR (dashed line) SED. The average is taken over the sample of 16 sources considered for this analysis.  
}
\end{figure}

\begin{figure*}[h]
\centering
\includegraphics[width=6.5cm,height=7.5cm,angle=90]{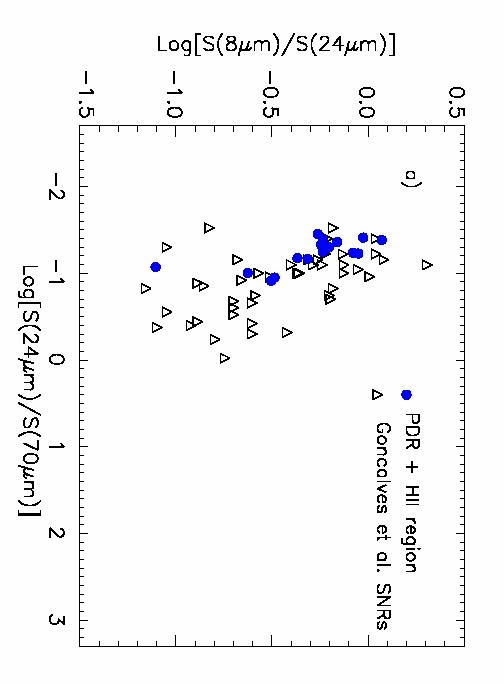}
\includegraphics[width=6.5cm,height=7.5cm,angle=90]{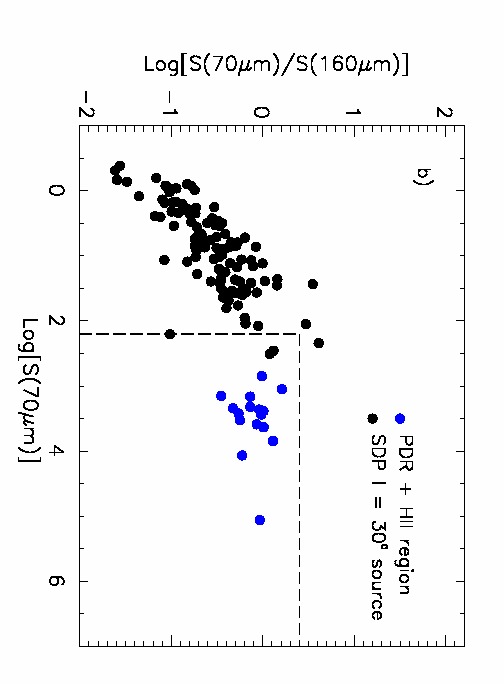}\\
\includegraphics[width=6.5cm,height=7.5cm,angle=90]{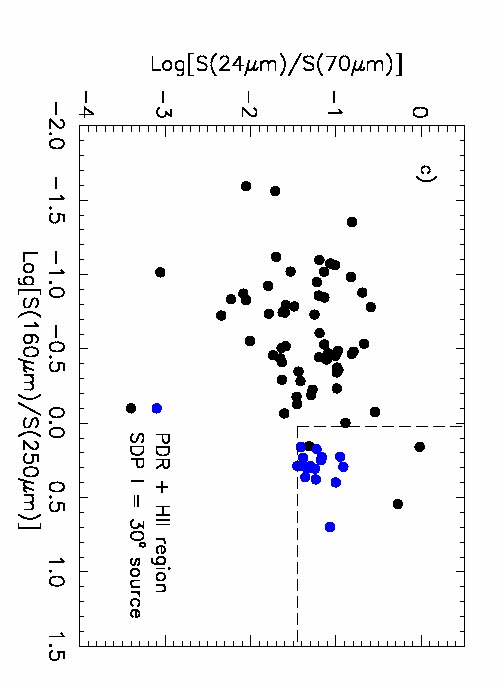}
\includegraphics[width=6.5cm,height=7.5cm,angle=90]{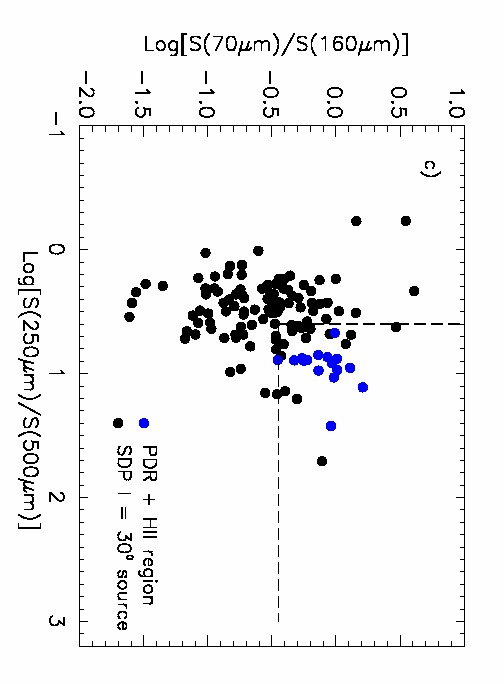}
\caption{Color-color diagrams obtained by combining PACS (70 $\mu$m and 160 $\mu$m) and SPIRE (250 $\mu$m, 350 $\mu$m and 500 $\mu$m) wavelengths, as well as IRAC 8 $\mu$m and
MIPS 24 $\mu$m. Shown in the plots are: the 16 evolved \ion{H}{2} regions + PDRs of this work (blue filled circles), the Hi-GAL point sources from Elia et al. (2010) (black filled circles) and 
the SNRs from Pinheiro-Gon\c{c}alves et al. (2010) (black triangles).}
\end{figure*}

\section{Conclusions}

The analysis of Spitzer IRAC and MIPS data combined with Herschel PACS and SPIRE data for 
a uniform sample of evolved \ion{H}{2} regions shows that the emission in the direction of these sources in the range 24 $\mu$m -- 500 $\mu$m 
is well represented (for all but three sources) by a 2-temperature dust component model with $\beta_{dust}$ =2. Of these two components, the cold one is in the range 20 - 30 K and the warm one  
in the range 50 - 90 K. For one of the sources, G30.1-0.2, a 1-temperature component model with a varying spectral emissivity index appears to be a better 
representation of the observed SED, although in this case the $\beta_{dust}$ (0.32) retrieved from the fit is outside the range of commonly accepted values.  
For two other sources, G30.3-0.0 and G30.6-0.1, neither a modified blackbody with $\beta_{dust}$ kept as a free parameter, nor 
a 2-temperature component model with fixed $\beta_{dust}$ agrees with the measurements in a satisfactorily fashion. We emphasize that, for all the sample sources, the temperatures obtained 
for the warm component are not well constrained due to a lack of measurements between 24 $\mu$m and 70 $\mu$m. Despite this 
limitation, 
these results highlight the importance of including bands shorter than  100 $\mu$m for studying sources, such as \ion{H}{2} regions, which are characterized by multiple 
temperature components. In fact, when the shorter wavelengths are not considered in the fitting process, the cold component, as well as the warm one, might not be correctly constrained. 

Our study strongly suggests that dust is present in the interior of \ion{H}{2} regions. Three complementary pieces of evidence are found in support of this conclusion: 

\begin{enumerate} 
\item{a statistically significant flux detection, after  background subtraction, for 8 $\mu$m $< \lambda <$ 500 $\mu$m (with an average S/N greater or of the order of 3);}
\item{a slope for the n$_{e}$ - d relation slightly steeper than -1.5;}
\item{values of IRE $>$ 1 for the entire sample \ion{H}{2} regions;}
\end{enumerate}

We emphasize that although the data seemd to favor the existence of dust grains in the interior of evolved \ion{H}{2} regions, the amount of dust in these environments 
is likely not significant. We also remind the reader that, based on our analysis, we cannot completely rule out
the possibility that the detected IR emission is due to background/foreground radiation along the
line of sight. 

We argue that partial dust depletion within \ion{H}{2} regions appears to be 
caused primarily by radiation-pressure drift, as recently proposed by Draine et al. (2011). In this light, we speculate that 
the observed 24 $\mu$m emission cannot be attributed to a dust population (either VSGs or BGs) co-eval with respect to the BGs and PAHs 
located in the surrounding PDR, but rather to a new generation of grains, in agreement with what was recently suggested by Everett $\&$ Churchwell (2010) for the case of WBB. 

Finally, we show that far-IR colors can be reliably used for identifying unresolved Galactic or extragalactic \ion{H}{2} regions with respect to younger populations
of sources, such as YSOs. On the contrary, distinguishing unresolved \ion{H}{2} regions from SNRs based solely on mid-IR colors appears to be more challenging, and 
may require the additional use of ancillary data, such as radio continnuum or RRL.

\acknowledgments
The authors would like to thank the anonymous referee for his/her comments which allowed us to greatly improve the content and presentation of the paper.

\begin{deluxetable}{cccccccccccccc}
\tabletypesize{\footnotesize}
\tablewidth{0pt}
\tablecaption{Sample of evolved \ion{H}{2} regions} 
\tablehead{
\colhead{Name} &
\colhead{l} &
\colhead{b} &
\colhead{S$_{6cm}$} &
\colhead{ {\em R}} &
\colhead{ {\em D}} &
\colhead{ {\em D$_{f}$}} &
\colhead{ {\em T$_{e}$}} & 
\colhead{ {\em EM$_{6cm}$} } &
\colhead{ {\em EM$_{20cm}$} } & 
\colhead{ {\em EM$_{shell}$} } &
\colhead{ {\em n$_{e,6cm}$} } & 
\colhead{ {\em n$_{e,20cm}$} } & 
\colhead{ {\em n$_{e,shell}$} }\\
\colhead{(deg)} &
\colhead{(deg)} &
\colhead{(Jy)} &
\colhead{(kpc)} &
\colhead{(kpc)} &
\colhead{(kpc)} &
\colhead{(K)} &
\colhead{(pc $\*$cm$^{-6}$)} &
\colhead{(pc $\*$cm$^{-6}$)} &
\colhead{(pc $\*$cm$^{-6}$)} &
\colhead{(cm$^{-3}$)} & 
\colhead{(cm$^{-3}$)} &
\colhead{(cm$^{-3}$)}
}
\startdata
G28.8-0.2  & 28.823 & -0.226 & 1.47  & 4.5 & 5.5 & $...$ & 5934 & 8.2e4 & 4.4e4 & $...$ & 152.7 & 95.9 & $...$\\
G29.0-0.6  & 28.983 & -0.603 & 1.01 & 5.8 & 3.4 & $...$ & 6417 & 5.3e4 & 2.6e4 & 7.0e6 & 152.7 & 91.3 & 5.9e3 \\
G29.1-0.0  & 29.136 & -0.042 & 1.85 & 5.8 & 11.5 &  $...$ & 6417 & 2.7e4 & 7.4e4 & $...$ & 43.9 & 91.7  & $...$\\
G29.1+0.4  & 29.139 & 0.431 & 2.68 & 7.4 & 13.6 & $...$ & 7012 & 3.3e4 & 2.3e4 & 4.8e6 & 41.8 & 32.0 & 1.7e3\\
G29.2-0.0  & 29.205 & -0.047  & 2.22 & 5.4 & 10.9 & $...$ & 6268 & 3.3e4 & 1.6e4 & $...$ & 49.2 & 29.1 & $...$\\
G30.1-0.2  & 30.069 & -0.160  & 0.95 & 4.4   & 8.5  & $...$ & 5897   & 2.8e4  & 2.9e4 & $...$ & 61.9 & 62.1 & $...$    \\
G30.2-0.1  & 30.227 &  -0.145  & 2.59  & 4.4  & 6.3* & 8.4*  & 9100  & 1.0e5  & 4.8e4 & $...$ & 147.2/127.5 & 83.7/72.5  & $...$\\
G30.3-0.0  & 30.277 &  -0.020  & 0.73 & 4.4  & 6.2  & $...$ & 5896   & 2.5e4  & 2.6e4 & 2.2e6 & 70.5 & 72.5  & 2.0e3 \\
G30.4-0.2  & 30.404 &  -0.238  & 3.59 & 4.3  & 8.0  & $...$ & 5859   & 7.0e4  & 9.9e4 & $...$ & 90.5 & 117.6  & $...$\\
G30.5-0.3  & 30.502 &  -0.290  & 2.96 & 4.3  & 7.3   & $...$ & 5860   & 4.3e4    & 7.9e4 & $...$ & 68.7 & 108.2  & $...$\\
G30.5+0.4  & 30.467 &  0.429   & 2.13 & 5.7  & 3.6* & 11.0*   & 6194   & 2.0e4  & 2.1e4 & $...$ & 59.8/34.2 & 61/34.8  & $...$\\
G30.6-0.1  & 30.602 &  -0.106  & 2.40 & 4.4  & 7.3   & $...$ & 6800  & 1.2e5  & 8.7e4 & $...$ & 158.2 & 123.5 $...$\\
G30.7-0.3  & 30.694 & -0.261   & 3.27 & 4.5  & 8.3   & $...$ & 6800  & 1.8e5   & 6.7e4 & $...$ & 185.4 & 87.3 $...$ \\
G30.8-0.0  & 30.776 & -0.029   & 62.17 & 4.6   & 5.7  & $...$ & 7030  & 8.6e5  & 1.4e5 & 4.2e8 & 345.2 & 88.6 & 2.9e4 \\
G31.0+0.5  & 31.050 &  0.480   & 1.75 & 7.1   & 12.9  & $...$ & 6901  & 6.3e4  & 2.1e4 & 5.5e6 & 77.8 & 34.6 & 2.2e3\\
G31.1+0.3  & 31.130 &  0.284   & 1.11 & 4.4   & 7.3   & $...$ & 5896   & 7.4e4    & 2.8e4 & 7.5e6 & 132.7 & 64.2 & 4.2e3\\
\enddata
\tablecomments{Quoted fluxes are from single-dish measurements by Altenhoff et al. (1979),
Downes et al. (1980) and Kuchar $\&$ Clark (1997). Galactocentric and solar distances are from AB09,
based on Lockman (1989) RRL observations. Solar distances denoted with * are computed from radial velocities 
in Lockman (1989) combined with the rotation curve by McClure$-$Griffiths $\&$ Dickey (2007). D$_{f}$ denotes the far distance solution for sources with unresolved distance ambiguity. 
Details on the derivation of T$_{e}$, EM and n$_{e}$ can be found in Section.~3. EM$_{6cm}$ and EM$_{20cm}$ are, respectively, the emission measure computed from single-dish and 
MAGPIS fluxes. n$_{e,6cm}$ and n$_{e,20cm}$ are the corresponding electron densities.  EM$_{shell}$ and n$_{e,shell}$ denote the emission measure and electron density estimated by assuming 
that the source is characterized by a shell-like geometry. 
}
\end{deluxetable}

\begin{deluxetable}{ccccccccc}
\tabletypesize{\footnotesize}
\tablewidth{0pt}
\tablecaption{Angular and linear diameters}
\tablehead{
\colhead{Name} &
\colhead{$\theta$} &
\colhead{ {\em d} } &
\colhead{${\theta_{maj}}$} &
\colhead{${\theta_{min}}$} &
\colhead{ {\em d$_{maj}$} } &
\colhead{ {\em d$_{min}$} } & 
\colhead{${\theta_{shell}}$} & 
\colhead{ {\em d$_{shell}$} } \\
\colhead{}   &
\colhead{(arcmin)} &
\colhead{(pc)} &
\colhead{(arcmin)} &
\colhead{(arcmin)} & 
\colhead{(pc)} &
\colhead{(pc)} & 
\colhead{(arcsec)} &
\colhead{(pc)} \\
}
\startdata
G28.8-0.2  & 2.2 &  3.5     &  3.0 & 3.0  & 4.8 & 4.8 & $...$ & $...$ \\
G29.0-0.6  & 2.3 &  2.3    &  3.5 & 3.0  & 3.4  & 3.0 & 6 & 0.1\\
G29.1-0.0  & 4.3 & 14.4     &  3.3 & 2.1 & 11.0  & 7.0 & $...$ & $...$\\
G29.1+0.4  & 4.8 & 19.0     & 9.7 & 3.4 & 38.4 & 13.4 & 12 & 0.8\\
G29.2-0.0  & 4.3 & 13.6     & 9.1 & 4.1  & 28.8  & 13.0 & $...$ & $...$\\
G30.1-0.2  & 3.0 & 7.4      & 3.9  & 2.3 & 9.6 & 5.7 & $...$ & $...$\\
G30.2-0.1  & 2.6 & 4.7/6.3  & 4.1  & 3.5 & 7.5/10.0 & 6.4/8.5 & $...$ & $...$\\
G30.3-0.0  & 2.8 & 5.0      & 3.6 & 2.1 & 6.5 & 3.8 & 9 & 0.3 \\
G30.4-0.2  & 3.7 & 8.6      & 4.6 & 2.1 & 10.7 & 4.9 & $...$ & $...$\\
G30.5-0.3  & 4.3 & 9.1      & 5.6 & 1.8 & 11.9 & 3.8 & $...$ & $...$\\
G30.5+0.4  & 5.4 & 5.6/17.3 & 5.6 & 5.1 & 5.9/17.9 & 5.3/16.3 & $...$ & $...$\\
G30.6-0.1  & 2.3 & 4.9      & 3.2 & 2.3 & 6.8 & 4.9 & $...$ & $...$\\
G30.7-0.3  & 2.2 & 5.3      & 4.4 & 3.0 & 10.6 & 7.2 & $...$ & $...$\\
G30.8-0.0  & 4.4 &  7.2     &  13.8 & 8.6 & 22.9 & 14.2 & 6 & 0.2 \\
G31.0+0.5  & 2.8 & 10.5     & 5.5 & 4.2 & 20.6 & 15.8 & 9 & 0.6 \\
G31.1+0.3  & 2.0 &  4.2     & 3.1 & 3.4 & 6.6 & 7.2 & 6 & 0.2 \\
\enddata
\tablecomments{The angular diameter $\theta$ and corresponding linear diameter, d, are obtained from the single-dish measurements by Altenhoff et al. (1979),
Downes et al. (1980) and Kuchar $\&$ Clark (1997), while $\theta_{maj}$, $\theta_{min}$, d$_{maj}$ 
and d$_{min}$ are derived from the MAGPIS data (Section~3 and Figure~7). All linear diameters are estimated with the heliocentric distances quoted in Table~1. 
$\theta_{shell}$ and d$_{shell}$ denote the shell angular size and linear thickness in the hypothesis of a shell-like geometry of the source. 
}
\end{deluxetable}

\clearpage
\begin{landscape}
\begin{deluxetable}{ccccccc}
\tabletypesize{\footnotesize}
\tablewidth{0pt}
\tablecaption{Free-free spectral indices}
\tablehead{
\colhead{Name} &
\colhead{$\beta_{ff,24}$} &
\colhead{$\beta_{ff,70}$} &
\colhead{$\beta_{ff,160}$} &
\colhead{$\beta_{ff,250}$} &
\colhead{$\beta_{ff,350}$} &
\colhead{$\beta_{ff,500}$} \\
\colhead{} &
\colhead{} &
\colhead{} &
\colhead{} &
\colhead{} &
\colhead{} &
\colhead{} 
}
\startdata
G28.8-0.2  &       1.67       &    0.60 & 0.40 & 0.34 &  0.30 & 0.27  \\
G29.0-0.6  &       1.39       &    0.56 & 0.38 &  0.32 &  0.29 &  0.27 \\
G29.1-0.0  &       1.39       &    0.56 & 0.38 &  0.32 &  0.29 &  0.27  \\
G29.1+0.4  &       1.18       &    0.52 & 0.36 &  0.31 &  0.28 &  0.26 \\
G29.2-0.0  &       1.47       &    0.57 & 0.39 &  0.33 &  0.30 &  0.27 \\
G30.1-0.2  &       1.69       &    0.60 & 0.40 &  0.34 &  0.30 &  0.27     \\
G30.2-0.1  &       0.80       &    0.43 & 0.32 &  0.28 &  0.25 &  0.23 \\
G30.3-0.0  &       1.69       &    0.60 & 0.40 &  0.34 &  0.30 &  0.27  \\
G30.4-0.2  &       1.72       &    0.60 & 0.40 &  0.34 &  0.31 &  0.28  \\
G30.5-0.3  &       1.72       &    0.60 & 0.40 &  0.34 &  0.31 &  0.28 \\
G30.5+0.4  &       1.51       &    0.58 & 0.39 &  0.33 &  0.30 &  0.27\\
G30.6-0.1  &       1.24       &    0.55 & 0.37 &  0.32 &  0.29 &  0.26\\
G30.7-0.3  &       1.24       &    0.53 & 0.37 & 0.32 &  0.29 &  0.26  \\
G30.8-0.0  &       1.17      &     0.52 & 0.36 &  0.31 &  0.28 &  0.26 \\
G31.0+0.5  &       1.21      &     0.52 & 0.37 & 0.31 &  0.28 &  0.26  \\
G31.1+0.3  &       1.69      &     0.60 & 0.40 & 0.34 &  0.30 &  0.27 \\
\enddata
\tablecomments{Free-free spectral indices ($\beta_{ff,\overline{\lambda}}$) at a fixed wavelength ${\overline{\lambda}}$ as a function of electron 
temperatures (see Table~1 Column 8). 
}
\end{deluxetable}

\begin{deluxetable}{cccccccccccccc}
\tabletypesize{\scriptsize}
\tablecaption{Results of photometric measurements}
\tablewidth{0pt}
\tablehead{
\colhead{Name} &
\colhead{S$_{24}$} &
\colhead{b$_{24}$} &
\colhead{S$_{70}$} &
\colhead{b$_{70}$} &
\colhead{S$_{160}$} &
\colhead{b$_{160}$} &
\colhead{S$_{250}$} &
\colhead{b$_{250}$} &
\colhead{S$_{350}$} &
\colhead{b$_{350}$} &
\colhead{S$_{500}$} &
\colhead{b$_{500}$} &
\colhead{Area} \\
\colhead{}                 &
\colhead{(Jy)}             &
\colhead{(MJy/sr)}         &
\colhead{(Jy)}             &
\colhead{(MJy/sr)}         &
\colhead{(Jy)}             &
\colhead{(MJy/sr)}         &
\colhead{(Jy)}             &
\colhead{(MJy/sr)}         &
\colhead{(Jy)}             &
\colhead{(MJy/sr)}         &
\colhead{(Jy)}             &
\colhead{(MJy/sr)}         &
\colhead{arcmin$^{2}$}
}
\startdata
G28.8-0.2  & 106$\pm$12 & 45$\pm$2  & 658$\pm$173 & 722$\pm$58 &  658$\pm$201 & 1848$\pm$227 &   568$\pm$123 & 1594$\pm$199 &   214$\pm$51 & 715$\pm$77 & 89$\pm$20 & 255$\pm$36 & 28\\
G29.0-0.6 & 126$\pm$13  & 30$\pm$1  & 1090$\pm$250  & 588$\pm$14 &  823$\pm$203 & 1049$\pm$56 &   525$\pm$87 & 692$\pm$65 &   181$\pm$35 & 356$\pm$23 & 71$\pm$13 & 126$\pm$11 & 10\\
G29.1-0.0 & 57$\pm$6  & 45$\pm$3  &  656$\pm$131  & 746$\pm$84  & 640$\pm$128 & 1831$\pm$305 &   240$\pm$36 & 1592$\pm$281 &   78$\pm$12 & 717$\pm$104 & 27$\pm$4 & 263$\pm$44 & 7\\
G29.1+0.4 & 52$\pm$5  & 25$\pm$1  & 791$\pm$158  & 278$\pm$16 & 1113$\pm$223 & 577$\pm$36 &   642$\pm$96 & 268$\pm$30 &   248$\pm$37  & 192$\pm$12  &  82$\pm$12  & 66$\pm$5 & 33\\
G29.2-0.0 & 102$\pm$10  & 39$\pm$3  &   1762$\pm$352  & 601$\pm$66 & 2581$\pm$516  & 1405$\pm$175  &  1234$\pm$185 & 1123$\pm$148 &   450$\pm$68 & 525$\pm$57 & 150$\pm$23 & 190$\pm$27 & 37\\
G30.1-0.2 & 28$\pm$3  & 74$\pm$4  &  300$\pm$60  & 1431$\pm$69 & 219$\pm$44 & 3290$\pm$313 &   95$\pm$16 & 2611$\pm$329 &   44$\pm$8 & 1070$\pm$132 & 21$\pm$4 & 398$\pm$64 & 9\\
G30.2-0.1  & 6$\pm$1  & 77$\pm$3  &     23$\pm$5  & 1265$\pm$69 & 21$\pm$4 & 2707$\pm$146 &   10$\pm$2 & 2171$\pm$110 &   4$\pm$1 & 927$\pm$49 & 2$\pm$1 & 333$\pm$19 &  14 \\
G30.3-0.0 & 43$\pm$4  & 66$\pm$2  &    457$\pm$91  & 1169$\pm$60  & 329$\pm$66 & 2722$\pm$181 &  136$\pm$21 & 2186$\pm$149 &   54$\pm$8 & 908$\pm$55  & 10$\pm$2 & 318$\pm$22 & 8\\
G30.4-0.2 & 209$\pm$21  & 55$\pm$5   & 1494$\pm$299  & 934$\pm$81 & 1450$\pm$290 & 2203$\pm$120 &  703$\pm$105 & 1779$\pm$69 &   259$\pm$39 & 772$\pm$26 & 85$\pm$13  & 282$\pm$11 & 10 \\
G30.5-0.3   & 69$\pm$7  & 69$\pm$5  & 1429$\pm$286 & 1137$\pm$117 & 1582$\pm$317 & 2538$\pm$198 &  769$\pm$115 & 1986$\pm$155 &   281$\pm$42 & 850$\pm$68 & 113$\pm$17 & 307$\pm$25 & 10\\
G30.5+0.4  & 50$\pm$12 & 34$\pm$2  &  1114$\pm$351  & 330$\pm$38 & 2508$\pm$877 & 934$\pm$122  &  1901$\pm$571 & 894$\pm$111 &   732$\pm$246 & 439$\pm$42 & 262$\pm$93 & 162$\pm$1 & 29\\
G30.6-0.1 & 62$\pm$6  & 89$\pm$7  &  1237$\pm$247  & 1554$\pm$38 & 1407$\pm$281 & 3915$\pm$125 &   689$\pm$103 & 3351$\pm$104  &   259$\pm$39 & 1364$\pm$41 & 52$\pm$8 & 497$\pm$18 &  7 \\
G30.7-0.3 & 299$\pm$30  & 75$\pm$4  &     2700$\pm$540  & 1473$\pm$62 & 1369$\pm$274 & 3398$\pm$164 &  456$\pm$69 & 2813$\pm$161 &   120$\pm$19 & 1153$\pm$72 & 42$\pm$7 & 419$\pm$28 & 13 \\
G30.8-0.0 & 2545$\pm$254  & 59$\pm$9  &     65867$\pm$13173  & 1148$\pm$84 &  53229$\pm$10647 & 3541$\pm$264 &   26784$\pm$4019 & 3162$\pm$256 &   9358$\pm$1406 & 1302$\pm$108 & 3150$\pm$475 & 474$\pm$44 & 119 \\
G31.0+0.5 & 71$\pm$7  & 32$\pm$1  &    1774$\pm$355  & 436$\pm$42 & 3198$\pm$640 & 926$\pm$100 &   2132$\pm$320 & 872$\pm$79 &    745$\pm$112 & 436$\pm$29 & 267$\pm$40 & 161$\pm$10 & 23\\
G31.1+0.3 & 41$\pm$5 & 48$\pm$1  &     451$\pm$124  & 808$\pm$41 &   511$\pm$198 & 2362$\pm$86 &   377$\pm$128 & 2284$\pm$91 &   143$\pm$54 & 995$\pm$31 & 53$\pm$20 & 368$\pm$13 & 11\\
\enddata
\tablecomments{For details on background level estimation and photometric uncertainties see Section 3. The area quoted in Column 14 provides an estimate of the extraction aperture derived from the MAGPIS 20 cm data. 
}
\end{deluxetable}
\clearpage
\end{landscape}

\begin{deluxetable}{ccccccccccccc}
\tabletypesize{\footnotesize}
\tablewidth{0pt}
\tablecaption{Best-fit SED parameters}
\tablehead{
\colhead{Name} &
\colhead{T$_{dust,2}$} &
\colhead{$\sigma{_{T_{dust,2}}}$} &
\colhead{$\beta_{dust}$} &
\colhead{$\sigma{_{\beta_{dust}}}$} & 
\colhead{${\chi_{1t}}^{2}$} &
\colhead{CDF$_{1t}$} &
\colhead{T$_{c,3}$} &
\colhead{$\sigma{_{T_{c,3}}}$} &
\colhead{T$_{w,3}$} &
\colhead{$\sigma{_{T_{w,3}}}$} &
\colhead{${\chi_{2t}}^{2}$} &
\colhead{CDF$_{2t}$} \\
\colhead{}         &
\colhead{(K)}         &
\colhead{(K)}         &
\colhead{}         &
\colhead{}         &
\colhead{}         &
\colhead{}         &
\colhead{(K)}         &
\colhead{(K)}         &
\colhead{(K)}         &
\colhead{(K)}         &
\colhead{}            &
\colhead{}
}
\startdata
G28.8-0.2 & 72.66 & 10.26 &  1.2$\times$10$^{-4}$ & 0.33 & 9.58 & 0.98 & 17.73 & 0.50 & 55.92 & 1.25 & 0.86 & 0.35\\
G29.0-0.6 & 65.81 & 9.87 & 3.6$\times$10$^{-1}$ & 0.36 & 8.93 & 0.97 & 21.27 & 0.80 & 54.17 & 0.96 & 0.95 & 0.38\\
G29.1-0.0 & 58.98 & 13.89 & 7.3$\times$10$^{-1}$ & 0.73 & 16.40  & 1.0 & 29.46 & 0.49 & 96.31 & 9.52 & 1.54 & 0.54\\
G29.1+0.4 & 60.96 & 7.65 & 1.8$\times$10$^{-1}$ & 0.41 & 22.93 & 1.0 & 23.40 & 0.90 & 57.12 & 15.59 & 0.35 & 0.16\\
G29.2-0.0 & 57.15 & 4.82 & 3.7$\times$10$^{-1}$ & 0.32 & 24.29 & 1.0 & 25.62 & 0.40 & 98.06 & 11.83 & 0.41 & 0.19\\
G30.1-0.2 & 65.04 & 2.14 & 3.2$\times$10$^{-1}$ & 0.09 & 2.91 & 0.59 & 19.81 & 0.89 & 51.49 & 0.72 & 2.34 & 0.69\\
G30.2-0.1 & 82.45 & 1.12 & 1.0$\times$10$^{-2}$ & 0.11 & 7.34  & 0.94 & 20.63 & 0.89 & 60.14 & 1.81 & 1.11 & 0.42\\
G30.3-0.0 & 55.01 & 8.79 & 1.1 & 0.50 & 24.49 & 1.0 & 31.91 & 1.01 & 98.37 & 10.54 & 7.16 & 0.97\\
G30.4-0.2 & 65.56 & 16.24 & 4.6$\times$10$^{-1}$ & 0.67 & 17.76 & 1.0 & 26.70 & 0.93 & 70.55 & 12.00 & 0.15 & 0.08\\
G30.5-0.3 & 56.93 & 12.37 & 4.2$\times$10$^{-1}$ & 0.62 & 11.83 & 0.99 & 24.65 & 1.10 & 53.00 & 11.69 & 0.92 & 0.37\\
G30.5+0.4 & 57.25 & 7.00 & 8.0$\times$10$^{-4}$ & 0.38 & 8.73 & 0.97 & 18.95 & 0.55 & 48.55 & 0.98 & 0.40  & 0.18\\
G30.6-0.1 & 22.60 & 1.36 & 2.75 & 0.21 & 102.20 & 1.0 & 27.95 & 0.38 & 95.07 & 11.84 & 0.79 & 0.98\\
G30.7-0.3 & 56.15 & 1.83 & 1.31 & 0.11 & 9.11 & 0.97 & 36.15 & 0.92 & 97.33 & 10.79 & 2.98 & 0.77\\
G30.8-0.0 & 51.61 & 4.63 & 8.2$\times$10$^{-1}$ & 0.25 & 10.29 & 0.98 & 25.94 & 1.08 & 50.33 & 16.81 & 0.40  & 0.18\\
G31.0+0.5 & 56.84 & 3.58 & 6.0$\times$10$^{-2}$ & 0.28 & 28.07  & 1.0 & 22.58 & 0.60 & 56.10 & 7.18 & 1.22 & 0.46\\
G31.1+0.3 & 68.16 & 18.55 & 3.0$\times$10$^{-5}$ & 0.71 & 4.28 & 0.77 & 19.72 & 0.64 & 52.33 & 1.02 & 0.23 & 0.11\\
\enddata
\tablecomments{Results obtained by fitting the 24 $\mu$m $<$ $\lambda$ $<$ 500 $\mu$m data points.
T$_{dust,2}$, $\sigma{_{T_{dust,2}}}$ and ${\chi_{1t}}^{2}$ are the best-fit parameters and corresponding $\chi^{2}$ for a 1-temperature component model with variable $\beta_{dust}$. 
T$_{c,3}$, $\sigma{_{T_{c,3}}}$, T$_{w,3}$, $\sigma{_{T_{w,3}}}$ and ${\chi_{2t}}^{2}$ are
the best-fit parameters and $\chi^{2}$ for a 2-temperature (warm and cold) component model with $\beta_{dust}$ equal to 2. PDF$_{1t}$ and 
PDF$_{2t}$ are the probability density functions associated to each model.} 
\end{deluxetable}

\begin{deluxetable}{ccccccc}
\tabletypesize{\footnotesize}
\tablewidth{0pt}
\tablecaption{Luminosities and IR excess}
\tablehead{
\colhead{Name}  &
\colhead{log(N$_{Lyc}$)} &
\colhead{log(L(Ly $\alpha$)/L$_{\odot}$)} &
\colhead{log(L$_{IR,HII}$/L$_{\odot}$)} &
\colhead{log(L$_{IR,PDR+HII}$/L$_{\odot}$)} &
\colhead{IRE$_{HII}$} &
\colhead{IRE$_{PDR+HII}$} \\
\colhead{}         &
\colhead{(s$^{-1}$)} &
\colhead{} &
\colhead{} &
\colhead{} &
\colhead{} &
\colhead{}
}
\startdata
G28.8-0.2   & 48.67 & 4.32 & 4.37 & 5.22 & 1.01 & 1.21\\
G29.0-0.6   & 48.07 & 3.7 & 4.5 & 4.95 & 1.21 & 1.34\\
G29.1-0.0  &  49.39 & 5.05 & 5.22 & 5.74 & 1.03 & 1.14\\
G29.1+0.4 &  49.69 & 5.34 &  5.55 & 6.14 & 1.04 & 1.15\\
G29.2-0.0 &  49.43 & 5.08 &  5.64 & 5.97 & 1.11 & 1.17\\
G30.1-0.2 &  48.86 & 4.51 &  4.71 & 5.13 & 1.04 & 1.14\\ 
G30.2-0.1 &  48.95/49.20 & 4.60/4.85 & 3.46/3.71 & 5.29/5.54 & 0.75/0.76 & 1.15/1.14\\
G30.3-0.0 &  48.47 & 4.13 &  4.52 & 4.92 &  1.09 & 1.19 \\
G30.4-0.2 &  49.38 & 5.04 &  5.96 & 6.29  & 1.18 & 1.25\\
G30.5-0.3 &  49.22 & 4.88 &  5.21 & 5.66 & 1.07 & 1.16\\
G30.5+0.4 &  48.46/49.43 & 4.11/5.08 &  4.61/5.57 & 4.73/5.69 & 1.21/1.10 & 1.15/1.12\\
G30.6-0.1 &  49.10 & 4.75 &  5.09 & 5.36 & 1.07 & 1.13\\
G30.7-0.3 &  49.35 & 5.00 &  5.55 & 5.96 & 1.11 & 1.19\\
G30.8-0.0 &   50.29 & 5.95 & 6.62 & 6.89 & 1.11 & 1.16\\
G31.0+0.5  &  49.46 & 5.11 & 5.86 & 6.01 & 1.15 & 1.17\\
G31.1+0.3 &   48.79 & 4.45 & 4.80 & 5.23 & 1.08 & 1.17\\
\enddata
\tablecomments{Lyman continuum flux (N$_{Lyc}$), Ly$\alpha$ luminosity (L(Ly $\alpha$)), IR luminosity (L$_{IR,HII}$ and L$_{IR,PDR+HII}$) and IR excess (IRE$_{HII}$ and IRE$_{PDR+HII}$)  
for the 16 \ion{H}{2} regions (and associated PDRs) in our sample.}
\end{deluxetable}

\newpage
\appendix
\renewcommand{\thefigure}{A\arabic{figure}}
\setcounter{figure}{0}
\section*{Appendix A}

\begin{figure*}[h]
\centering
\includegraphics[width=9cm,height=7.4cm,angle=0]{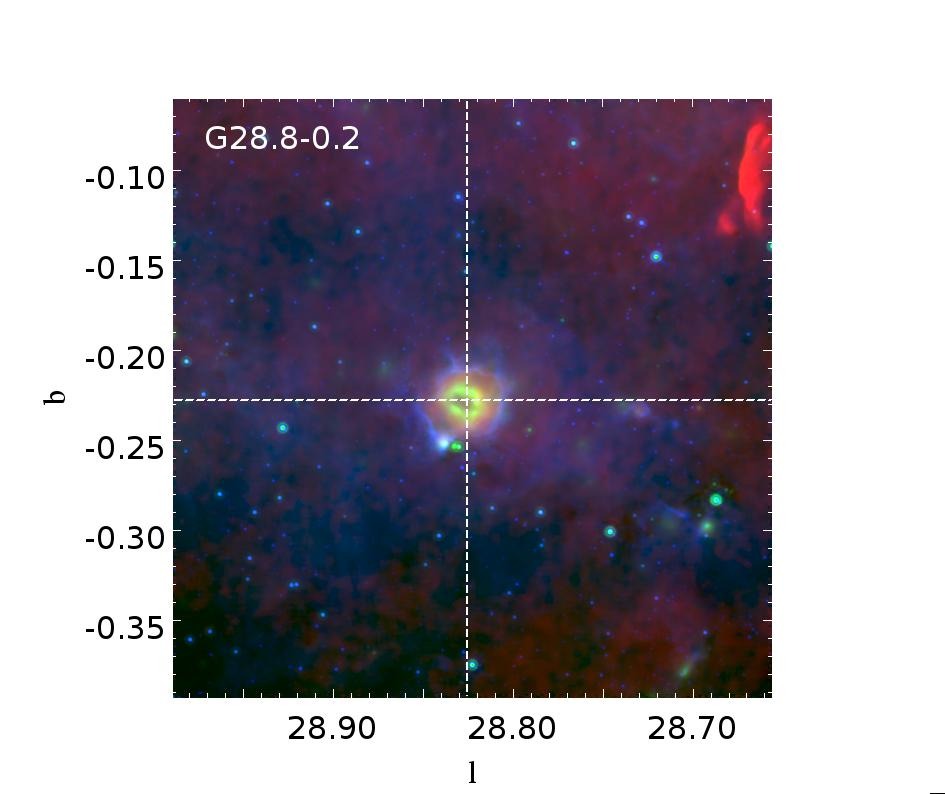}
\hspace*{-1.5truecm}
\includegraphics[width=9cm,height=7.4cm,angle=0]{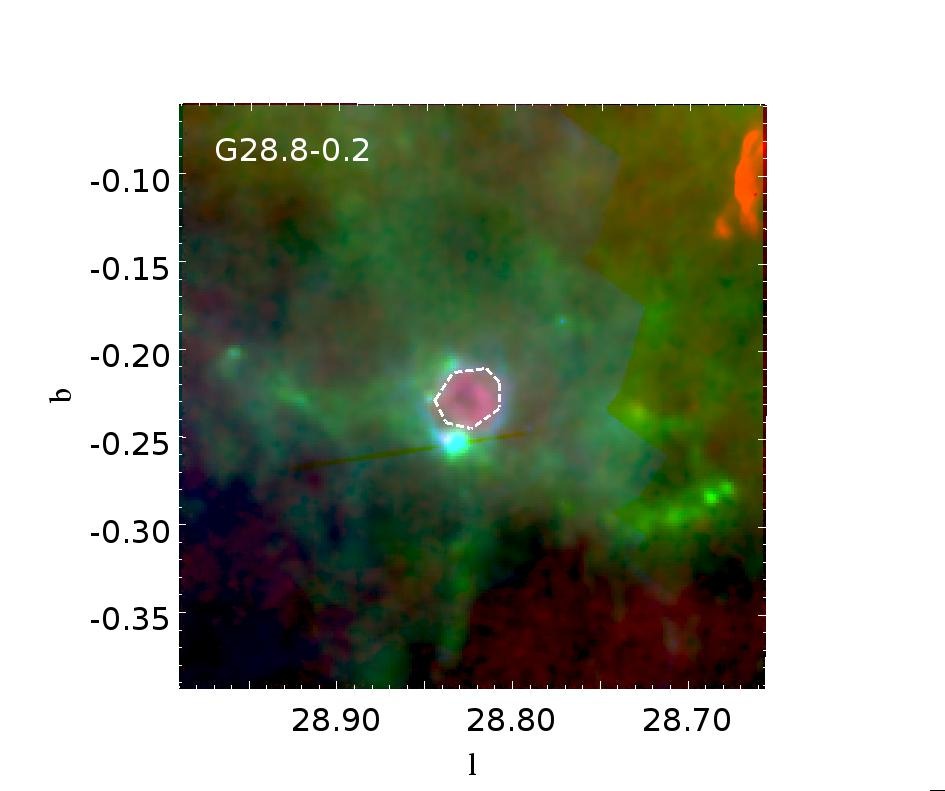}\\
\includegraphics[width=9cm,height=7.4cm,angle=0]{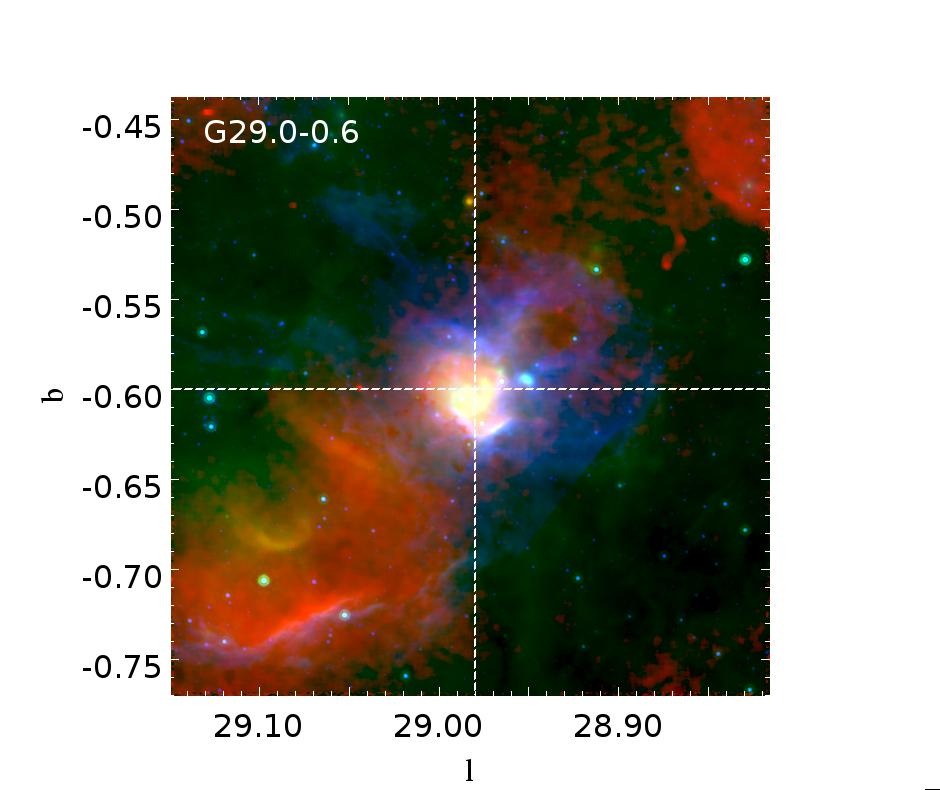}
\hspace*{-1.5truecm}
\includegraphics[width=9cm,height=7.4cm,angle=0]{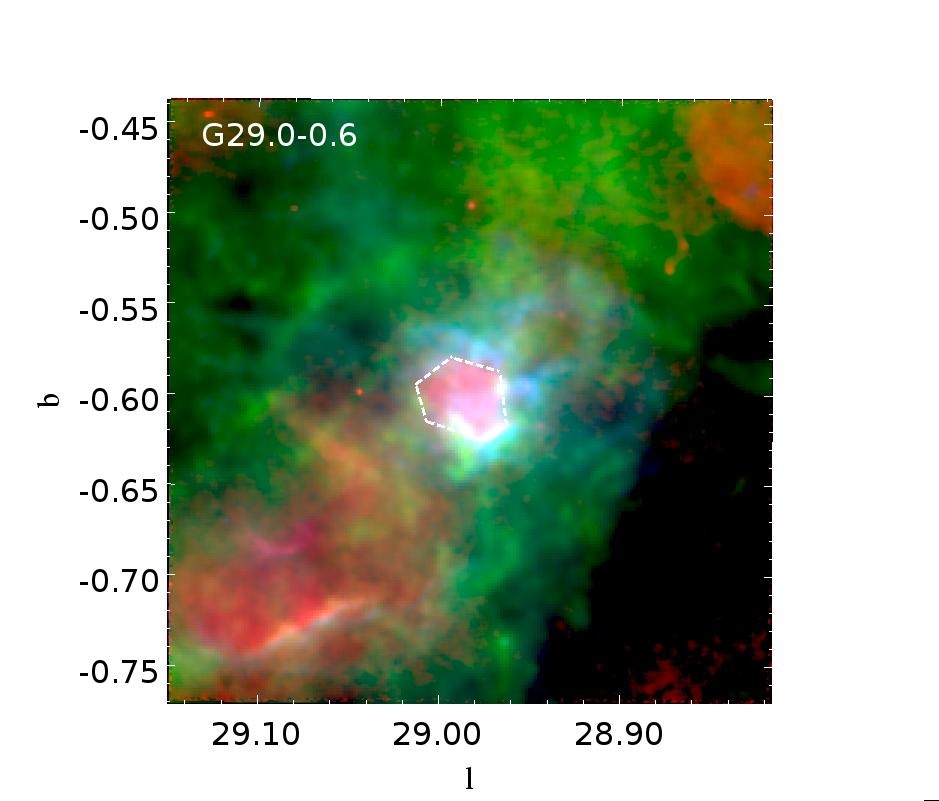}\\
\includegraphics[width=9cm,height=7.4cm,angle=0]{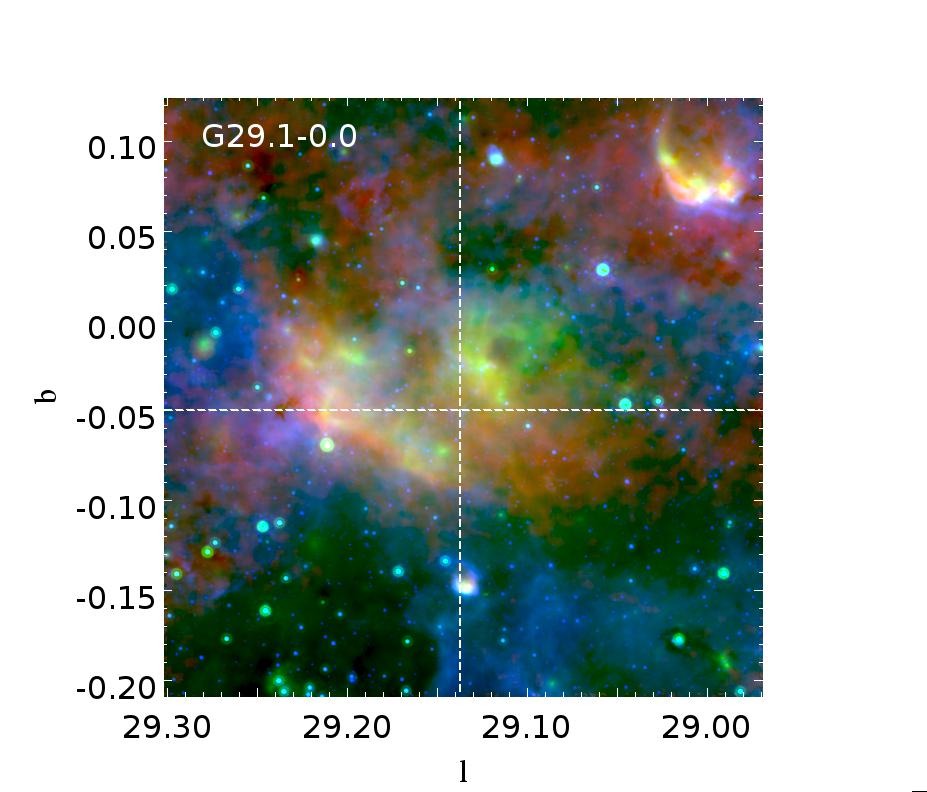}
\hspace*{-1.5truecm}
\includegraphics[width=9cm,height=7.4cm,angle=0]{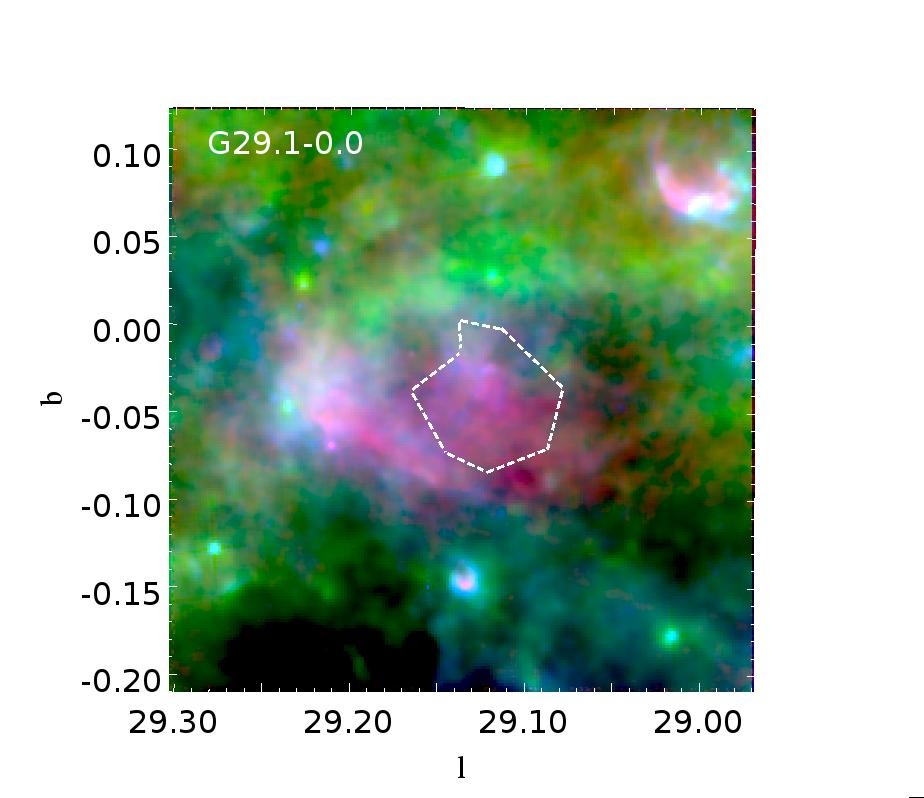}\\
\setcounter{figure}{0}
\caption{3-color images of the sample evolved \ion{H}{2} regions. Left panel: MAGPIS 20 cm (red), MIPS 24 $\mu$m (green), IRAC 8 $\mu$m (blue) data. Dashed lines indicate 
cuts in latitude and longitude operated to create the profiles in Appendix~B. 
Right panel: MAGPIS 20 cm (red), SPIRE 250 $\mu$m (green), PACS 70 $\mu$m (blue) data. G29.1-0.0 and G29.2-0.0 are
shown in the same panel (third row). Areas enclosed by white dashed lines denote approximate apertures using MAGPIS 20 cm data.}
\end{figure*}

\newpage

\begin{figure*}[h]
\centering
\includegraphics[width=9cm,height=7.4cm,angle=0]{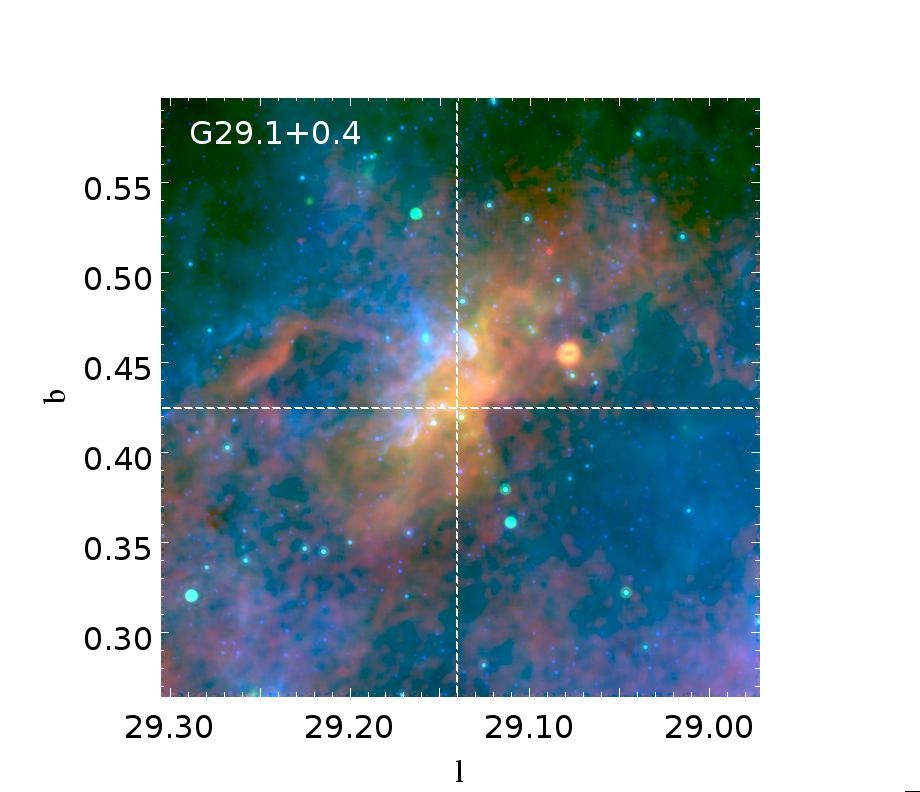}
\hspace*{-1.5truecm}
\includegraphics[width=9cm,height=7.4cm,angle=0]{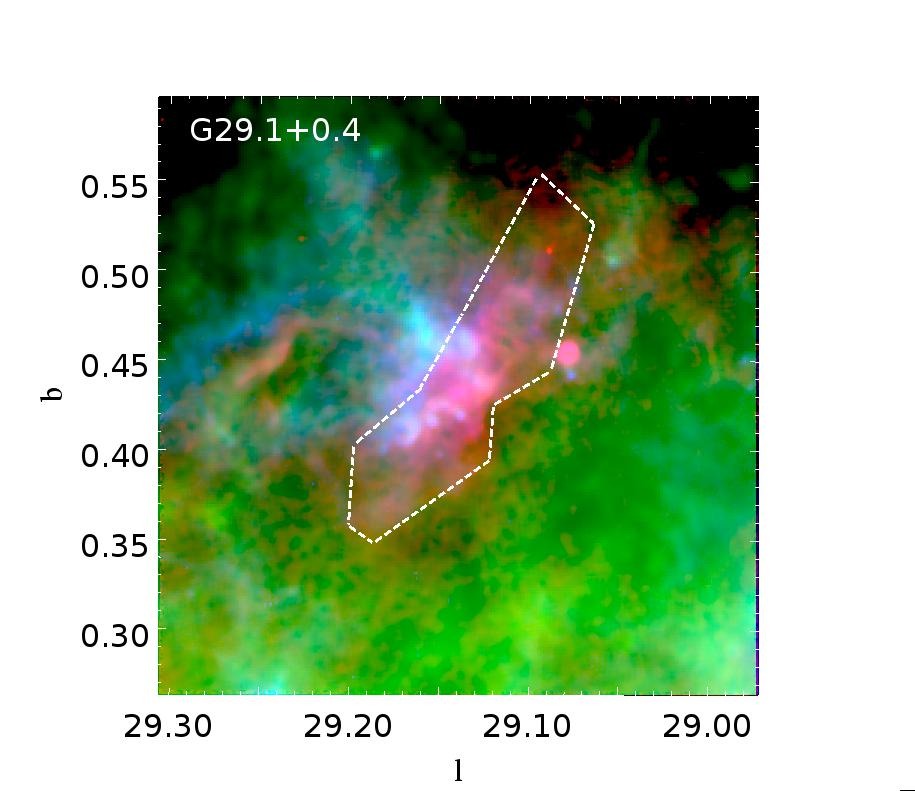}\\
\includegraphics[width=9cm,height=7.4cm,angle=0]{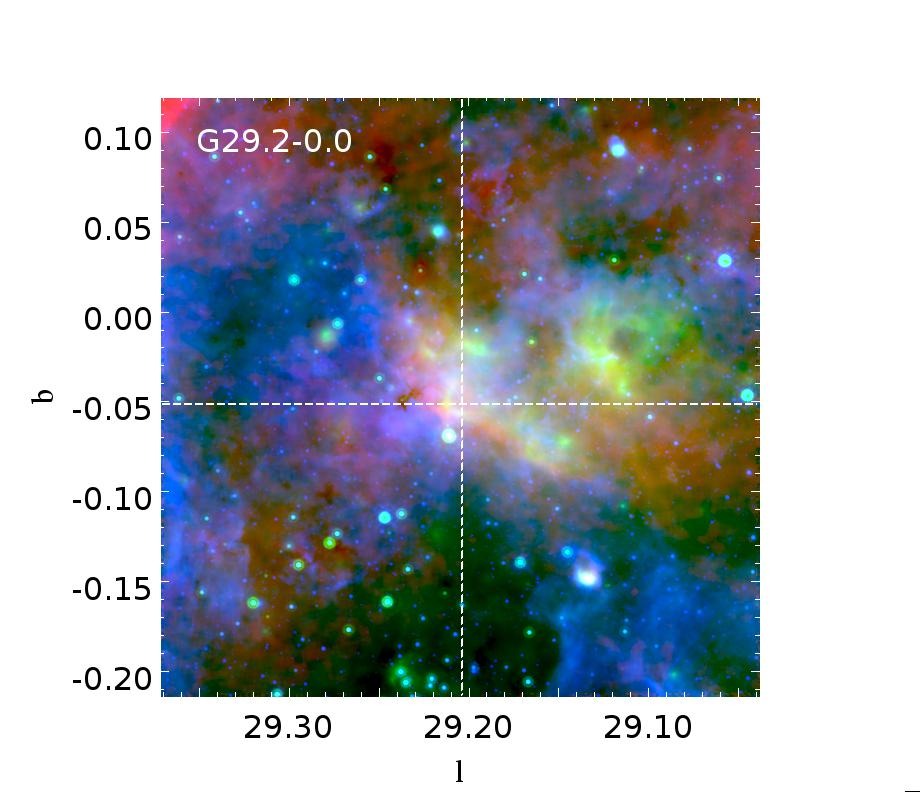}
\hspace*{-1.5truecm}
\includegraphics[width=9cm,height=7.4cm,angle=0]{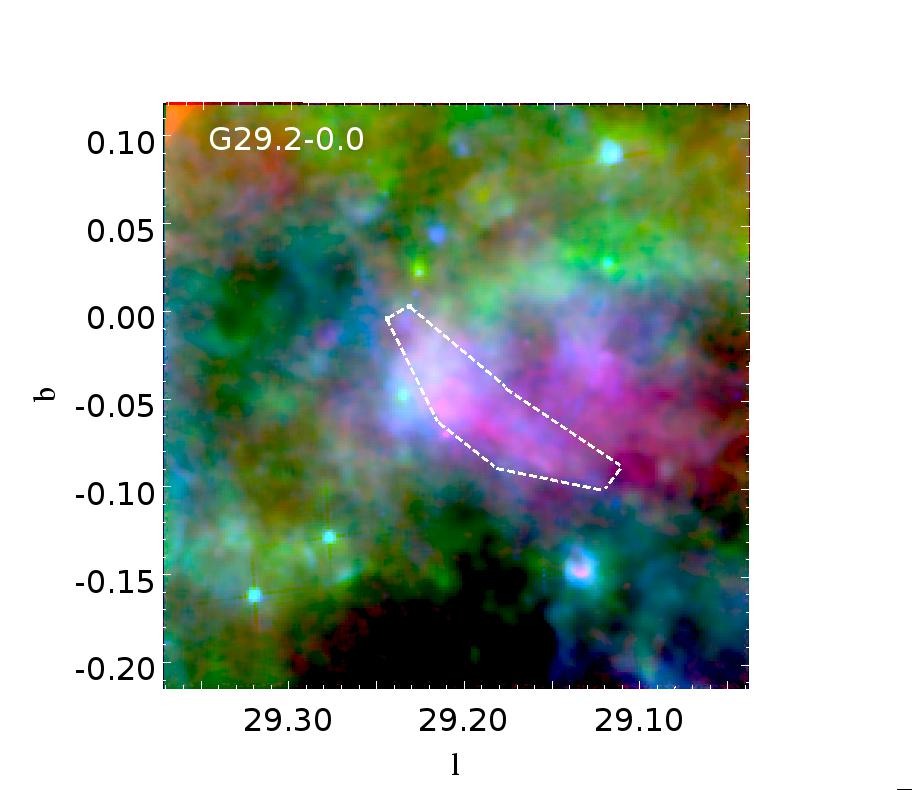}\\
\includegraphics[width=9cm,height=7.4cm,angle=0]{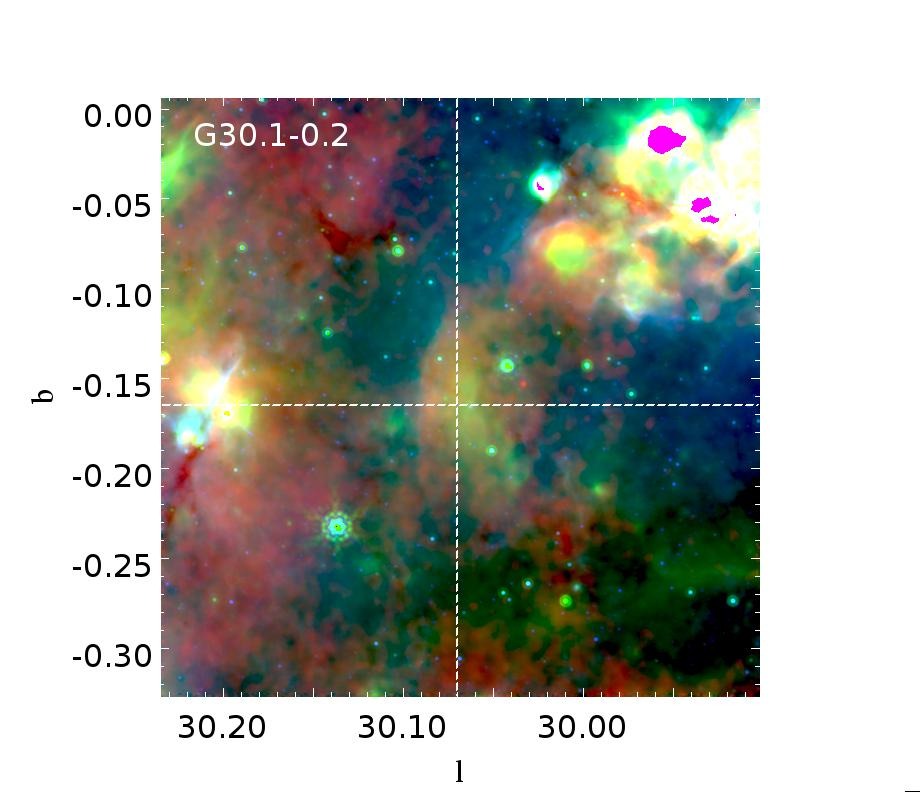}
\hspace*{-1.5truecm}
\includegraphics[width=9cm,height=7.4cm,angle=0]{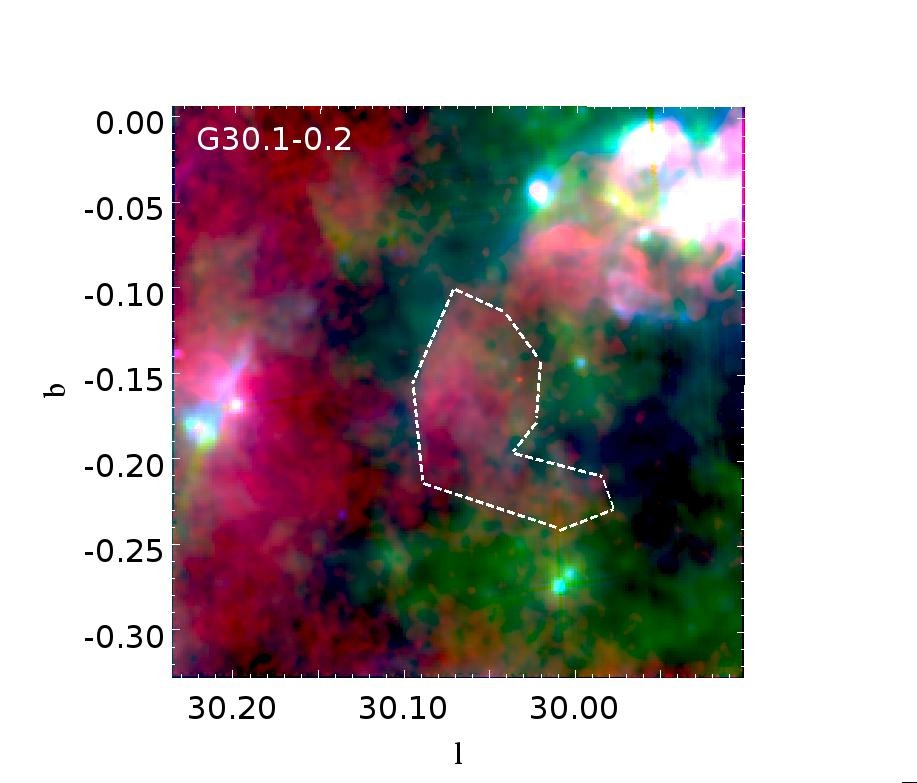}\\ 
\setcounter{figure}{0}
\caption[]{continued}
\end{figure*}

\newpage

\begin{figure*}[h]
\centering
\includegraphics[width=9cm,height=7.4cm,angle=0]{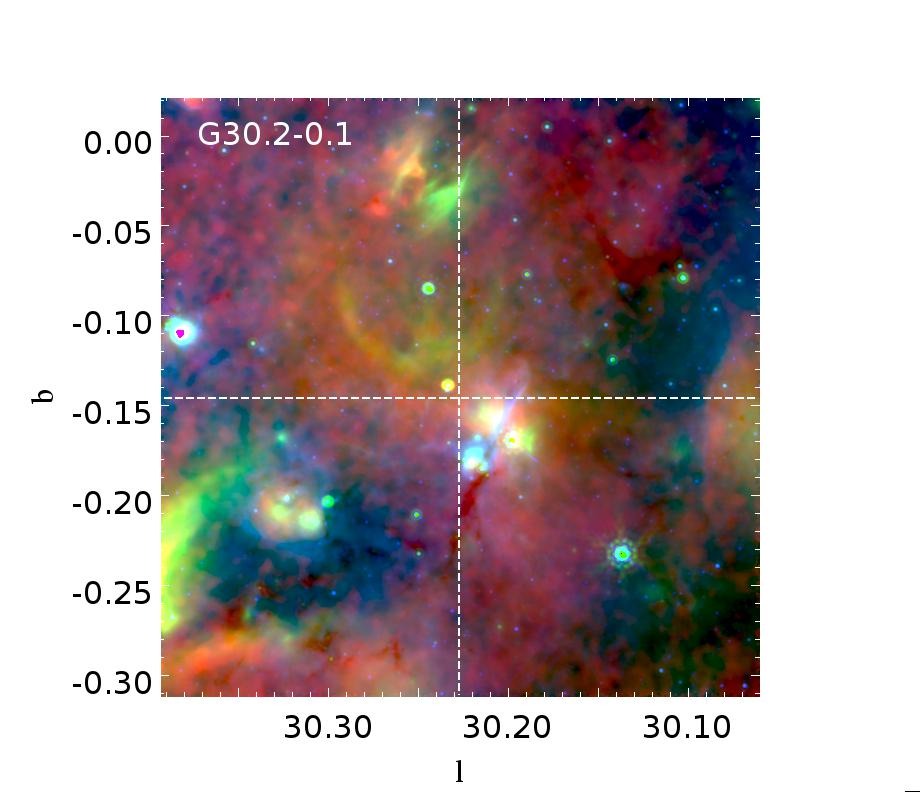}
\hspace*{-1.5truecm}
\includegraphics[width=9cm,height=7.4cm,angle=0]{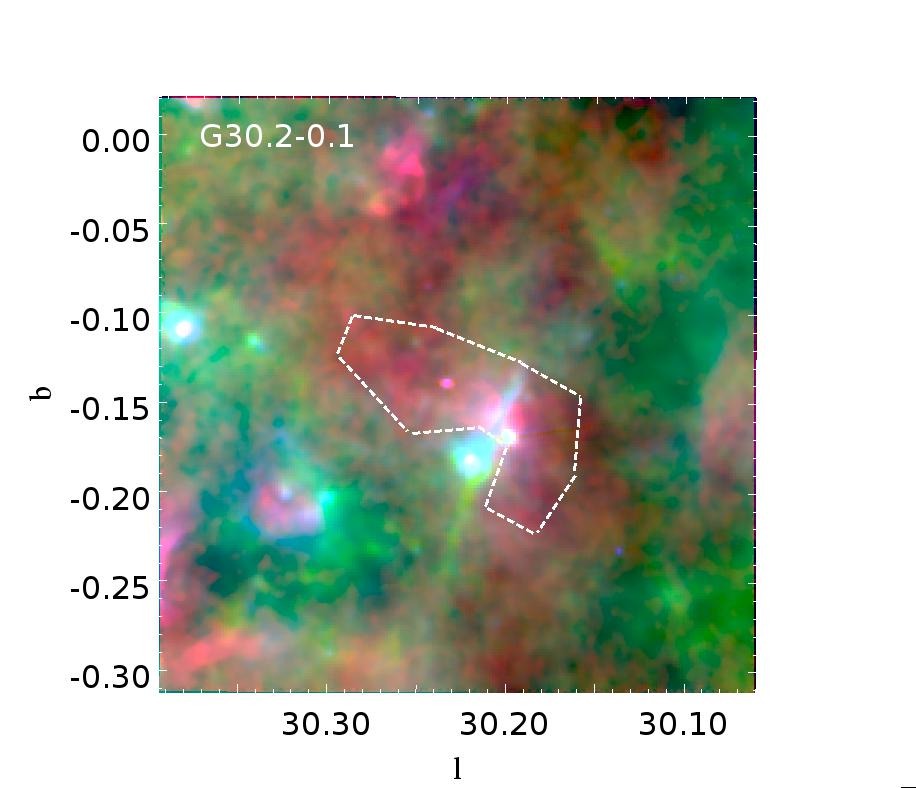}\\
\includegraphics[width=9cm,height=7.4cm,angle=0]{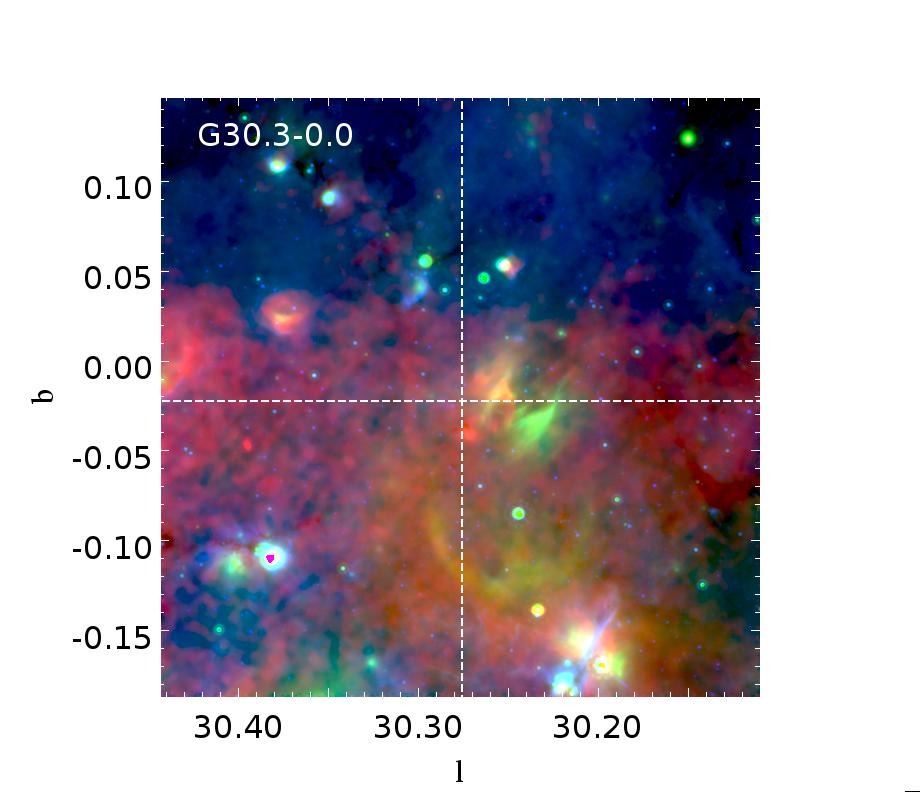}
\hspace*{-1.5truecm}
\includegraphics[width=9cm,height=7.4cm,angle=0]{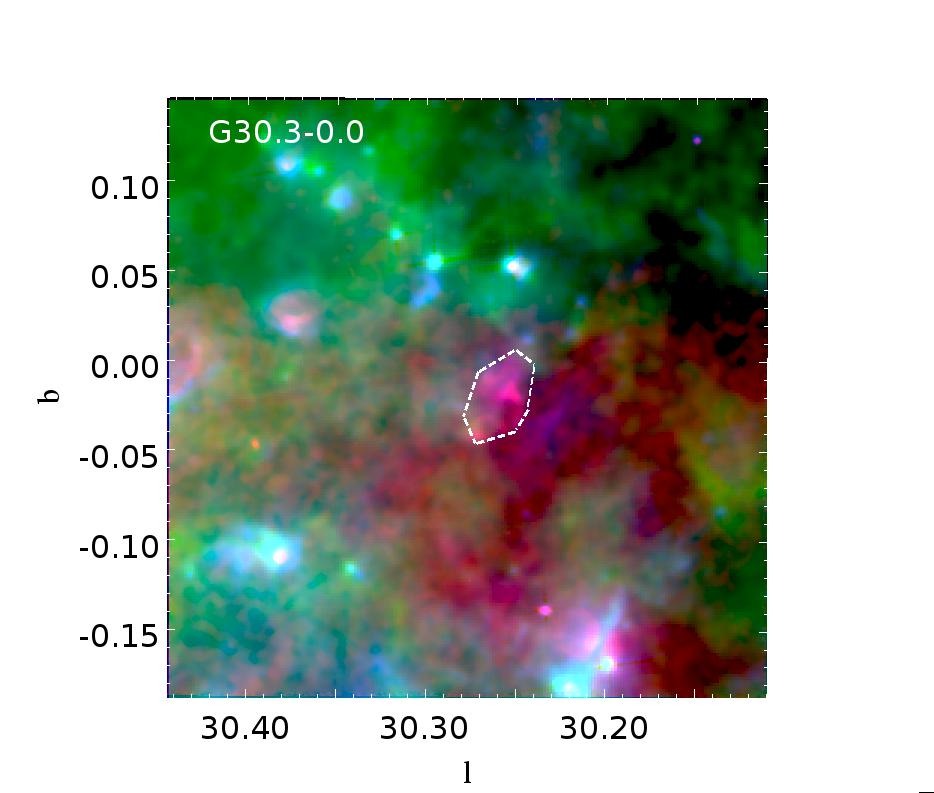}\\
\includegraphics[width=9cm,height=7.4cm,angle=0]{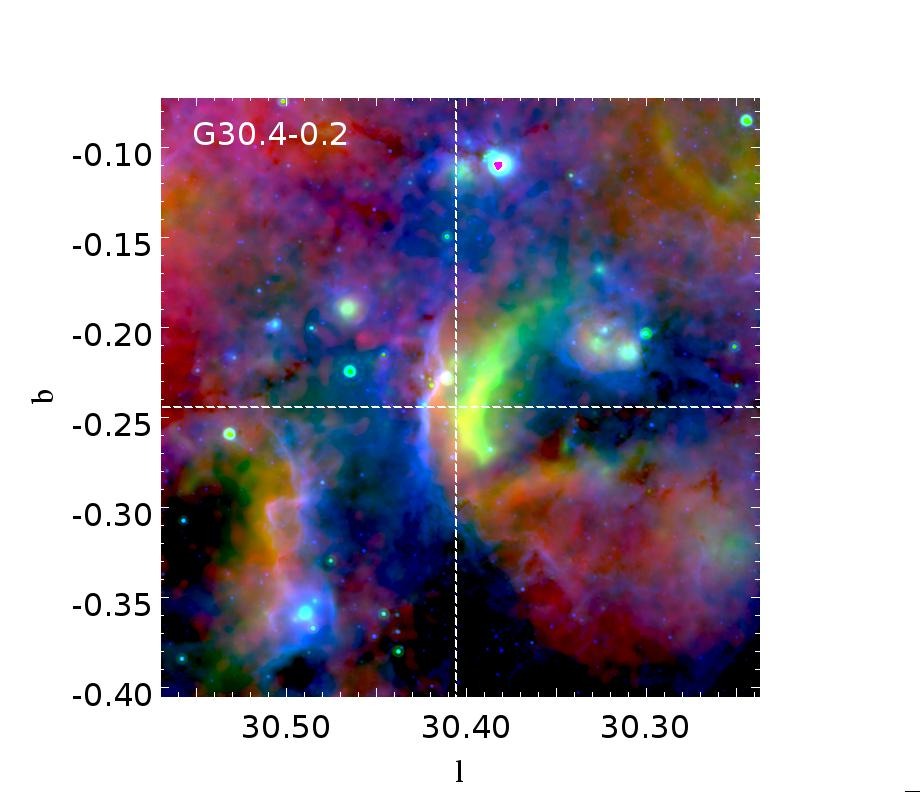}
\hspace*{-1.5truecm}
\includegraphics[width=9cm,height=7.4cm,angle=0]{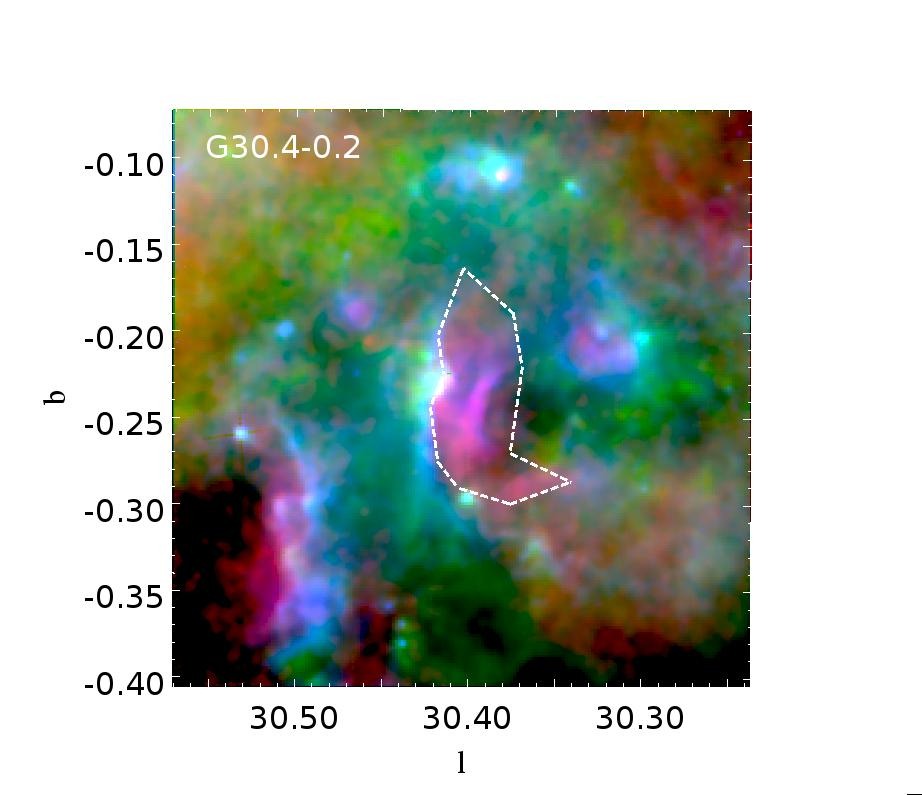}\\
\setcounter{figure}{0}
\caption[]{continued}
\end{figure*}

\newpage

\begin{figure*}[h]
\centering
\includegraphics[width=9cm,height=7.4cm,angle=0]{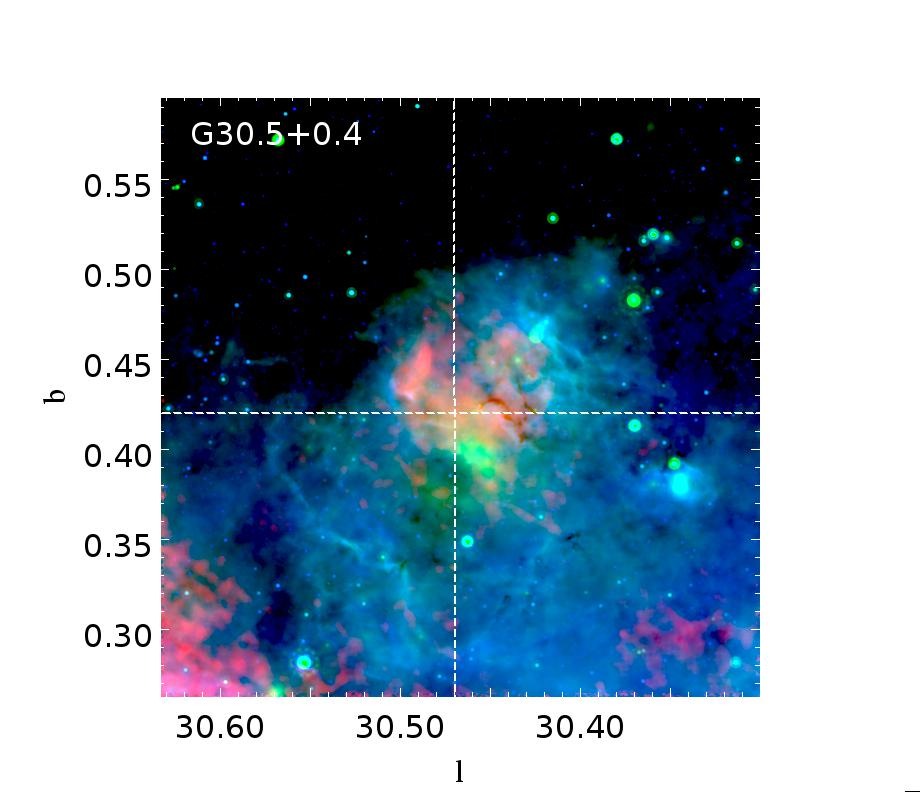}
\hspace*{-1.5truecm}
\includegraphics[width=9cm,height=7.4cm,angle=0]{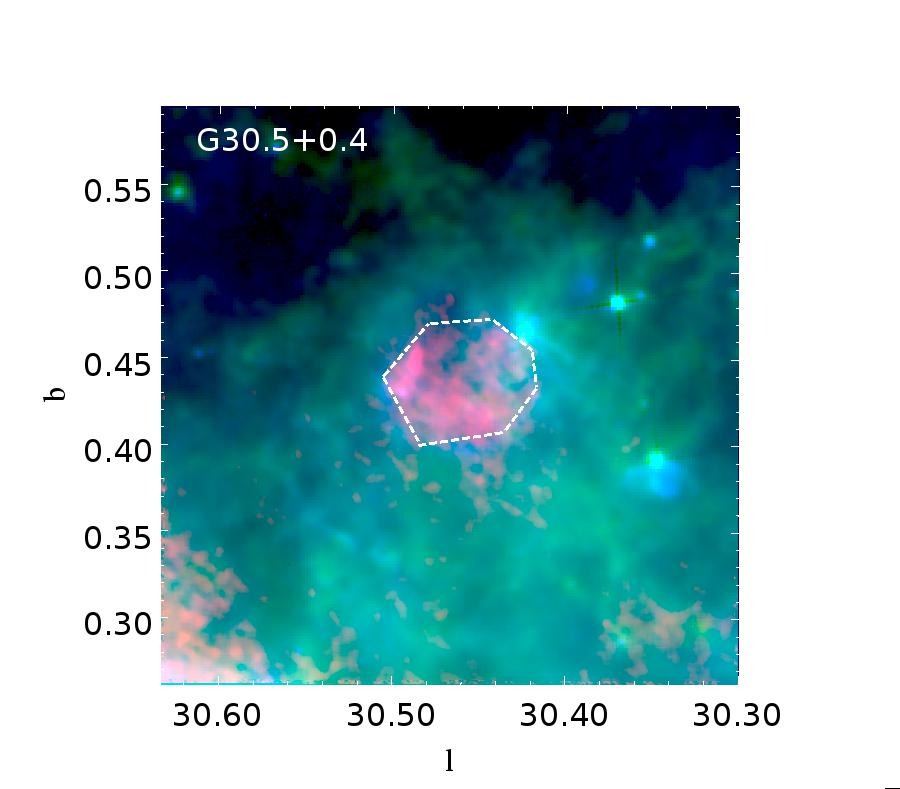}\\
\includegraphics[width=9cm,height=7.4cm,angle=0]{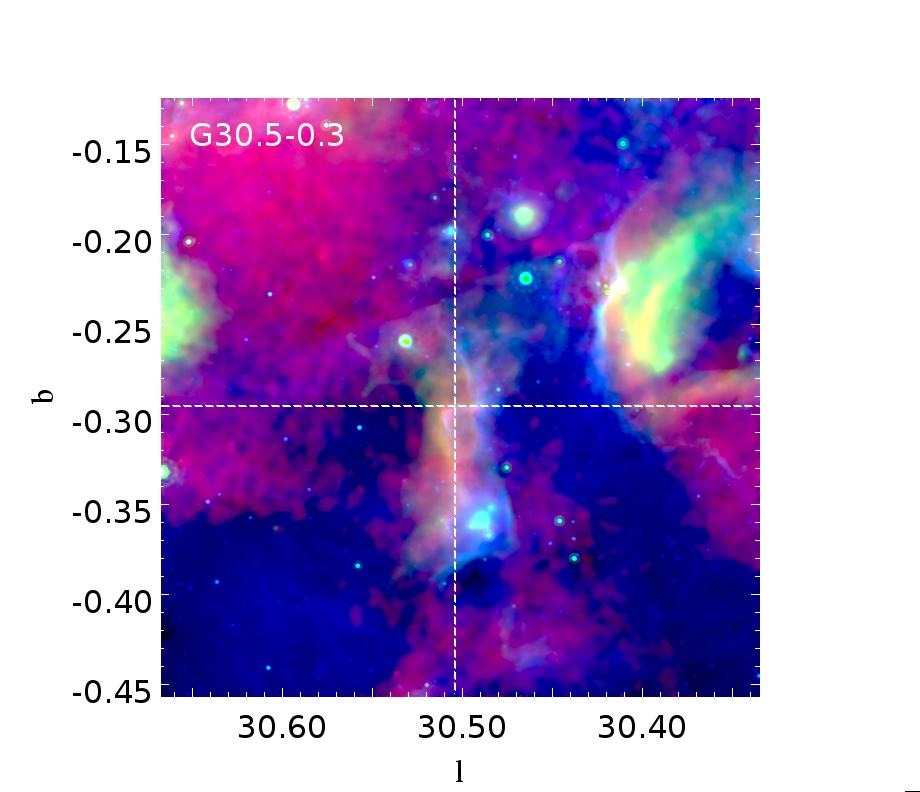}
\hspace*{-1.5truecm}
\includegraphics[width=9cm,height=7.4cm,angle=0]{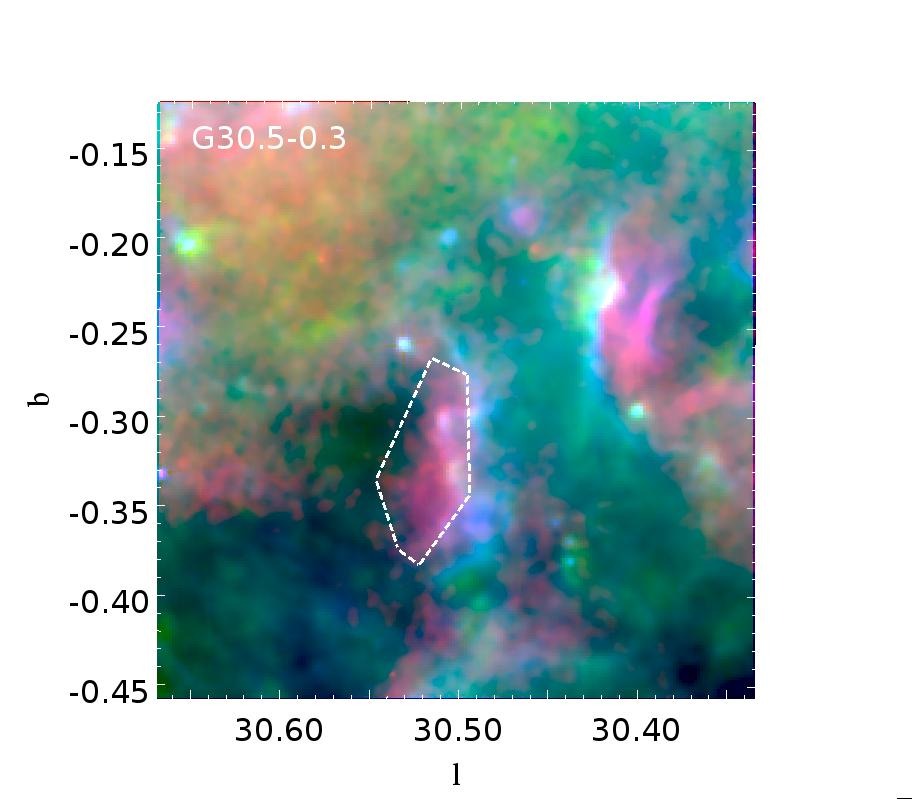}\\
\includegraphics[width=9cm,height=7.4cm,angle=0]{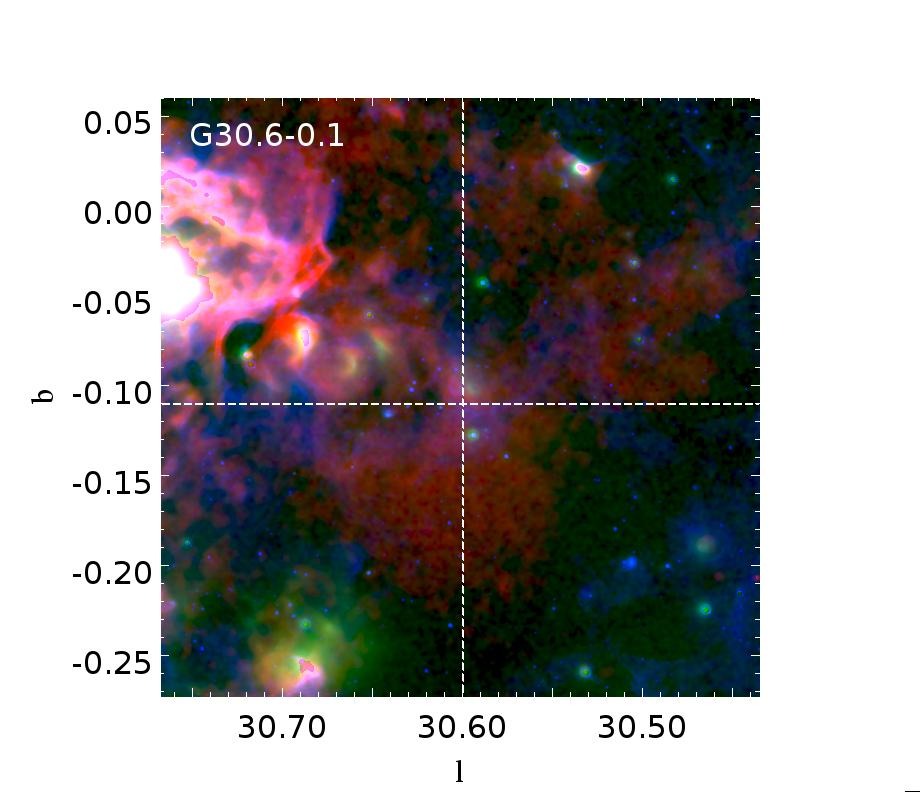}
\hspace*{-1.5truecm}
\includegraphics[width=9cm,height=7.4cm,angle=0]{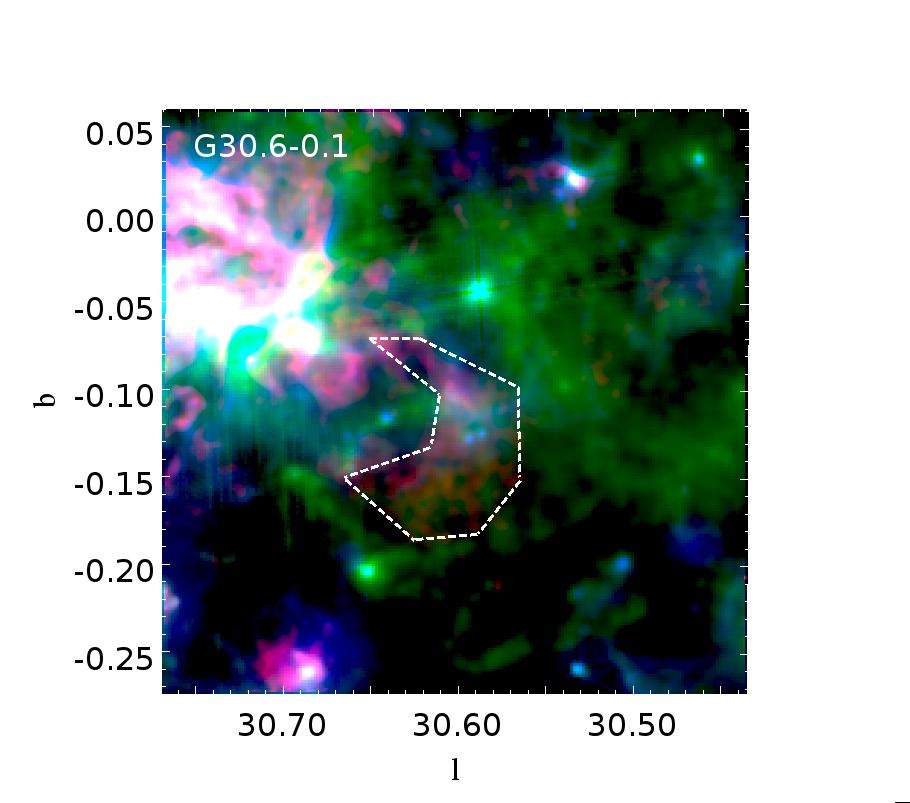}\\
\setcounter{figure}{0}
\caption[]{continued}
\end{figure*}

\newpage

\begin{figure*}[h]
\centering
\includegraphics[width=9cm,height=7.4cm,angle=0]{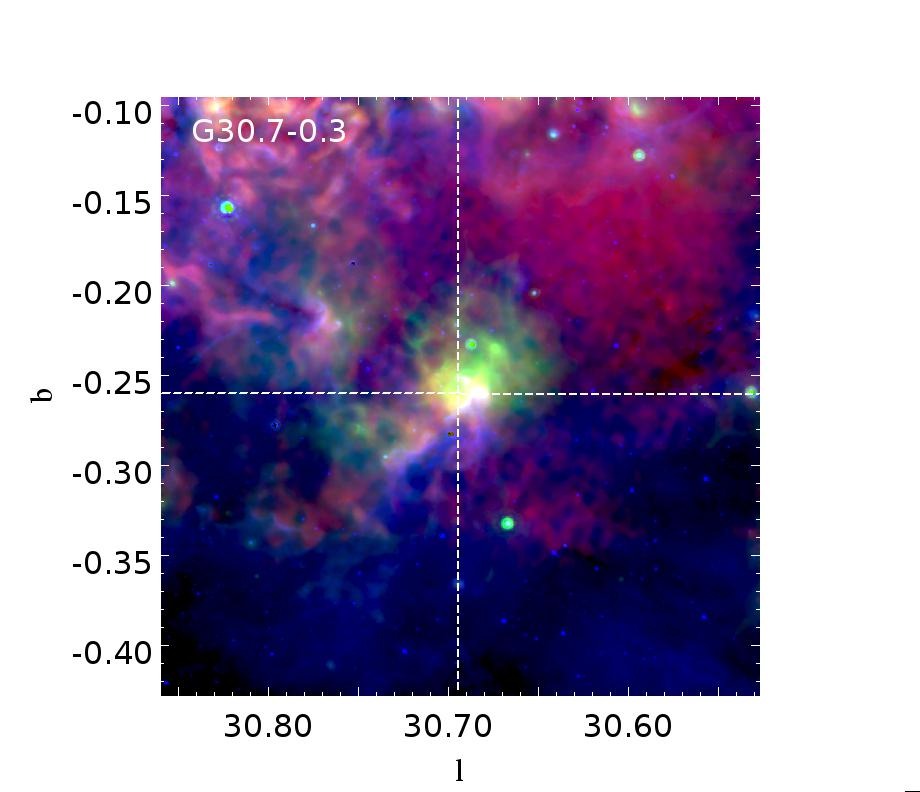}
\hspace*{-1.5truecm}
\includegraphics[width=9cm,height=7.4cm,angle=0]{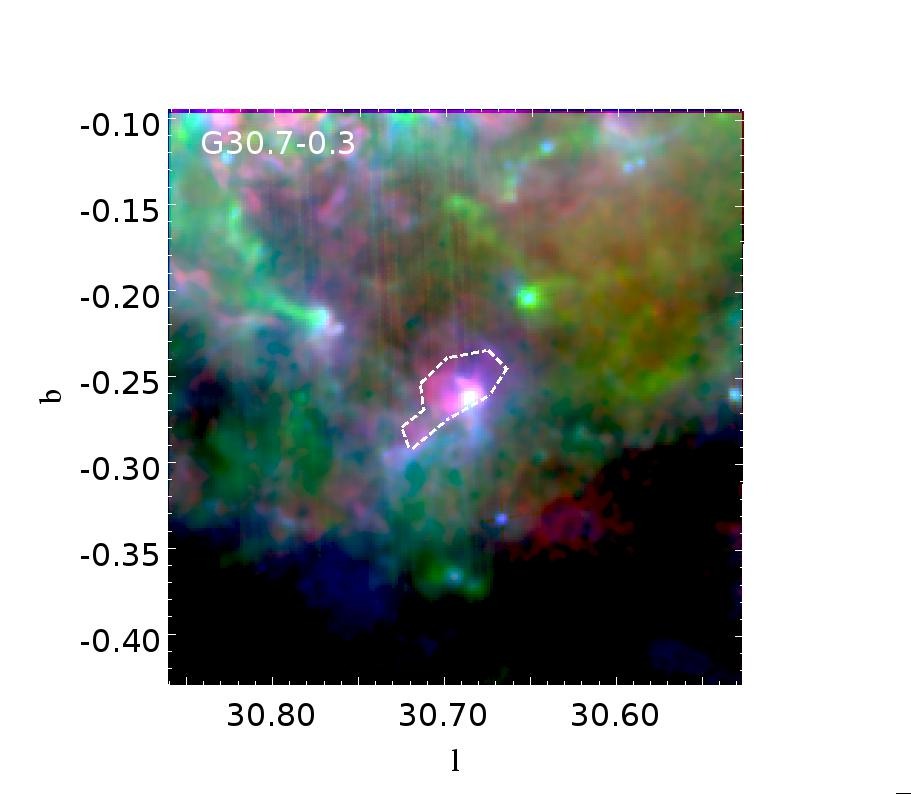}\\
\includegraphics[width=9cm,height=7.4cm,angle=0]{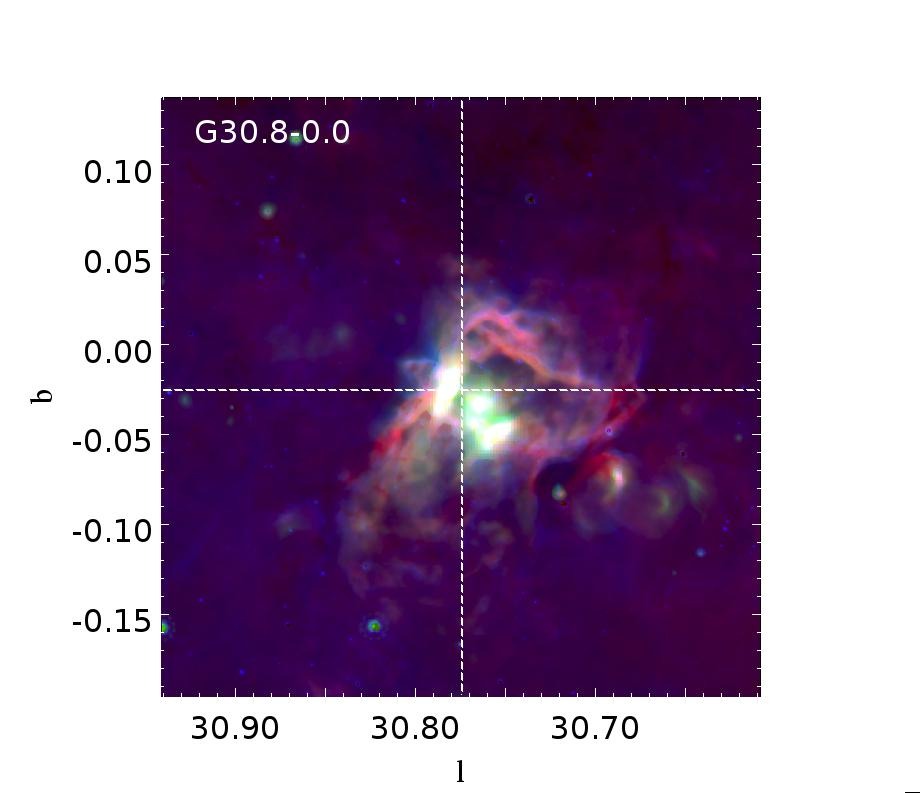}
\hspace*{-1.5truecm}
\includegraphics[width=9cm,height=7.4cm,angle=0]{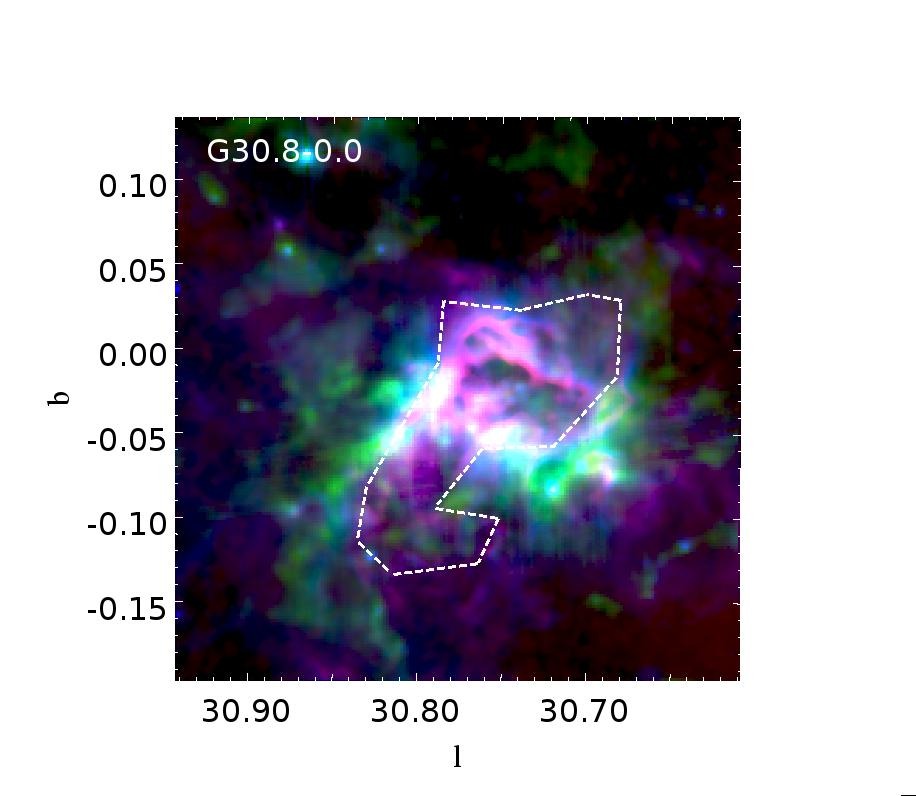}\\
\includegraphics[width=9cm,height=7.4cm,angle=0]{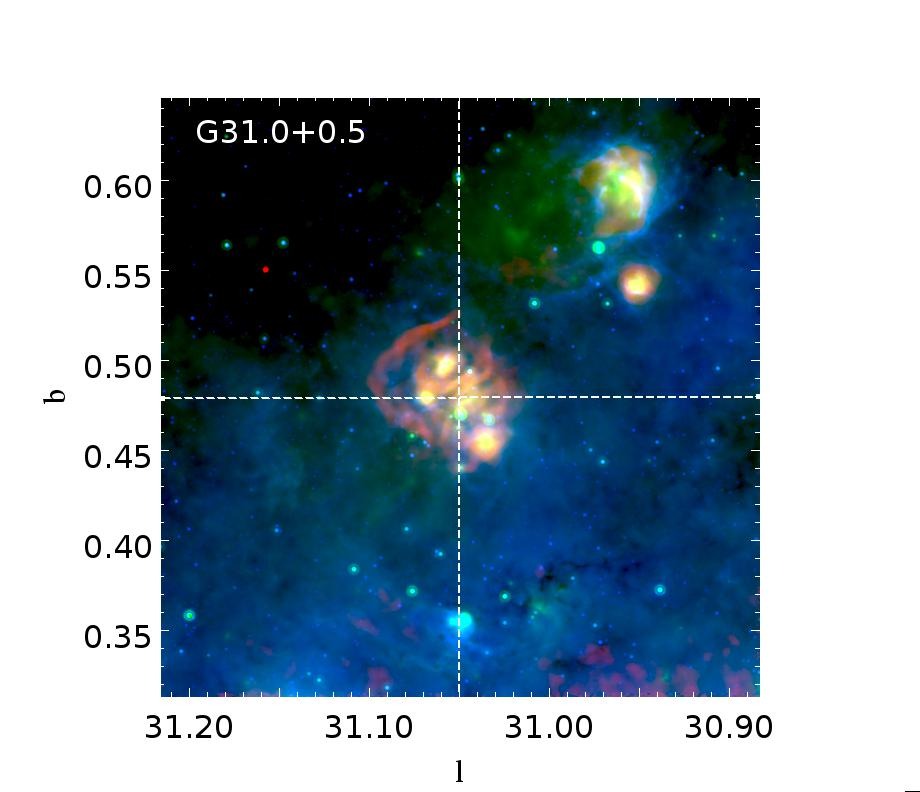}
\hspace*{-1.5truecm}
\includegraphics[width=9cm,height=7.4cm,angle=0]{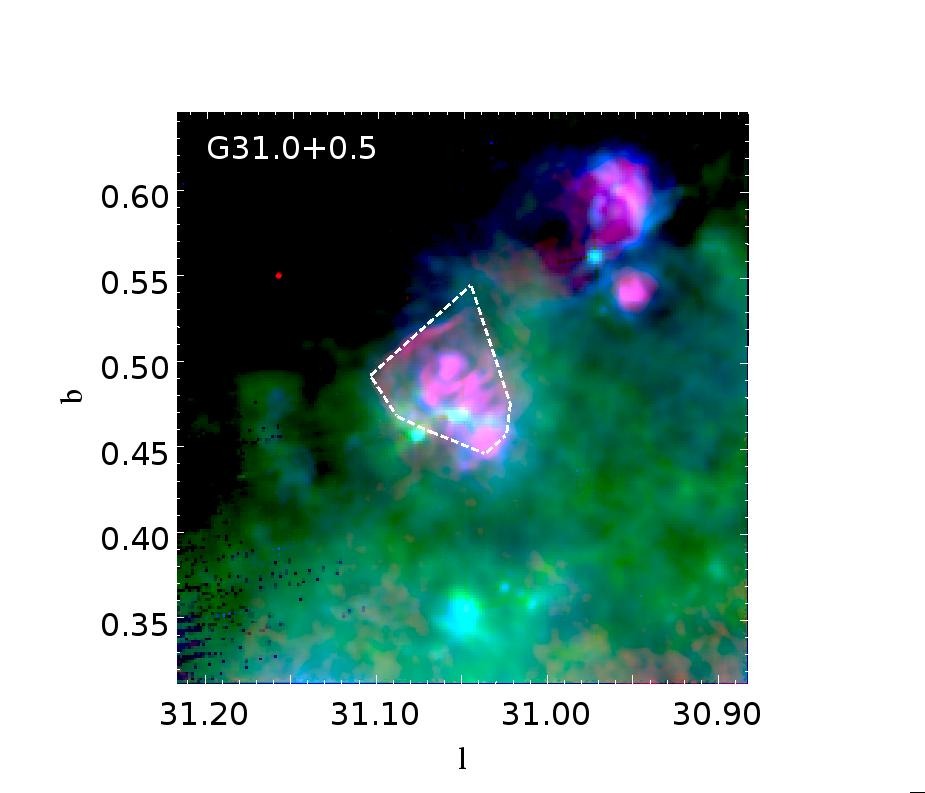}\\
\setcounter{figure}{0}
\caption[]{continued}
\end{figure*}

\begin{figure*}[h]
\centering
\includegraphics[width=9cm,height=7.4cm,angle=0]{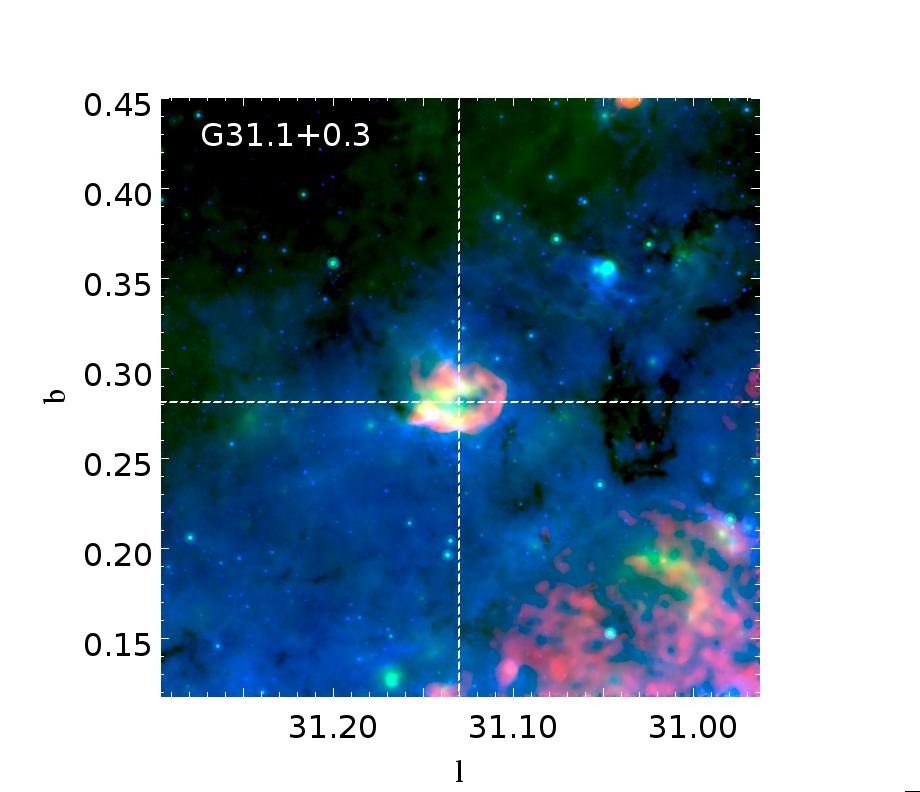}
\hspace*{-1.5truecm}
\includegraphics[width=9cm,height=7.4cm,angle=0]{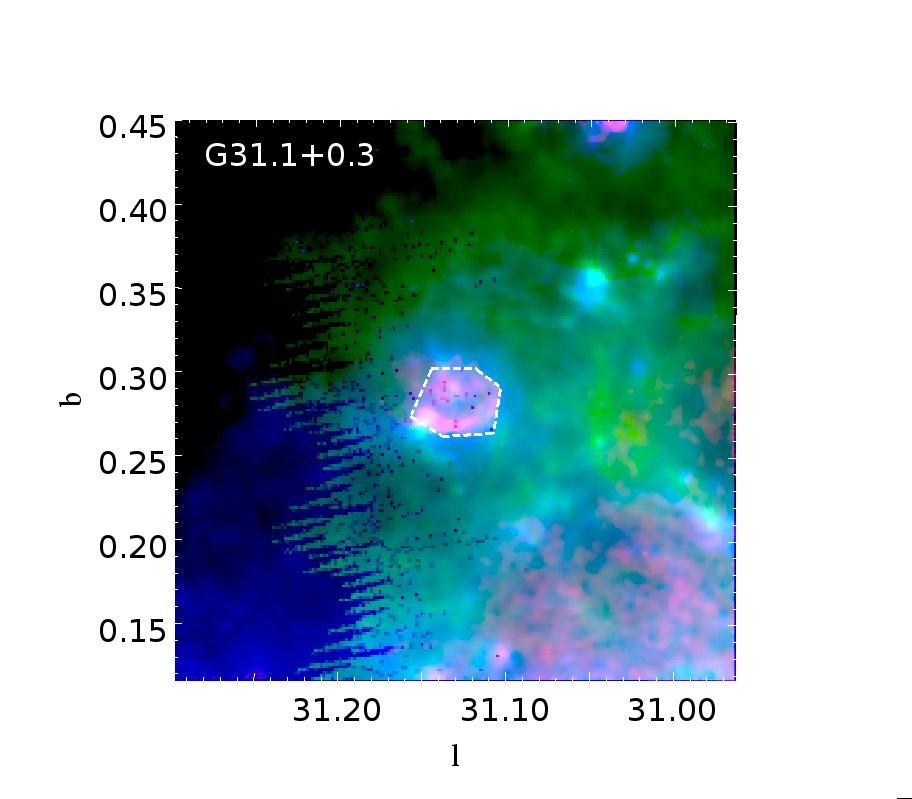}\\
\setcounter{figure}{0}
\caption[]{continued}
\end{figure*}

\newpage
\appendix
\renewcommand{\thefigure}{B\arabic{figure}}
\setcounter{figure}{0}
\section{Appendix B}

\begin{figure*}[h]
\centering
\includegraphics[width=6cm,height=7cm,angle=90]{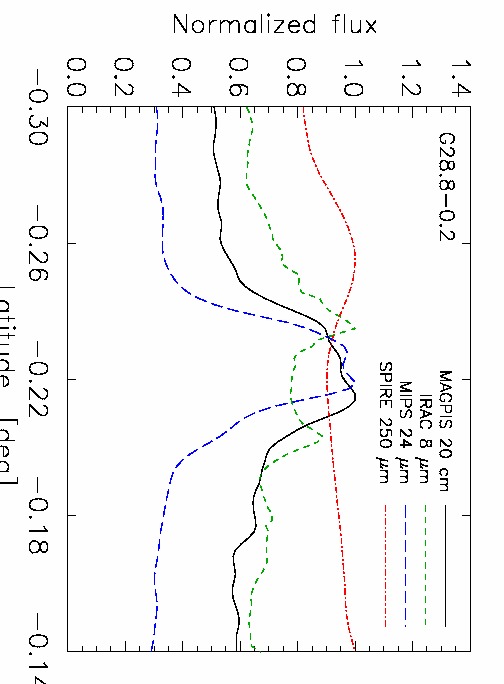}
\includegraphics[width=6cm,height=7cm,angle=90]{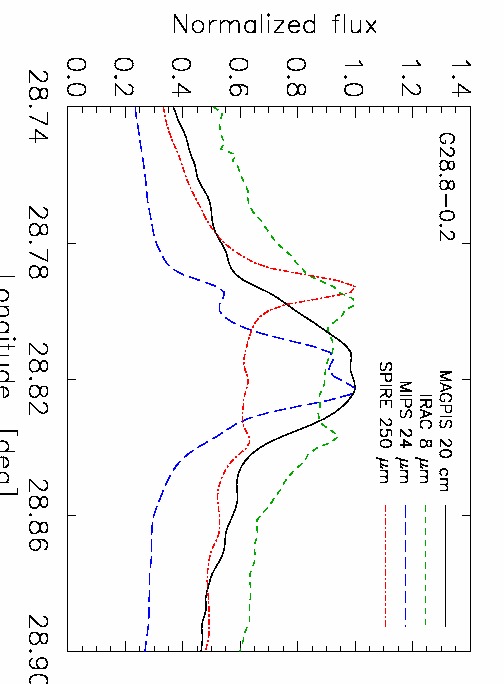}\\
\includegraphics[width=6cm,height=7cm,angle=90]{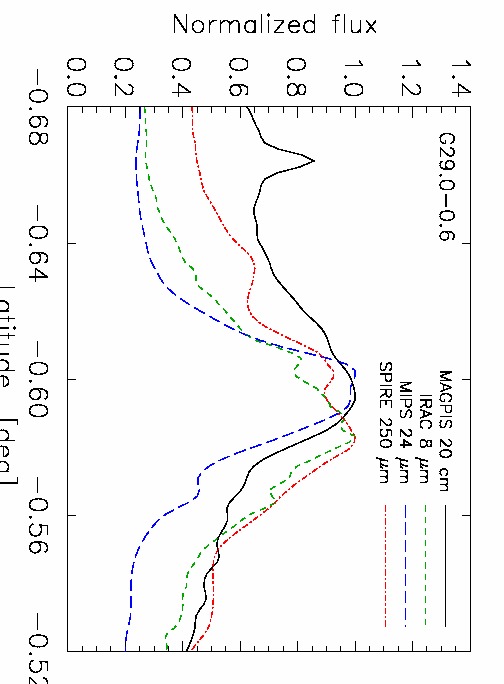}
\includegraphics[width=6cm,height=7cm,angle=90]{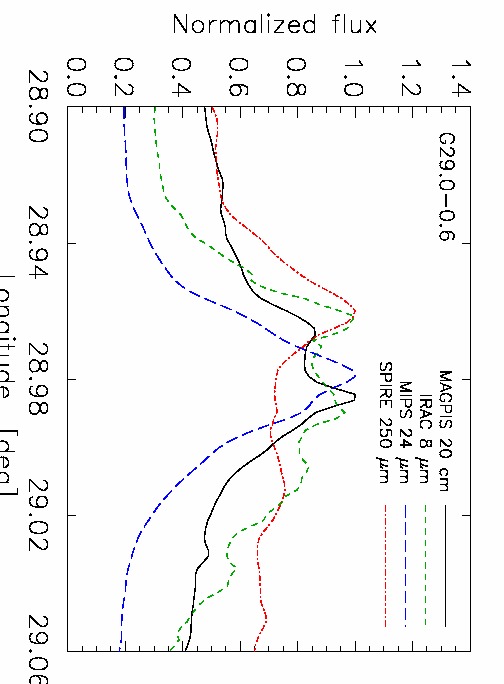}\\
\includegraphics[width=6cm,height=7cm,angle=90]{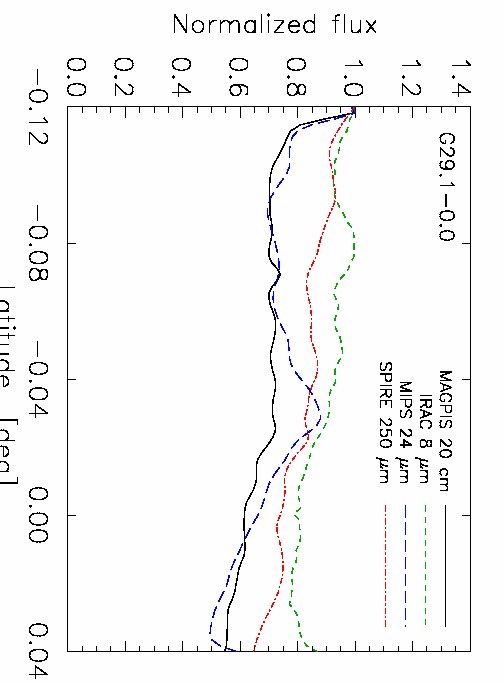}
\includegraphics[width=6cm,height=7cm,angle=90]{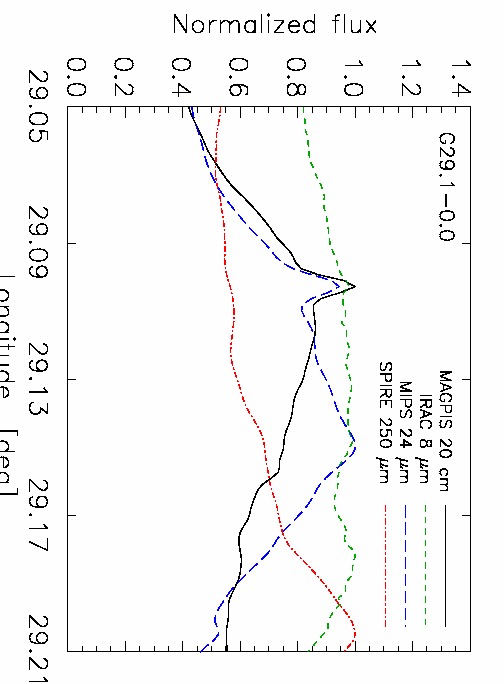}\\
\includegraphics[width=6cm,height=7cm,angle=90]{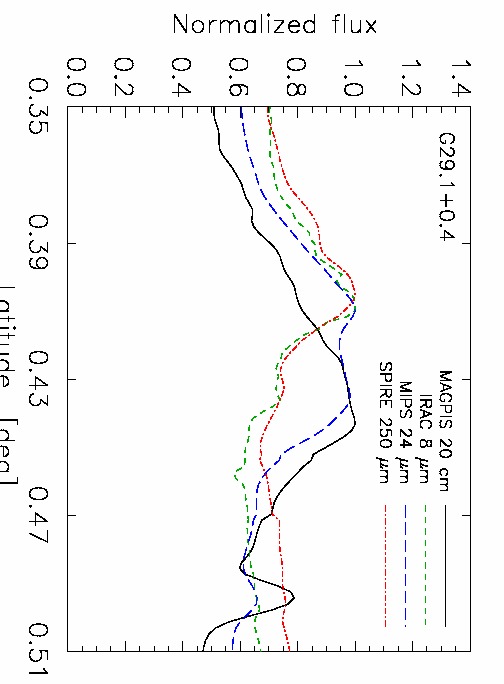}
\includegraphics[width=6cm,height=7cm,angle=90]{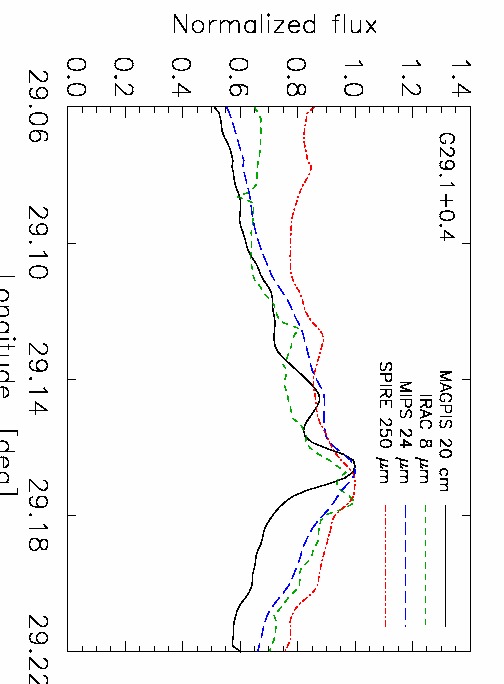}\\
\setcounter{figure}{0}
\caption{Longitude (right) and latitude (left) profiles, obtained by {\em{slicing}} MAGPIS 20 cm, IRAC 8 $\mu$m, MIPS 24 $\mu$m and SPIRE 250 $\mu$m postage stamp images
of the sources. All images are convolved to SPIRE 250 $\mu$m resolution (18") prior to {\em{slicing}}.}
\end{figure*}

\begin{figure*}[h]
\centering
\includegraphics[width=6cm,height=7cm,angle=90]{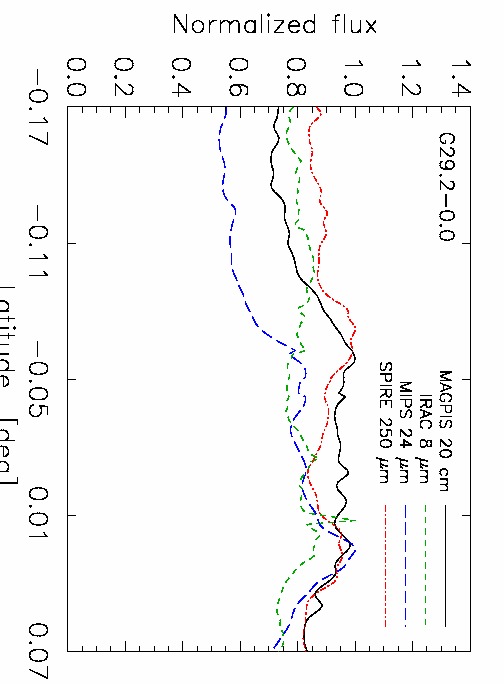}
\includegraphics[width=6cm,height=7cm,angle=90]{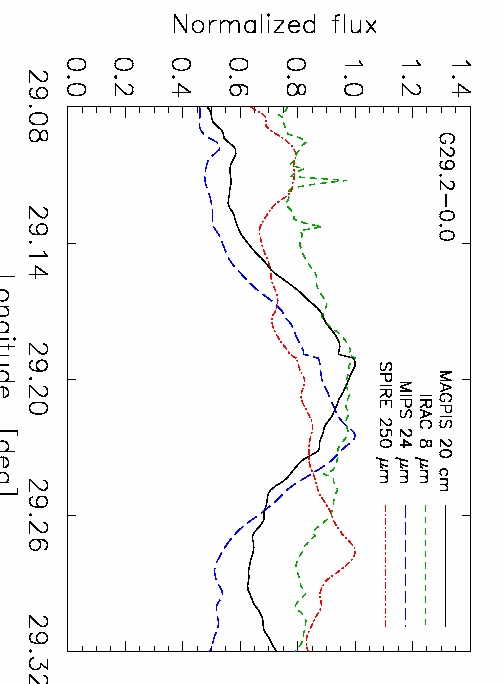}
\includegraphics[width=6cm,height=7cm,angle=90]{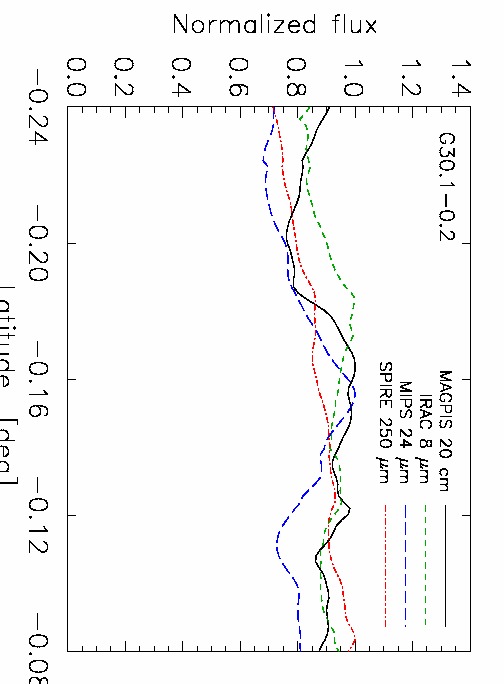}
\includegraphics[width=6cm,height=7cm,angle=90]{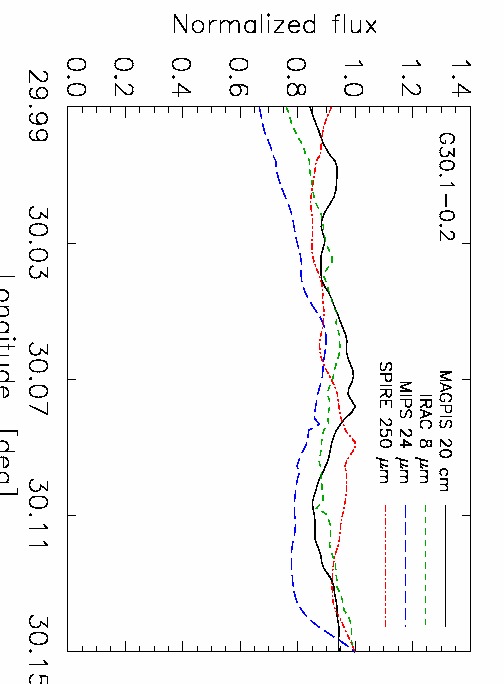}\\
\includegraphics[width=6cm,height=7cm,angle=90]{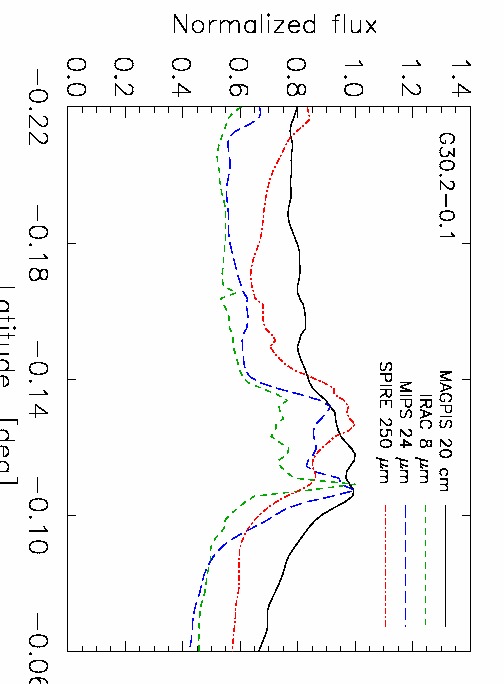}
\includegraphics[width=6cm,height=7cm,angle=90]{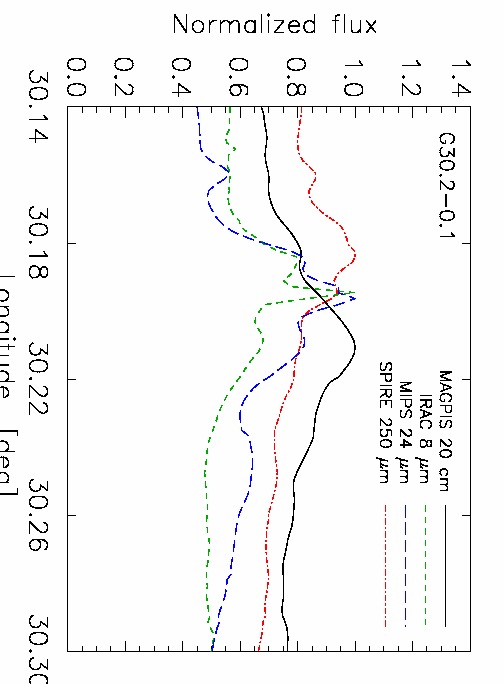}
\includegraphics[width=6cm,height=7cm,angle=90]{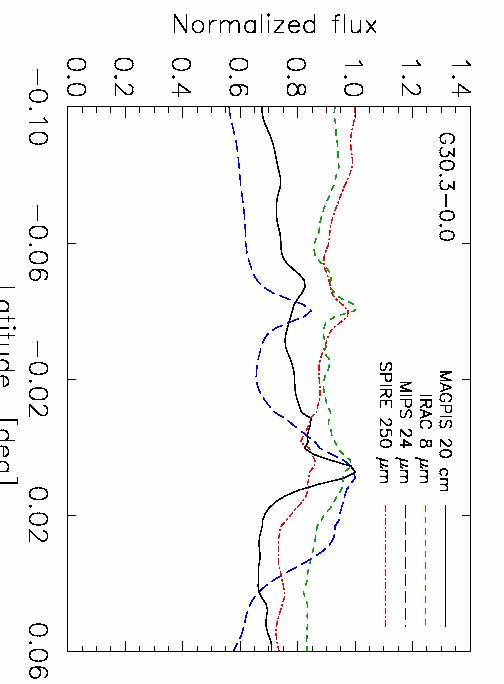}
\includegraphics[width=6cm,height=7cm,angle=90]{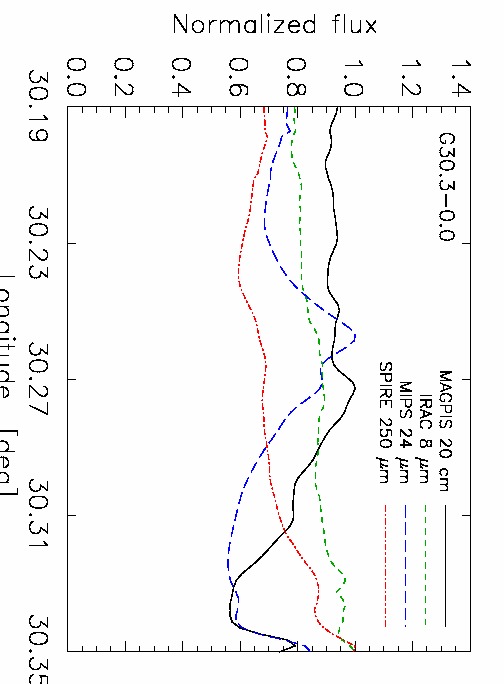}\\
\setcounter{figure}{0}
\caption[]{continued}
\end{figure*}

\begin{figure*}[h]
\centering
\includegraphics[width=6cm,height=7cm,angle=90]{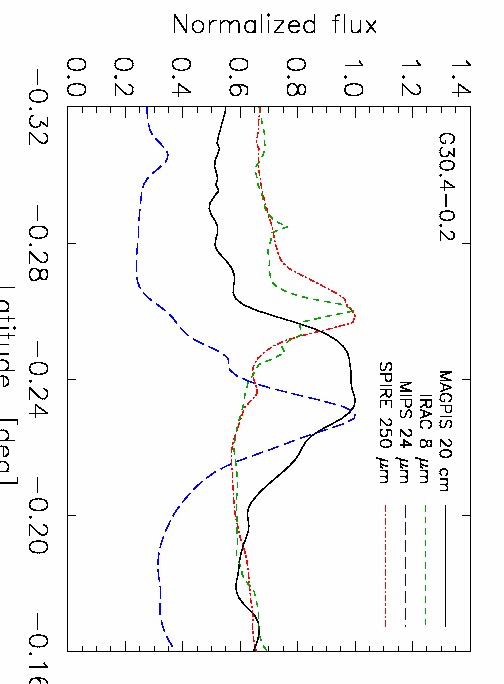}
\includegraphics[width=6cm,height=7cm,angle=90]{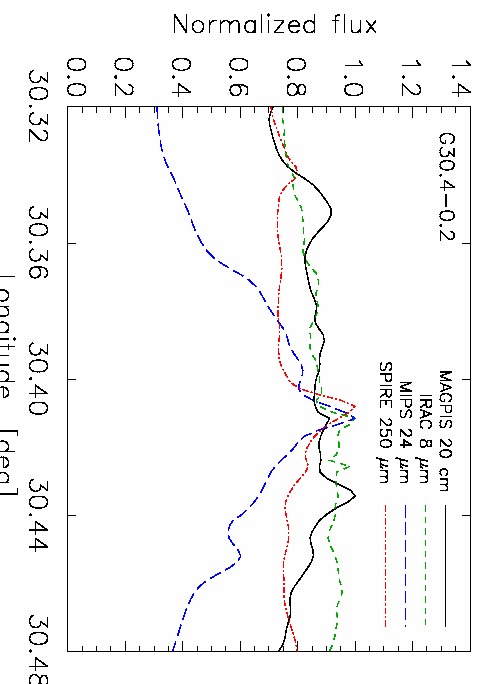}\\
\includegraphics[width=6cm,height=7cm,angle=90]{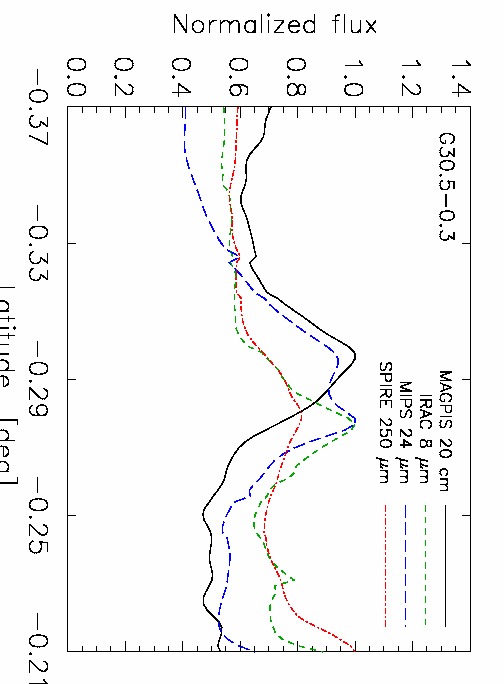}
\includegraphics[width=6cm,height=7cm,angle=90]{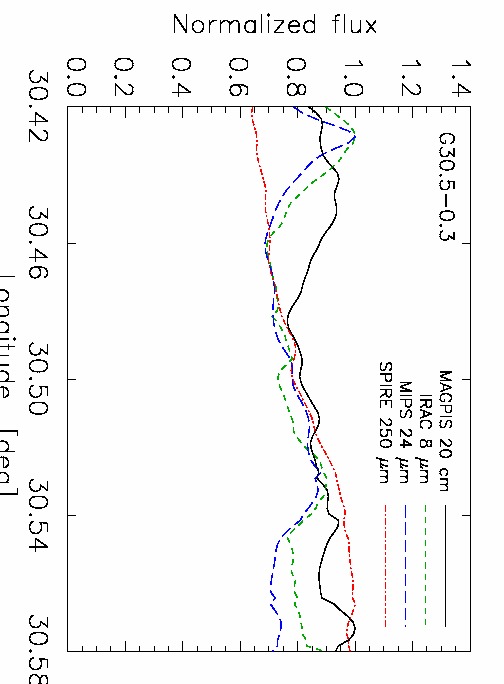}\\
\includegraphics[width=6cm,height=7cm,angle=90]{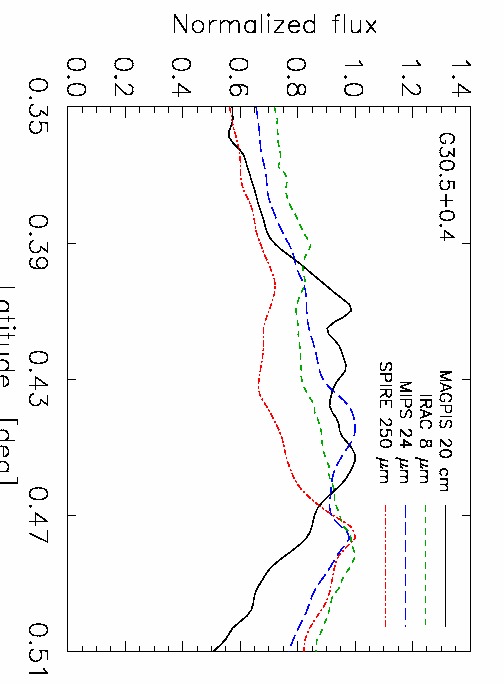}
\includegraphics[width=6cm,height=7cm,angle=90]{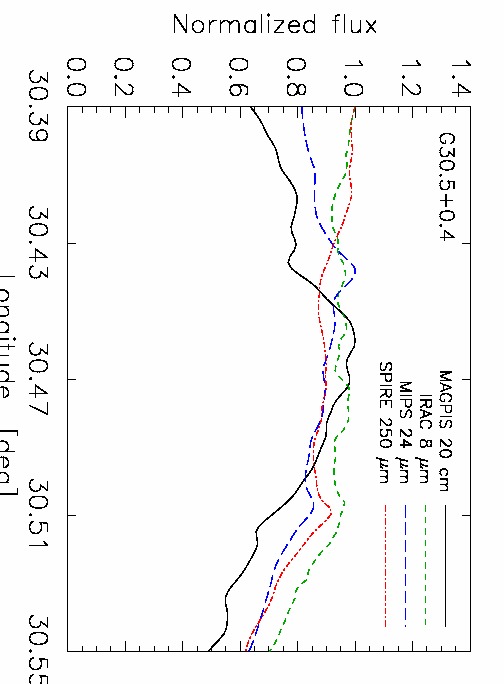}
\includegraphics[width=6cm,height=7cm,angle=90]{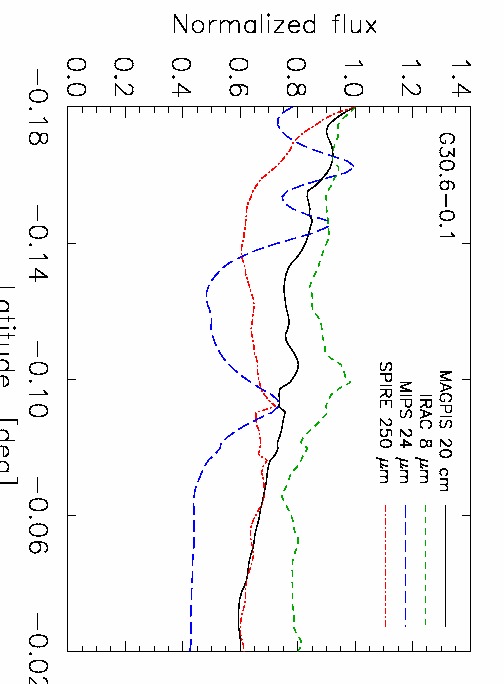}
\includegraphics[width=6cm,height=7cm,angle=90]{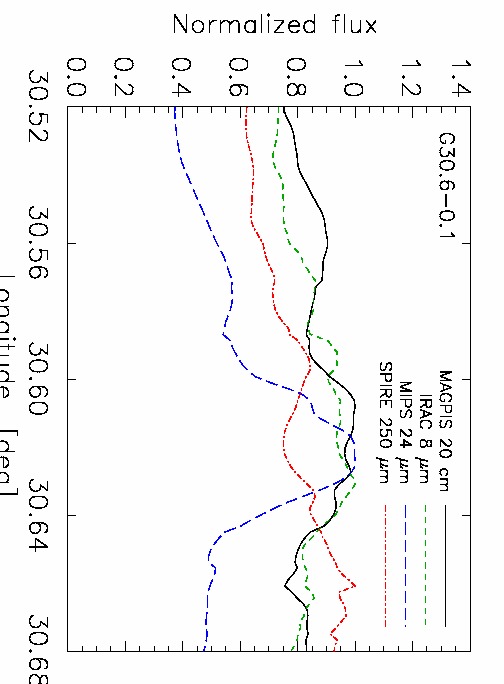}\\
\setcounter{figure}{0}
\caption[]{continued}
\end{figure*}

\begin{figure*}[h]
\centering
\includegraphics[width=6cm,height=7cm,angle=90]{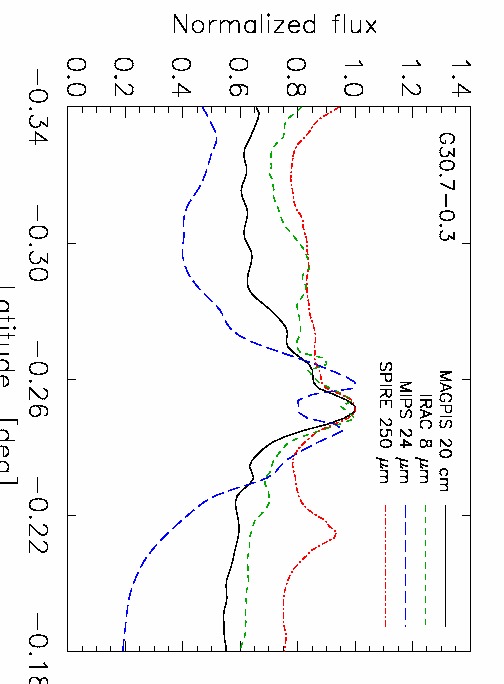}
\includegraphics[width=6cm,height=7cm,angle=90]{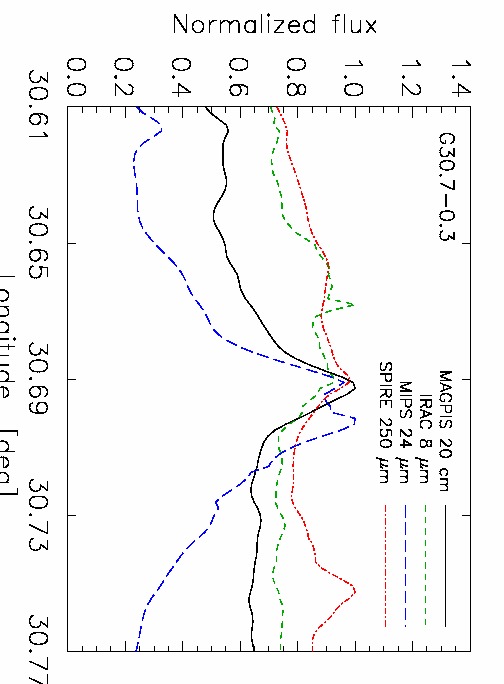}
\includegraphics[width=6cm,height=7cm,angle=90]{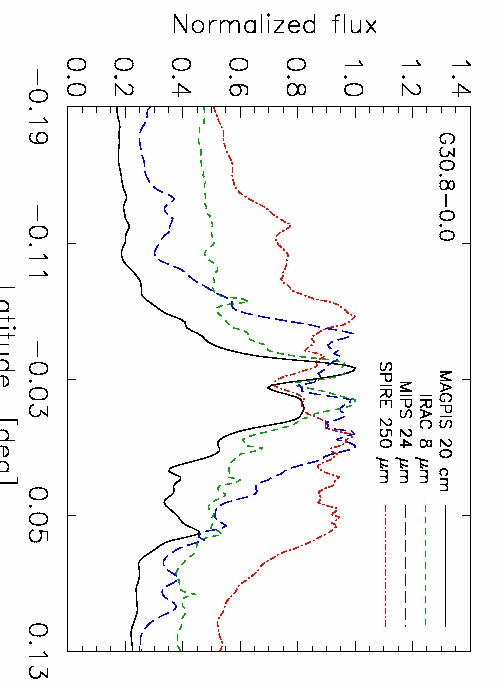}
\includegraphics[width=6cm,height=7cm,angle=90]{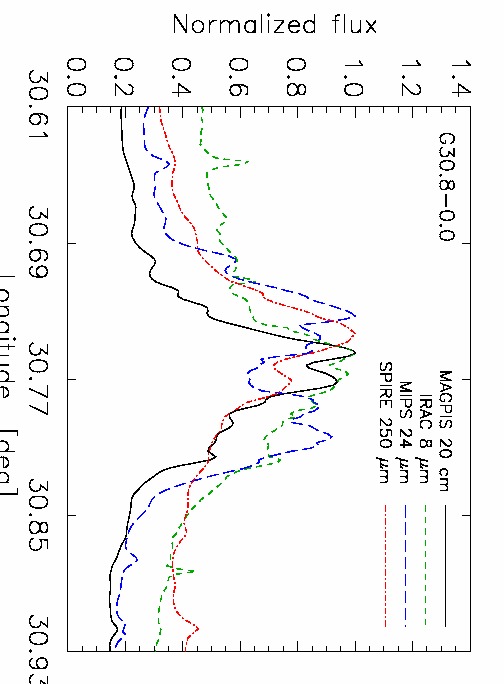}\\
\includegraphics[width=6cm,height=7cm,angle=90]{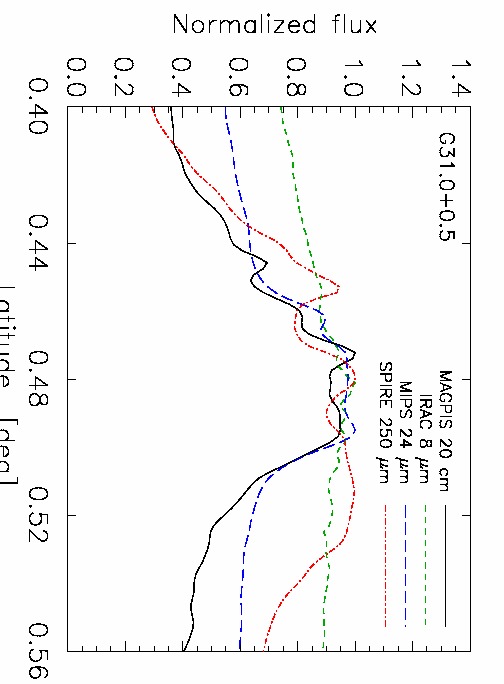}
\includegraphics[width=6cm,height=7cm,angle=90]{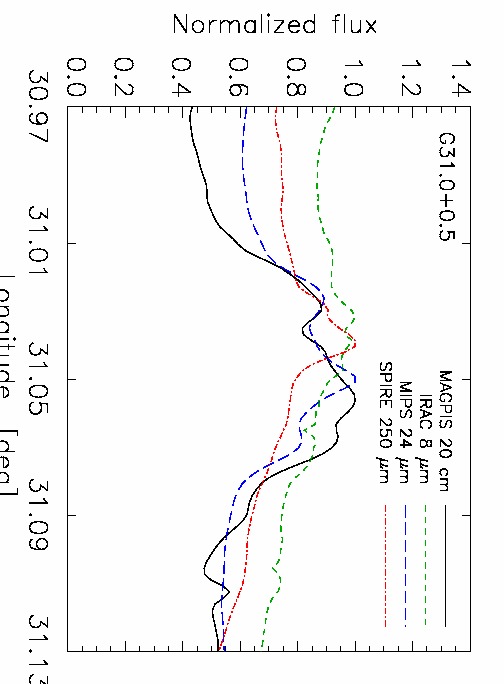}\\
\includegraphics[width=6cm,height=7cm,angle=90]{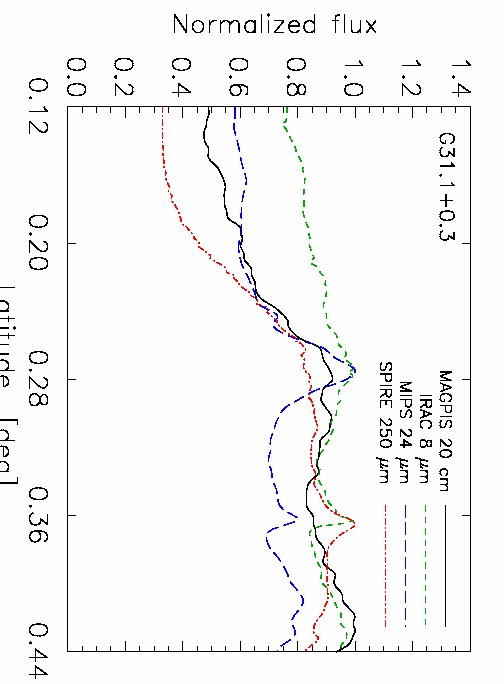}
\includegraphics[width=6cm,height=7cm,angle=90]{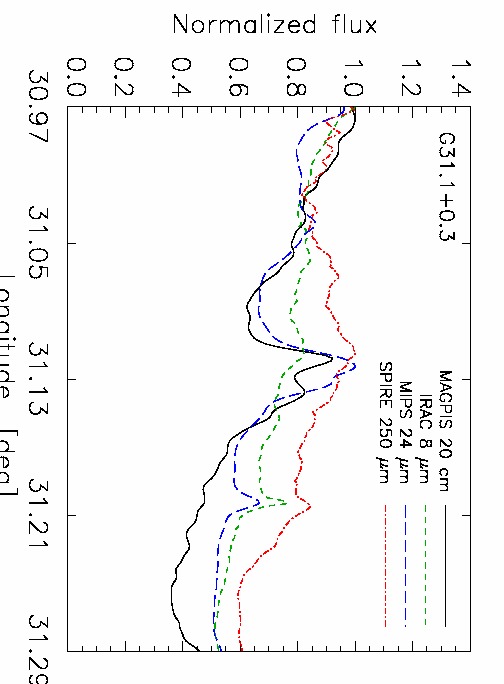}
\setcounter{figure}{0}
\caption[]{continued}
\end{figure*}

\end{document}